\documentclass[aps,prx,twocolumn,10pt]{revtex4-2}

\usepackage{array}[=2016-10-06]

\usepackage[utf8]{inputenc}
\usepackage[english]{babel}
\usepackage[T1]{fontenc}
\usepackage{xcolor}
\usepackage{microtype}
\usepackage{enumitem}
\usepackage[normalem]{ulem}
\usepackage[pdftex]{graphicx}
\usepackage{hyperref}
\usepackage{amsmath}
\usepackage{amssymb}
\usepackage{mathtools}
\usepackage{mathrsfs}
\usepackage{multirow}
\usepackage{bbm}
\usepackage{bbold}
\usepackage{braket}
\usepackage{siunitx}
\usepackage[only,llbracket,rrbracket]{stmaryrd}
\usepackage[caption=false]{subfig}
\usepackage{float}
\usepackage{makecell}
\usepackage[capitalize]{cleveref}

\hypersetup{%
    hypertexnames=false,
    colorlinks=true,
    allcolors=blue,
    unicode=true
}

\graphicspath{{figures/}}

\newcommand{\phantomsubfloat}[1]{%
    {%
        \captionsetup[subfloat]{farskip=0pt,captionskip=0pt}
        \captionsetup[subfigure]{labelformat=empty}
        \subfloat{#1}
    }
}

\crefname{section}{Sec.}{Sec.}
\crefname{appendix}{App.}{App.}
\newcommand{\Subsecref}[1]{Subsection~\labelcref{#1}}

\DeclarePairedDelimiter{\norm}{\lVert}{\rVert}
\DeclarePairedDelimiter\ceil{\lceil}{\rceil}

\newcommand{\hc}{\mathrm{h.c.}}

\DeclareMathOperator{\tr}{Tr}

\begin{document}

\title{End-to-End Optimization of Partial Fault Tolerance in Neutral-Atom Arrays}

\author{Stefano Veroni}
\email{stefano.veroni@physics.ox.ac.uk}

\author{Aydin Deger}

\author{Andrew J. Daley}

\affiliation{Department of Physics, Clarendon Laboratory, University of Oxford, Oxford OX1 3PU, United Kingdom}

\date{\today}

\begin{abstract}  
Fault-tolerant (FT) quantum computation offers a pathway to quantum advantage, but its realization is hindered by its significant overhead.
The partially FT ``STAR'' architecture reduces the cost of universal computation 
at the price of relaxing FT requirements for non-Clifford operations, increasing exposure to hardware noise. 
Therefore, maximizing its performance and applicability demands careful co-design with the underlying hardware.
In this work we investigate the co-design of the STAR architecture with neutral-atom platforms, connecting the underlying atomic physics to logical error modeling and application-level performance.
At the physical level, we identify a favorable noise bias in $^{171}\mathrm{Yb}$ Rydberg gates, and show how it can be enhanced through error-selective pulse optimization.
At the logical level, we identify an intrinsic error suppression mechanism, yielding vanishing logical coherent noise in leading order.
Detailed circuit-level simulations of surface code implementations show more than an order of magnitude reduction in logical error relative to previous estimates. 
As a concrete target application, we focus on 
the Fermi-Hubbard model with periodic boundaries, for which we design a simulation circuit particularly suited to early implementations.
We quantify the resource tradeoff of partially and fully FT approaches to gate synthesis, with the former showing approximately an order of magnitude reduction in state-injection depth, and an even larger reduction in Clifford resources.
Comparing with classical and fully FT approaches, we identify regimes where partial fault tolerance provides both lower overheads and lower error rates. In our estimates, this is accessible with moderate error rates $p\sim 10^{-3}$ and a few thousand qubits, underlining that for these types of problems, partial FT approaches provide a compelling pathway to early quantum advantage. 
\end{abstract}

\maketitle

\section{Introduction}
\label{sec:introduction}

Large-scale quantum advantage is expected to require quantum error correction (QEC) to overcome the inherently noisy nature of quantum devices. 
QEC encodes logical information across multiple physical qubits, and repeatedly detects physical faults by means of syndrome measurements~\cite{terhal2015}.
Remarkable progress in QEC has been demonstrated across different architectures, such as superconducting qubits~\cite{google2023, gupta2024, caune2024-riverlane, google2024, lacroix2024}, trapped ions~\cite{pogorelov2024, ryan-anderson2024-quantinuum, dasilva2024, dasu2026-quantinuum}, nitrogen vacancies in diamonds~\cite{abobeih2022}, and neutral-atom arrays~\cite{bluvstein2022, bluvstein2024, reichardt2024, bedalov2024, rodriguez2024, bluvstein2025, rines2025-infleqtion-logic}.
Critical to the success of large-scale quantum computation is the concept of fault tolerance.
A protocol is fault-tolerant (FT) if it constrains the propagation of errors, thus suppressing noise, provided that the hardware error rate is lower than a threshold value.
Fault tolerance is essential not only to error correction, but also to the execution of logical operations.

Transversal logical gates, which do not couple physical qubits within the same code block, prevent uncontrolled error propagation and are naturally FT.
However, no QEC code admits a gate set that is both transversal and universal~\cite{eastin2009-notransversal}. 
Several methods have therefore been proposed to enable the FT realization of a universal gate set~\cite{bravyi2005-magicstate,litinski2019-magicstate,gidney2024-cultiv,anderson2014-codeswitch}, but they come with significant space-time overhead.
In particular, arbitrary operations are synthesized by tens of non-Clifford gates~\cite{kliuchnikov2023_Tsynth}, each considerably more demanding than transversal operations.
To reduce this overhead, \textit{partially} fault-tolerant computation has recently been proposed \cite{akahoshi2024, choi2023a, toshio2025, ismail2025, zeng2025_pftbias, ismail2026}.
In the so-called ``Space-Time efficient Analog Rotation'' (STAR) architecture, the execution of Clifford gates remains FT, whereas arbitrary-angle rotations are natively implemented rather than synthesized via tens of FT gates.
This overhead reduction comes at the expense of reduced resilience against specific protocol-dependent faults, making accurate modeling and co-design with the underlying hardware indispensable for its success.
Among the most promising hardware approaches, neutral-atom architectures have recently demonstrated a rich collection of desirable features, including mid-circuit measurements and reconfigurable non-local connectivity~\cite{menssen2026_strategicplanneutralatom}.
These enable fast transversal logical gates, hence a significant reduction in QEC overhead~\cite{zhou2024-algorithmicFT, cain2024, cain2025, serraperalta2025-decodetransversal}, and have recently been incorporated within the framework of partial fault tolerance, in the so-called transversal STAR architecture~\cite{ismail2025}.

\begin{figure*}[ht!]
    \includegraphics[width=\linewidth]{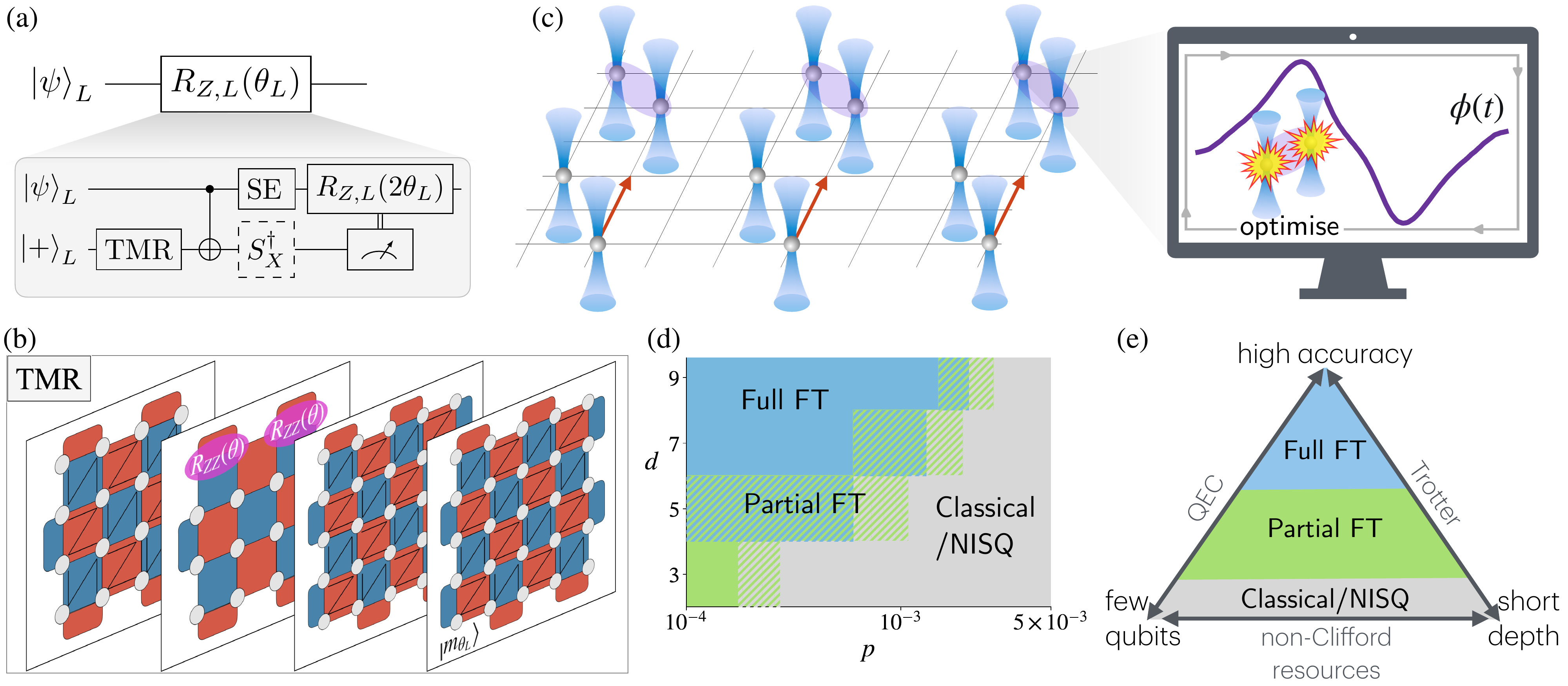}
    \phantomsubfloat{\label{fig:sub:teleportation}}
    \phantomsubfloat{\label{fig:sub:TMR}}
    \phantomsubfloat{\label{fig:sub:rydberg-intro}}
    \phantomsubfloat{\label{fig:sub:pFT-in-pd-diagram-sketch}}
    \phantomsubfloat{\label{fig:sub:general-tradeoff-sketch}}
    
    \vspace*{-2em}
    \caption{
        Overview.
        (a) Partially fault-tolerant realization of arbitrary-angle $R_{Z,L}(\theta_L)$ rotations.
        An angle-dependent resource state $\ket{m_{\theta_L}}_L$ is encoded via the Transversal Multi-Rotation (TMR) protocol, then gate teleportation is performed and eventually repeated until the target measurement is obtained.
        The Clifford $S_X^{\dagger}$ is required if an even number of transversal rotations is applied in the TMR stage.
        (b) TMR encoding of the $\ket{m_{\theta_L}}_L$ resource state in the rotated surface code. 
        First, a logical patch is encoded with code distance $d_s$; then $r$ multi-qubit transversal rotations by a suitable angle are applied; finally, other rounds of syndrome extraction are performed, growing the code distance if needed.
        Post-selection is used upon nontrivial syndromes.
        In this example, $d_s=4$, $r=2$ and the final code distance is $d=5$.
        (c) Operation and optimization of a neutral atom device.
        Shuttling enables non-local connectivity and transversal gates.
        Pulse-level optimization of gates under realistic noise models is performed to increase the logical performance of the scheme.
        (d) Regimes of advantage of different computational paradigms, in terms of code distance $d$ and physical error rate $p$, taking time evolution in the Fermi-Hubbard model as concrete benchmark. 
        Green and blue denote regions where partial and full fault tolerance, respectively, yield lower error, and are capable of simulating larger systems than classical/NISQ alternatives with a contained error mitigation overhead.
        The striped region highlights the extended near-term regime, opened by the improvements detailed in the manuscript, in which the transversal STAR architecture is not only a lower-overhead alternative to full fault-tolerance, but also yields the lowest errors. 
        The background color behind the green stripes indicates the alternative method that would have yielded lower errors under previous estimates~\cite{ismail2025}.
        %clf data here
        (e) Trilemma of quantum computation for Trotterised evolution, where high accuracy, short depth, and low qubit counts are mutually exclusive. 
        A balance is struck by tuning QEC approach, Trotterisation, and availability of non-Clifford resources.
    }
    \vspace*{-1em}
    \label{fig:intro}
\end{figure*} 

In our work, we study the co-design of partially FT computation with neutral-atom architectures~\cite{menssen2026_strategicplanneutralatom}, spanning gate-pulse optimization, analysis of logical operations, and application-layer design, taking the Fermi-Hubbard model as representative case-study.
Altogether, our work opens a near-term regime in which partially FT computation surpasses the reach of classical and NISQ simulations, as well as that of fully FT implementations at comparable hardware size.
More precisely, we show that it improves over standard FT approaches not only by yielding shorter circuit depth, but also by achieving lower logical noise, thereby enabling simulations of larger system sizes.
The STAR architecture thus emerges not only as a lower-overhead alternative, but as an all-round compelling route for early quantum advantage.
Concretely, we first uncover intrinsic error-suppression mechanisms in partially FT computation, elucidating further noise reduction approaches.
Second, we tailor partially FT computation to neutral-atom architectures by identifying a favorable noise bias in the relevant gates, and increasing it through pulse-level optimization.
Third, focusing on time-evolution in the Fermi-Hubbard model with periodic boundaries, we design a simulation circuit particularly suited to early implementations.
Then we assess the performance of partially FT computation, both by investigating its resilience to logical noise, and by quantifying its tradeoff between physical resources and circuit depth.

The manuscript is organized as follows. 
\cref{sec:overview-summary-results} provides a concise outline of our approach and highlights our main findings. 
In \cref{sec:partial-ft}, we introduce partially fault-tolerant computation, discuss its dominant error mechanisms, and develop a detailed model of error propagation in the protocol. 
In \cref{sec:gate-optim}, we specialize the analysis to neutral-atom architectures, characterize the noise properties of Rydberg gates, and improves their performances with pulse-level optimization.
In \cref{sec:performance}, we present simulations under circuit-level noise of partially FT computation with the rotated surface code, and quantify the resulting logical performance. 
In \cref{sec:applications}, we compare the robustness and resource requirements of the partially and fully FT approaches, and examine the consequences for near-term quantum simulation of the Hubbard model.

\section{Overview and Key Results}
\label{sec:overview-summary-results}

This section provides a summary of our approach and results in the co-design of partial fault-tolerance with neutral-atom architectures, spanning the physical, logical and application layers.

First, we develop an accurate model of error propagation in partially FT operations, independent of the underlying hardware.
In the STAR architecture, logical rotations $R_{Z,L}$ are natively implemented via injection of a suitably encoded resource state.
We provide a detailed enumeration of leading noise sources and their impact on the injection and state preparation, shown in~\cref{fig:sub:teleportation,fig:sub:TMR}, 
revealing that the protocol is naturally resilient to physical coherent errors---often the dominant contributions in non-Clifford operations. 
Moreover, we find that the resulting logical noise has a vanishing coherent component in leading order.
These observations are critical to a correct characterization of noise in partially FT operations, and the development of suitable mitigation approaches.

Secondly, we analyze and improve the robustness of the protocol in the concrete setting of neutral-atom arrays, schematically depicted in~\cref{fig:sub:rydberg-intro}, with a particular focus on $^{171}\rm Yb$ implementations.
We show that Rydberg gates---to which partially FT computation is particularly sensitive in this setting---exhibit noise scaling and bias that are particularly favorable for this approach. 
Then, as sketched in~\cref{fig:sub:rydberg-intro}, we optimize these gates at the pulse level to reduce the rate of undetectable errors, yielding a further two-fold improvement in logical performance with no additional overhead.

Third, we perform full circuit-level simulations of the logical protocol in the rotated surface code.
For this purpose, we develop a tailored simulation method that captures the effects of non-Clifford operations and noise in the protocol, and propagates them through multiple iterations of the scheme.
These simulations demonstrate more than one order of magnitude reduction in logical error relative to previous estimates~\cite{toshio2025, ismail2025}. 

Finally, we assess the implications for near-term quantum simulation.
As a concrete setting, we focus on time evolution in the Hubbard model on a square lattice with periodic boundaries, with evolution time growing linearly with lattice length.
We evaluate implementations with the rotated surface code in neutral-atom architectures, accounting for the cost of non-Clifford resources, as well as of transversal Clifford operations.
The latter are found to have non-negligible impact at moderate distances, hence we design a new circuit implementation tailored to early FT approaches by lowering the number of Clifford gates, exploiting a previously proposed construction~\cite{hastings2015_FH, flannigan2022_FH}.
First, our analysis quantifies the accumulation of logical errors.
Quantum simulations are assumed to be supported by error mitigation, hence are considered practically feasible when the sum of gates' error rates is below unity.
The reach of classical methods for simulating the Hubbard model is understood to be somewhat below a system size of a $10 \times 10$ lattice for dynamics that spread information over the whole system~\cite{papic2026_fermionic,flannigan2022_FH}.

We identify regimes of advantage of partially FT, fully FT and classical computation, as summarized in~\cref{fig:sub:pFT-in-pd-diagram-sketch}. 
For comparison purposes, assuming periodic boundaries and evolution time comparable to the length of the lattice, we place the limit of current classical methods at $6\times6$
sites, as indicated by Refs.~\cite{papic2026_fermionic, alam2025_fermionicquantinuum}.
Remarkably, the improvements obtained in our framework open up a near-term regime (striped green) characterized by moderate code distance and error rate $p\sim10^{-3}$, in which the transversal STAR architecture yields lower errors than all alternatives considered, in contrast to the conclusion following from previous modeling~\cite{ismail2025}.
Notably, we estimate partial fault-tolerance to enable simulations beyond the reach of both current classical methods and NISQ approaches already on devices with only $5,000$ physical qubits. 
As the code distance is increased, partially FT computation eventually becomes limited by its fundamental constraints, and fully FT approaches achieve lower error rates.
Second, we extend the analysis to the broader tradeoffs of quantum computation---accuracy, qubit count, and depth---as sketched in~\cref{fig:sub:general-tradeoff-sketch}.
We show that partially FT computation exhibits a much more favorable spacetime tradeoff relative to fully FT alternatives.
As a result, it achieves approximately an order of magnitude reduction in non-Clifford depth, and an even greater reduction in Clifford resources for gate synthesis, up to three orders of magnitude depending on ancillary availability.
%clifford depth modeling assumptions here
Altogether, our work establishes partially FT computation as a compelling route toward near-term demonstrations of practical quantum advantage.

\section{Partial Fault-tolerance}
\label{sec:partial-ft}

Two-dimensional stabilizer codes only admit transversal gates belonging to the Clifford group~\cite{bravyi2013-konig}. 
Thus, several methods have been proposed to enable the FT realization of specific gates, e.g. $T$ or $CCZ$ gates, to be combined with Clifford operations to form a universal digital gate set.
These include magic state distillation~\cite{bravyi2005-magicstate, campbell2012-magicstate, campbell2017b-magicstate, gorman2017-magicstate, litinski2019-magicstate, wills2025_magicstate} or cultivation~\cite{gidney2024-cultiv, sahay2025-cultiv, vaknin2026_Tsynth}, code switching~\cite{anderson2014-codeswitch, beverland2021-codeswitch, butt2024-codeswitch, butt2024-mf}, and code deformation~\cite{yoder2017-baconshor-ccz, veroni2025}.
However, these approaches are considerably more demanding than transversal operations, and arbitrary operations must additionally be synthesized by tens of these non-Clifford gates~\cite{kliuchnikov2023_Tsynth}.

Partial fault tolerance reduces the overhead of universal logical computation, by avoiding both the cost of realizing fully FT non-Clifford gates and the need to approximate arbitrary operations with long gate sequences.
In the transversal STAR architecture~\cite{ismail2025}, the logical gate set contains Clifford operations together with arbitrary-angle single qubit rotations $R_L(\theta_L)$.
Clifford gates are applied transversally---with qubit permutations when needed---and therefore have low overhead and typically higher fidelity. 
By contrast, arbitrary-angle rotations are implemented in a \textit{partially} FT fashion, in which the significant reduction in overhead comes at the price of an increased exposure to physical faults.

The STAR architecture, originally proposed in Ref.~\cite{akahoshi2024}, and further improved in subsequent works~\cite{choi2023a, toshio2025, ismail2025}, relies on the preparation and injection of a suitable resource state
\begin{equation}
    \ket{m_{\theta_L}}_L = R_{Z,L}(\theta_L) \ket{+}_L 
    = \cos \frac{\theta_L}{2} \ket{+}_L - i \sin \frac{\theta_L}{2} \ket{-}_L,
    \label{eq:mstate}
\end{equation}
which encodes the arbitrary-angle logical rotation
\begin{equation}
    R_{Z,L}(\theta_L)= e^{- i \frac{\theta_L}{2} Z_L}.
    \label{eq:rz-convention}
\end{equation}
The resource state is then consumed to inject the rotation onto a target state $\ket{\psi}_L$ via gate teleportation, as shown in~\cref{fig:sub:teleportation}.
Gate teleportation is probabilistic: with probability $1/2$, upon a $+1$ measurement outcome, the output is the correctly rotated state $R_{Z,L}(\theta_L) \ket{\psi}_L$, while the opposite outcome produces the inversely rotated state $R_{Z,L}(-\theta_L) \ket{\psi}_L$.
In the latter case, one repeats the resource-state preparation and injection with a doubled rotation angle, until a successful +1 outcome is observed.
This \textit{repeat-until-success} (RUS) procedure is successful on average in two trials.

In the remainder of this section, we review the state preparation and injection protocol~(\cref{subsec:partial-ft-design}), discuss the error mechanisms that limit its performance~(\cref{subsec:partial-ft-error-budget}), and finally quantify error propagation within the broader RUS procedure~(\cref{subsec:partial-ft-rus}).
Remarkably, we show that the RUS nature of the scheme leads to the beneficial cancellation of coherent error sources, as well as leading-order suppression of coherent logical noise.

\subsection{Transversal Multi-rotation Protocol}
\label{subsec:partial-ft-design}

The resource state preparation is referred to as the Transversal Multi-Rotation (TMR) protocol~\cite{toshio2025}, for reasons that will soon become apparent, and is illustrated in~\cref{fig:sub:TMR}.
Let us denote by $P^{\otimes w}_q$ the tensor product of the operator $P$ over $w$ consecutive qubits, starting at index $wq$.
Then, let us define a set of $r$ transversal $Z^{\otimes w}$-rotations
\begin{equation}
    \bigotimes_{q=0}^{r-1} R_{Z^{\otimes w}_q} (\theta) = 
    \bigotimes_{q=0}^{r-1} \left[ \cos \frac{\theta}{2}\,I^{\otimes w}_q - i \sin \frac{\theta}{2} \, Z^{\otimes w}_q
    \right],
    \label{eq:physical-tmr-rotations}
\end{equation}
where $R_{Z_q^{\otimes w}}(\theta) = e^{-i \theta Z^{\otimes w}_q / 2}$.
Next, consider a logical qubit in the state $\ket{+}_L$, with code distance $d$ and logical operator $Z_L = Z^{\otimes d}$.
Assuming $d=wr$, applying the transversal rotations on the support of $Z_L$ yields
\begin{align}
    \bigotimes_{q=0}^{r-1} R_{Z^{\otimes w}_q} (\theta) \ket{+}_L =&\, 
    \cos^{r} \frac{\theta}{2} \ket{+}_L + (-i)^r \sin^r \frac{\theta}{2} \ket{-}_L 
    \nonumber \\ &+ \text{(Z-error terms)},
    \label{eq:physical-tmr-rotations-applied}
\end{align}
where the Z-error terms encompass all terms that project the state outside the logical subspace.
Defining a logical rotation angle $\theta_L$ satisfying
\begin{equation}
    \tan \frac{\theta_L}{2} = \tan^r {\frac{\theta}{2}},  
\end{equation}
with $0 \leq \theta, \theta_L \leq \pi/2$, the above expression is written in terms of the desired resource state in~\cref{eq:mstate} as
\begin{widetext}
\begin{equation}
    \bigotimes_{q=0}^{r-1} R_{Z^{\otimes w}_q} (\theta) \ket{+}_L = 
    \begin{cases}
    \sqrt{p_{\theta_L}}\,\cdot \ket{m_{(-1)^{\lfloor r/2 \rfloor} \theta_L}}_{L} + (\text{$Z$-error terms})
    & (r\text{  odd}), \\[5pt]
    \sqrt{p_{\theta_L}}\,\cdot e^{i\frac{\pi}{4}}\,S_X\,\ket{m_{(-1)^{\lfloor r/2 \rfloor} \theta_L}}_{L} + (\text{$Z$-error terms})
    & (r\text{ even}), 
    \end{cases}
    \label{eq:big-tmr-result}
\end{equation}
\end{widetext}
where
\begin{equation}
    p_{\theta_L}(\theta_L,r) = \cos^{2r} \frac{\theta}{2} + \sin^{2r} \frac{\theta}{2},
    \label{eq:pthetal}
\end{equation}
is the probability that, upon a following syndrome measurement, the state collapses onto the desired resource state.
$S_X = R_{X,L}(\pi/2)$ can be removed by applying $S_X^\dagger$ before the teleportation measurement, as in~\cref{fig:sub:teleportation}.
We highlight that in the rotated surface code this can be implemented with only transversal gates and qubit permutations~\cite{chen2024_transversalS}.
Finally, we note that no additional correction needs to be applied to reverse the sign of the target angle $\theta_L$ when this is negative, as when $\lfloor r/2 \rfloor$ is odd, due to the random nature of gate teleportation.

In summary, the TMR protocol encodes the desired resource state $\ket{m_{\theta_L}}_L$ via the following steps: initialize $\ket{+}_L$, apply the transversal $w$-qubits physical rotations $R_{Z^{\otimes w}}$, and then perform syndrome measurements with suitable post-selection.

Although error detection and post-selection can strongly suppress faults, several error mechanisms can still prevent successful implementation of the protocol; these must therefore be quantified.
In what follows, $p$ denotes the physical circuit-level error rate, whereas the noise $\mathcal{E}$ acting on each transversal rotation is best modeled by its chi-matrix $\chi_{\scriptscriptstyle P,P'}$, such that 
\begin{equation}
    \mathcal{E}(\rho) = \sum_{P,P'\in \mathcal{P}^{\otimes w}} \chi_{\scriptscriptstyle P,P'} P \rho P',
\end{equation}
where $\mathcal{P}^{\otimes w}$ denotes the set of $w$-qubit Pauli operators.

\subsection{Error sources of the TMR protocol}
\label{subsec:partial-ft-error-budget}

\begin{figure*}[ht!]
    \includegraphics[width=\linewidth]{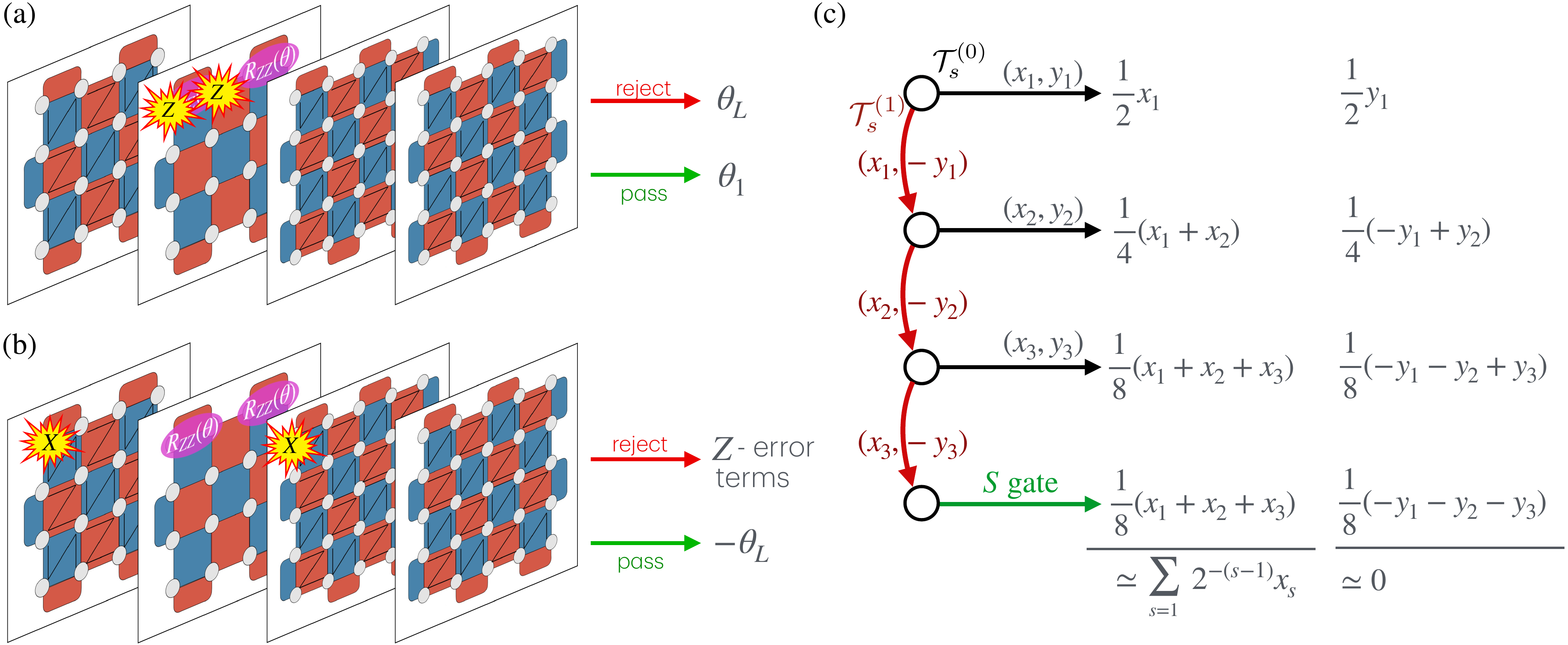}
    \phantomsubfloat{\label{fig:sub:tmr-noise-phaseflips}}
    \phantomsubfloat{\label{fig:sub:tmr-noise-bitflips}}
    \phantomsubfloat{\label{fig:sub:rus-yerrors-diagram}}
    \vspace*{-2em}
    \caption{
        Overview of noise sources and propagation. 
        (a) Correlated phase-flips, occurring shortly before or after the physical rotations, cause the rejection of the correctly rotated state, while passing a wrongful rotation by angle $\theta_1$ (see~\cref{eq:theta1-main-text}). 
        (b) Two repeating bit-flips, occurring shortly before and after the physical rotations respectively, lead to the inverse of the target $\theta_L$ rotation.
        $Z$-error terms are correctly rejected upon post-selection. 
        (c) Schematic depiction of leading-order error propagation in the Clifford-terminated RUS (ctRUS) procedure.
        Subscripts denote the step within the RUS process, in this example from $1$ to $3$.
        Black (red) arrows represent gate-teleportation channels $\mathcal{T}^{(\mu)}_{\rm s}$ with ideal (opposite) measurement outcome $\mu=0(1)$.
        Each has probability $1/2$ in leading order.
        The green arrow represents the action of the terminating Clifford $S$, here noiseless for simplicity.
        Each arrow points to the leading-order estimate of the incoherent $x$ and coherent $y$ noise on the resulting state, weighted by the probability such state is produced. 
        Incoherent noise $x_s$ adds up, whereas coherent noise $y_s$ vanishes in leading order, owing to the opposite sign this noise takes upon opposite measurement outcomes.
    }
    \vspace{-1em}
    \label{fig:noise-tmr-and-rus-diagram}
\end{figure*} 

All error-corrected schemes of logical computation suffer from noise in syndrome measurements and transversal logical gates. However, two main error mechanisms limit the performance of the TMR protocol:
(i) undetected correlated phase-flips, modifying the logical rotation angle as depicted in~\cref{fig:sub:tmr-noise-phaseflips};
(ii) undetected repeating bit-flips, inverting the sign of the logical rotation angle as depicted in~\cref{fig:sub:tmr-noise-bitflips}.
In this subsection, we provide a concise overview of the former, while deferring detailed modeling to~\cref{subsec:app:error-budget:noise-from-qec,subsec:app:error-budget:noise-from-correl-phase-flips}.
The latter error source, despite its potential to ultimately limit the performance of the protocol, has a negligible contribution in the setting we consider, hence its treatment is deferred to~\cref{subsec:app:error-budget:noise-from-repeat-flips}. 

The first error mechanism originates from $Z^{\otimes w}$ faults occurring shortly before or after the physical rotations~\cref{eq:physical-tmr-rotations}, as illustrated in~\cref{fig:sub:tmr-noise-phaseflips}.
To understand its origin and effect, it should first be observed that the error terms in~\cref{eq:big-tmr-result} that are closest to the logical subspace---hence dominating the error budget---take the form
\begin{equation}
    -i
    \sum_{q=0}^{r-1} \sqrt{p_{\theta_1}}\,Z_q^{\otimes w}\, \ket{m_{(-1)^{\lfloor r/2 \rfloor} \theta_1}}_L,
    \label{eq:z-error-terms-explicit-main-text}
\end{equation}
where their effective rotation angle $\theta_1(\theta_L,r)$ satisfies
\begin{equation}
    \tan {\frac{\theta_1}{2}} 
    = - \tan^{(1-\frac{2}{r})} \frac{\theta_L}{2}
    = - \tan^{r-2} \frac{\theta}{2}, 
    \label{eq:theta1-main-text}
\end{equation}
and the probability $p_{\theta_1}(\theta_L,r)$ of collapsing onto each one of these terms is
\begin{equation}
    p_{\theta_1} 
    = \cos^{2(r-1)} \frac{\theta}{2} \sin^{2} \frac{\theta}{2}
    + \sin^{2(r-1)} \frac{\theta}{2} \cos^2 \frac{\theta}{2}.
    \label{eq:p1-main-text}
\end{equation}
These terms lie outside the logical subspace, thus noiseless syndrome extraction and error detection successfully rejects them.
However, a \(Z_q^{\otimes w}\) fault exchanges the syndrome subspace of the desired term and the corresponding error component.
Ideal post-selection therefore rejects the desired term affected by this fault while accepting the component proportional to \(p_{\theta_1}\), producing the wrong logical angle \(\theta_1\).
The probability of occurrence of one of these faults on any of the physical rotations is
\begin{equation}
    p_{Z^{\otimes w}} \simeq r [ a_{Z}\,p^w + \chi_{\scriptscriptstyle Z^{\otimes w},Z^{\otimes w}}(\theta)],
    \label{eq:pZZ-main-text}
\end{equation}
where $\chi_{\scriptscriptstyle Z^{\otimes w},Z^{\otimes w}}$ denotes the probability of its occurrence in one of the transversal rotations, and $a_{Z}>0$ is a combinatorial prefactor.

\subsection{Repeat Until Success}
\label{subsec:partial-ft-rus}

\begin{table*}[ht!]
    \centering
    \caption{Summary of the key parameters characterizing the Transversal Multi-Rotation (TMR) protocol at the $s^\mathrm{\rm th}$ step of the Repeat-Until-Success (RUS) scheme, ultimately realizing a rotation by the angle $\theta_L$. 
    The expressions detail their analytical dependence on the number $r$ of physical rotations and the rotation angles.}
    \label{tab:rus_parameters}
    \renewcommand{\arraystretch}{2}
    \begin{tabular*}{\textwidth}{c l c}
        \hline\hline
        \textbf{Symbol} & \textbf{Description} & \textbf{Definition} 
        \\ 
        \hline
        $\theta_{L,s}$ 
        & Target logical rotation angle 
        & $\displaystyle 2^{s-1} \theta_L - \frac{\pi}{2} \underset{k \in \mathbb{Z}}{\operatorname{argmin}} \left| 2^{s-1} \theta_L - \frac{k\pi}{2} \right| $ 
        \\ 
        $\theta_s$ & Physical transversal rotation angle & $\displaystyle 2 \arctan \left( \tan^{1/r} \frac{\theta_{L,s}}{2} \right)$ 
        \\ 
        $p_{\theta_{L,s}}$ & Probability of collapse onto the desired resource state & $\displaystyle \cos^{2r} \frac{\theta_s}{2} + \sin^{2r} \frac{\theta_s}{2}$ 
        \\ 
        $\theta_{1,s}$ & Logical rotation angle of leading $Z$-error terms & $\displaystyle 2 \arctan \left( -\tan^{r-2} \frac{\theta_s}{2} \right)$ 
        \\ 
        $p_{\theta_{1,s}}$ & Probability of collapsing onto one of the leading $Z$-error terms & $\qquad \displaystyle \cos^{2(r-1)} \frac{\theta_s}{2} \sin^{2} \frac{\theta_s}{2} + \sin^{2(r-1)} \frac{\theta_s}{2} \cos^2 \frac{\theta_s}{2}$ \vspace{0.5em} \\
        \hline\hline
    \end{tabular*}
\end{table*}

The previous subsection established the main ingredients required to accurately model noise in the TMR protocol.
In this subsection, we present their combined effect in the whole RUS scheme.
In particular, we derive a remarkable property of the procedure, neglected in previous works : an intrinsic suppression of logical coherent errors.

The RUS scheme consists of a repetition of gate-teleportation attempts, each succeeding with half probability.
In principle, each $s^{\rm th}$ step, with $s=1,2,\ldots$ targets a rotation angle twice as large as the previous, i.e. $\theta_{L,s}=2\,\theta_{L,s-1}$, with $\theta_{L,1}=\theta_L$.
However, as will be shown shortly, smaller angles yield smaller error rates.
Then, as $\pi/2$ rotations are FT Clifford gates, the preferred logical rotation angle at every step is the \textit{least absolute remainder} with respect to $\pi/2$, i.e. the one that has the smallest absolute value up to multiples of $\pi/2$,
\begin{equation}
    \theta_{L,s} = 2^{s-1} \theta_L - \frac{\pi}{2} \underset{k \in \mathbb{Z}}{\operatorname{argmin}} \left| 2^{s-1} \theta_L - \frac{k\pi}{2} \right| .
\end{equation}
In general, we denote with $s$ subscript the quantities pertaining to each $s^{\rm th}$ step, conveniently summarized in~\cref{tab:rus_parameters}.
Moreover, we assume that the RUS scheme can be terminated at the $C^\mathrm{\rm th}$ step with full certainty, by application of the Clifford $S$ gate.
As the target angle is doubled at every step of the scheme, this is the case for any target angle of the form
\begin{equation}
        \theta_L = \frac{\pi}{2} \frac{2k+1}{2^C},
        \label{eq:app:dyadic-thetaL-main-text}
\end{equation}
for any integer $k$.
It should be noted that this form provides a dense set of angles, which can approximate any other with arbitrary precision.

To model noise propagation, we first observe that the error sources specific to the TMR protocol amount to incoherent and coherent noise along the $Z_L$ axis.
Each teleportation attempt within the RUS procedure is well described by the ansatz
\begin{align}
    \mathcal{T}_s(\Psi, \theta_{L,s};\,x_s,y_s) 
    =\;& 
    \,(1-x_s) \Psi_{\theta_{L,s}} + x_s\, Z_L \Psi_{\theta_{L,s}} Z_L 
    \nonumber \\
    &- i y_s\, Z_L \Psi_{\theta_{L,s}} + i y_s^*\Psi_{\theta_{L,s}} Z_L,
    \label{eq:ansatz-noise-process-main-text}
\end{align}
where $\Psi$ is the input state; $\Psi_{\bullet}$ denotes $ R_Z(\bullet)\Psi\,R_{Z}^\dagger(\bullet)$ such that $\Psi_{\theta_{L,s}}$ is the ideal output; $x_s$ and $y_s$ represent, respectively, the incoherent and coherent error contributions, both dependent on the rotation angle $\theta_s$ and the number $r$ of physical rotations.
Importantly, upon opposite teleportation outcomes, the signs of both the encoded angle $\theta_{L,s}$ and the coherent error $y_s$ are flipped, as explained in~\cref{subsec:app:error-budget:noise-whole-teleport}.

A step-by-step derivation of error accumulation in the RUS realization of the protocol, including in the absence of Clifford termination, is deferred to~\cref{subsec:app:error-budget:noise-whole-RUS}, while here we proceed to state the result.
Assuming first that the terminating Clifford is noiseless, the result of the Clifford-terminated repeat-until-success (ctRUS) is described by the channel
\begin{align}
    \rho_\mathrm{ctRUS}(\Psi, \theta_L)
    &= \,(1-p_{\mathrm{ctRUS}}) \Psi_{\theta_L} + p_{\mathrm{ctRUS}}\, Z_L \Psi_{\theta_L} Z_L 
    \nonumber \\
    &\quad+O(x^2_s + x_s y_s + y_s^3),
    \label{eq:rus-error-ansatz-main-text}
\end{align}
yielding the target output $\Psi_{\theta_L}$ under stochastic $Z$-noise of strength $p_{\mathrm{ctRUS}} = \sum_{s=1}^{C} 2^{-(s-1)} x_s$. 
Remarkably, owing to the difference in sign of the coherent noise upon opposite teleportation outcomes, the leading order contribution to the overall logical coherent noise \textit{vanishes}.
Further intuition on the RUS error accumulation or cancellation is provided by the diagram in~\cref{fig:sub:rus-yerrors-diagram}, which illustrates the propagation of leading order errors in the ctRUS scheme.

This result does not change upon inclusion of the noise from the $S$ gate, which has negligible coherent noise owing to the applicability of Pauli twirling to Clifford operations~\cite{dur2005-twirling, emerson2007-twirling}.
Moreover, the noise from the $S$ gate is suppressed exponentially with code distance and is further reduced by the probability $\sim 2^{-C}$ of its application, thus is generally negligible.
The dominant noise terms $x_s$ are determined by the error mechanisms presented in the previous subsection.
Only considering the effect of undetected correlated phase-flips---the dominant error mechanism---yields
\begin{align}
    x_s &\simeq
    \frac{p_{\theta_{L,s}}}{n_{\theta_{L,s}}} \,p_{\theta_{1,s}} \,p_{Z^{\otimes w}}\,\sin^2\frac{\theta_{1,s}-\theta_{L,s}}{2}  
    \label{eq:rus-remaining-xs-terms}
\end{align}
where $p_{Z^{\otimes w}}$ generally depends on $\theta_{L,s}$ and $r$ as per~\cref{eq:pZZ-main-text}, and 
with normalization factor $n_{\theta_{L,s}}$ defined in~\cref{{eq:teleport-resource-state-channel-coefficients}} and close to one in the small angle scenario. 
A more complete expression taking into account also the effect of undetected bit-flips is provided in~\cref{subsec:app:error-budget:noise-whole-teleport}.

The results outlined in this section highlight some important points.
First, the RUS realization of logical rotations benefits from intrinsic cancellation of coherent noise in leading order, avoiding the need for the additional mitigation scheme devised in~\cite{toshio2025}, and making partially FT computation more amenable to typical quantum error mitigation techniques~\cite{cai2023_qem}.
Second, noise is expected to decrease with the rotation angle $\theta_L$, which is desirable given the expected applications of the protocol.
Third, by increasing the size $w$ of the transversal rotations, the contribution from physical noise can in principle be reduced, namely that coming from correlated phase flips in~\cref{eq:pZZ-main-text}.
In practice, however, gates beyond two qubits have been rarely implemented, and with lower fidelity than two-qubit gates~\cite{evered2023}. 
For this reason, we assume $w=2$ throughout the rest of this paper.
Finally, it is evident that a reliable estimate of the protocol performance requires an accurate characterization of the noise in the transversal $R_{Z^{\otimes w}}$ rotations.

Accordingly, in the next section we perform a detailed analysis of $R_{ZZ}$ gates in realistic neutral-atom settings.
We show that the undetectable error $\chi_{\scriptscriptstyle ZZ,ZZ}$ exhibits a favourable dependence on the rotation angle, and is comparable in magnitude to $p^2$, despite not scaling as such. 
Both features contrast with assumptions made in previous works~\cite{toshio2025, ismail2025}.
We also optimize the gate at the pulse level to further improve the performance of the protocol.
Then, in~\cref{sec:performance} we simulate and discuss the resulting behavior of the logical protocol.

\section{Rydberg Gate Characterization and Optimization}
\label{sec:gate-optim}

As explained in the previous section, partial fault tolerance significantly reduces the overhead of universal logical computation, at the price of exposure to physical faults.
Specifically, the protocol is particularly sensitive to $ZZ$-type errors in the application of physical $R_{ZZ}(\theta)$ gates, as they modify the logical angle of rotation while going undetected, as explained in~\cref{subsec:partial-ft-error-budget}.
Therefore, it is critical to accurately investigate the magnitude of these undetectable faults in the realistic scenario of interest, and to reduce it if possible.
In this regard, we note that Ref.~\cite{zeng2025_pftbias} has recently proposed an error reduction technique exploiting the structure of dissipative noise, particularly suitable to superconducting qubits.
Our work considers a different platform and approach.

We address the exposure to critical faults of partial fault-tolerance in the context of neutral-atom architectures, one of the most promising platforms for digital quantum computation, with recent demonstrations showcasing a broad range of capabilities across different implementations~\cite{menssen2026_strategicplanneutralatom}.
These include high-fidelity gates~\cite{evered2023,tsai2024, muniz2025, peper2024,senoo2025,radnaev2024}, mid-circuit measurements~\cite{ma2023, singh2023, scholl2023, norcia2023, radnaev2024, bluvstein2024, bluvstein2025, muniz2025-ancillae}, scalability beyond a thousand atoms~\cite{pause24, manetsch2024, chiu2025}, continuous reloading of the array~\cite{singh2022, norcia2024,gyger2024,chiu2025}, and reconfigurable non-local connectivity via atomic shuttling~\cite{bluvstein2022,bluvstein2024,reichardt2024,zhang2025}.

In this section, we focus on the modeling and optimization of $R_{ZZ}(\theta)$ Rydberg gates implemented with ${}^{171}\mathrm{Yb}$ atoms~\cite{ma2023,peper2024,senoo2025}, noting that our considerations readily extend to other alkali-earth atoms, such as strontium~\cite{tsai2024}.
Unlike the construction proposed in Ref.~\cite{zeng2025_pftbias}, which to our knowledge has not yet been tested in neutral-atom platforms, our study builds directly on experimentally demonstrated gate implementations.
We first introduce Rydberg gates and their dominant noise sources, then optimize the pulse specifically against faults that remain undetected by the TMR protocol, and finally quantify the error channel of the gates.

\subsection{Rydberg Gates}
\label{subsec:rydberg-gates}

\begin{figure*}[!th]
    \includegraphics[width=\linewidth]{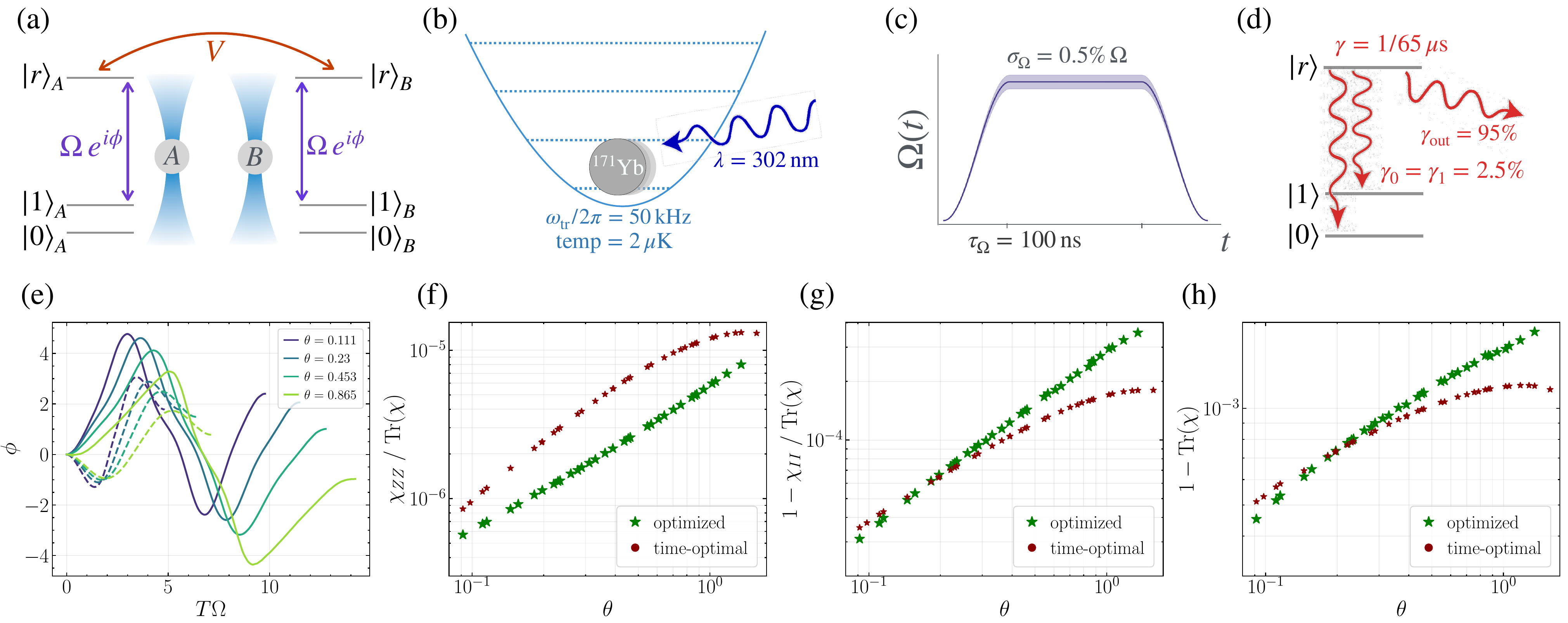}
    \phantomsubfloat{\label{fig:sub:rydberg-gate}}
    \phantomsubfloat{\label{fig:syb:rydberg-noise-atomic-motion}}
    \phantomsubfloat{\label{fig:sub:rydberg-noise-amplitude}}
    \phantomsubfloat{\label{fig:sub:rydberg-noise-decay}}
    \phantomsubfloat{\label{fig:sub:rydberg-optim-pulses}}
    \phantomsubfloat{\label{fig:sub:rydberg-optim-chi}}
    \phantomsubfloat{\label{fig:sub:rydberg-optim-inf}}
    \phantomsubfloat{\label{fig:sub:rydberg-optim-leak}}
    \vspace*{-2em}
    \caption{
        Rydberg gate and associated noise, with relevant parameters.
        (a) The two-qubit Rydberg gate exploits the blockade interaction $V \ket{rr}\bra{rr}$, arising when the $\ket{1}$ state of both atoms is coupled to the highly-excited Rydberg state $\ket{r}$.
        (b) We consider ${}^{171}\mathrm{Yb}$ atoms trapped at low yet finite temperature, thus undergoing atomic motion coupled with recoil from the driving photons.
        (c) The amplitude of the driving laser has finite ramping time, modeled by a cosine-tapered envelope, as well as gaussian fluctuations in its value. 
        (d) The Rydberg state has finite lifetime $1/\gamma$, thus experiences decay both outside the computational subspace, as well as back to the computational states.
        (e) Examples of time-optimal (dashed) and noise-optimized (solid) pulses for different physical angles of rotation.
        (f,g,h) Comparison of relevant error metrics between time-optimal and noise-optimized pulses at different angles of rotation: (f) probability of undetectable error post-selected upon leakage; (g) error rate post-selected upon leakage; (h) leakage probability.
    }
    \vspace*{-1em}
    \label{fig:rydberg-gate-and-noise}
\end{figure*}

Multi-qubit $Z$-type rotations can be natively implemented in neutral atoms.
Typically, these are two-qubit gates relying on the repulsive van der Waals $V$ interaction, which arises when the atoms are in the same Rydberg state $\ket{r}$, a highly excited atomic state. 
Specifically, Rydberg gates are implemented by state-selectively coupling the atomic qubit of each atom, conventionally from its $\ket{1}$ state, to the Rydberg level via a laser, as illustrated for two atoms in~\cref{fig:sub:rydberg-gate}.
The laser amplitude $\Omega$, phase $\phi$, and detuning $\Delta$ can be modulated in time to yield the desired gate, ranging from the maximally-entangling $CZ$ gate~\cite{jaksch2000,levine2019,jandura2022}, to arbitrary angle $R_{ZZ}(\theta)$ rotations~\cite{evered2023, fromonteil2024}.
In practice, only one between laser phase and detuning is typically modulated, as they are related up to a change of reference frame, as $\dot\phi=\Delta$~\cite{fromonteil2024}.

General $N_q$-qubits Rydberg gates are realized by the Hamiltonian
\begin{align}
    \label{eq:rydberg-hamiltonian}
    H(t) = H_\mathrm{drv}(t) + H_\mathrm{int}(t),
\end{align}
with
\begin{align}
    \label{eq:rydberg-hamiltonian-defs}
    H_\mathrm{drv}(t)/\hbar = & 
    \sum_{q}^{N_q} \left( \frac{\tilde\Omega(t)}{2} \ket{r} \bra{1}_q + \mathrm{\hc} \right)
    - \Delta(t) \ket{r} \bra{r}_q, \nonumber \\
    H_\mathrm{int}(t)/\hbar = & \sum_{q<q'}^{N_q} V_{qq'} \ket{rr}\bra{rr}_{qq'},
\end{align}
where $\tilde\Omega(t) = \Omega(t) e^{-i \phi(t)}$ and $V_{qq'}=C_6/R_{qq'}^6$, where $C_6$ is a coefficient of the chosen Rydberg state, and $R_{qq'}$ is the distance between atoms $q$ and $q'$.

State-of-the-art realizations have achieved $CZ$ gates fidelity exceeding $ 99.5\%$, and native three-qubit $CCZ$ gates have also been demonstrated, with fidelity $\sim98\%$~\cite{evered2023}.

\subsection{Noise on Rydberg Gates}
\label{subsec:rydberg-noise}

In this subsection, the noise model considered for the Rydberg gates is outlined.
Specifically, we consider (i) atomic motion due to finite temperature and photon recoil, (ii) driving imperfections, including shot-to-shot fluctuations and finite amplitude rise- and fall- times, and (iii) Rydberg decay, both into and out of the computational subspace.
These noise sources are schematically depicted in~\cref{fig:rydberg-gate-and-noise}, together with the relevant parameters adopted in our simulations.
The laser phase is modulated in time and the laser detuning set to zero.

For modeling of atomic motion, we assume that the atoms sit in harmonic traps with radial frequency $\omega_\mathrm{tr}$, inverse temperature $\beta$, and initial thermal state $\rho \propto \exp \big( - \beta \hbar \omega_\mathrm{tr} a^\dagger a^{\phantom{\dagger}} \big)$, where $a$ ($a^\dagger$) is the annihilation (creation) operator of the relevant motional modes, described by the usual Hamiltonian $H_\mathrm{mot}=\hbar \omega_{\rm tr} a^\dagger a $.
Then, the driving term in the Hamiltonian is modified to include the momentum transfer of the laser field, with wavelength $\lambda$,
\begin{align}
        H_\mathrm{drv}(t)/{\hbar}  & =  \frac{1}{2}  \sum_{q}^{N_q} \! \left( \tilde \Omega(t) e^{i \eta ( a^{\phantom{\dagger}}_q + a^\dagger_q  ) } \ket{r} \! \bra{1}_q  + \mathrm{\hc}  \right),
        \label{eq:recoil}
\end{align}
with Lamb-Dicke parameter $\eta = 2 \pi x_\mathrm{osc} / \lambda$, where $x_\mathrm{osc} = \sqrt{ \hbar / ( 2 m  \omega_\mathrm{tr}  ) }$.
Moreover, the van der Waals potentials are expanded to first order in the interatomic distance fluctuations $\delta R_{qq'} = x_q - x_{q'}$, where $x_q = x_\mathrm{osc} ( a^\dagger_q + a_q )$, such that
\begin{align}
       H_{\mathrm{int}}/{\hbar}  \simeq \sum_{q<q'}^{N_q} V_{qq'} \! \left( 1 -  \frac{6 ( x_q -  x_{q'} ) }{R_{qq'}} \right)  \ket{rr} \! \bra{rr}_{qq'}. \label{eq:fluct_potentials}
\end{align}
In~\cref{subsec:app:rydberg-axial-fluctuations} we argue that axial fluctuations, ignored in our computation, indeed only have a small impact in the considered set-up.

The driving amplitude $\Omega=|\tilde\Omega|$, rather than being constant throughout the protocol, is assumed to have a finite ramping time $\tau_{\Omega}$, with cosine-tapered shape~\cite{senoo2025}.
That is, the amplitude is initially ramped-up with profile $\Omega(t<\tau_{\Omega}) = \Omega_{\rm max} \sin^2 [\pi t /(2\tau_{\Omega})]$, and similarly ramped-down in the final stage.
The laser phase is accordingly adapted. 
Moreover, the amplitude has shot-to-shot gaussian fluctuations $\delta \Omega$ with standard deviation $\sigma_{\Omega}$.
Due to the off-resonant coupling of the laser to other states in the Rydberg manifold, these deviations lead to self-light shifts, i.e detuning fluctuations correlated with those on the driving amplitude, as $\delta \Delta_q = 2 \kappa \, \Omega\delta \Omega$.
With regards to alkali-earth implementations, $2 \pi \kappa \approx 10\, \mathrm{kHz/MHz^2}$ was observed in strontium~\cite{madjarov2020}, and a twice-as-large upper bound was determined in ytterbium~\cite{zhang2025}.
In our simulations, we set $\kappa = 2\, \mathrm{kHz/MHz^2}$.

Finally, Rydberg gates suffer from decay from the Rydberg state, often one of the major sources of infidelity~\cite{evered2023, radnaev2024, tsai2024, peper2024, muniz2025, senoo2025}.
Rydberg decay outside of the computational subspace can be modeled by the addition of a non-Hermitian effective Hamiltonian
\begin{equation}
    H_\mathrm{out} /\hbar= - i \frac{ \gamma_\mathrm{out}}{2} \sum_{q=1}^{N_q} \ket{r} \! \bra{r}_q,
    \label{eq:decay}
\end{equation}
whereas decay to the computational state $\ket{0}$ and $\ket1$ is modeled by jump operators $L_q^{(i)} = \sqrt{\gamma_\mathrm{i}} \ket{i} \bra{r}_q$, where $\gamma_{0/1}$ is the decay rate to $\ket{0/1}$.  
It has been observed that in the metastable qubit of ${}^{171} \mathrm{Yb}$ here considered, $\gamma_\mathrm{out}$ accounts for 95\% of the total Rydberg decay rate $\gamma \approx 1/(65\,\mu \mathrm{s})$~\cite{ma2023}.
This results in leakage to levels whose population can be continuously monitored without affecting the qubit, thus detectable \textit{erasure} errors~\cite{sahay2023}.
In the context of the transversal rotations on the TMR's resource state, the state can be discarded upon detection of erasures, thus leakage does not contribute to the logical error rate.
Finally, we note that by using alternative leakage-detection units, most leakage events can be detected in absence of built-in erasure capabilities, at the price of additional overhead~\cite{cong2022}.

\subsection{Gate Optimization}
\label{subsec:rydberg-optim}

State-of-the-art implementations adopt the so-called time-optimal gate proposed in Ref.~\cite{jandura2022}, or an approximation thereof~\cite{evered2023, radnaev2024, muniz2025, tsai2024, peper2024, senoo2025}.
This is the laser pulse realizing the desired gate in the shortest time in a noiseless scenario, up to additional single-qubit $Z$-rotations.
However, gates can also be optimized to increase their robustness to noise, as previously demonstrated for Rydberg gates in different scenarios~\cite{pelegri2022, fromonteil2023, jandura2023, giudice2025, jandura2024}.

Given the partially FT nature of the TMR protocol, specifically its sensitivity to $ZZ$-type errors in the application of physical $R_{ZZ}(\theta)$ gates, we optimize the latter at the pulse level, to further enhance the logical performance of the protocol.
Therefore, we perform \textit{error-selective} pulse optimization, not to suppress the overall infidelity, as commonly done, but specifically the contribution of those undetectable $ZZ$-type errors.
Such error-selective approaches have only recently attracted attention in neutral atoms~\cite{jandura2024, pecorari2025-noise-selective-optim} and superconducting architectures~\cite{tasler2025}, albeit within the context of quantum memories rather than logical computation.

In the remainder of the section, we focus on two-qubit Rydberg gates.
The laser field is assumed to be global, in line with most experimental implementations~\cite{bluvstein2025, muniz2025, tsai2024}, with constant amplitude $\Omega$, null detuning $\Delta$, and time-modulated phase $\phi(t)$.
We adopt the cost function
\begin{equation}
    \mathrm{cost} = \chi_{\scriptscriptstyle ZZ,ZZ} + (1-F)\,\eta_F,
    \label{eq:cost-function}
\end{equation}
where $\chi_{\scriptscriptstyle ZZ,ZZ}$ denotes the probability of a $ZZ$-error following the gate, whereas $(1-F)$ is the gate's infidelity, entering the cost function with weight $\eta_F$. 
We choose $\eta_F=10^{-3}$, to make the $ZZ$-error the focus of the optimization process, while ensuring the gate maintains sufficiently high fidelity.
The cost function is computed under the effect of the noise sources mentioned in the previous subsection, although appropriately approximated to be efficiently estimated, as detailed in~\cref{subsec:app:rydberg-noise-approximation}, while the computation of $\chi_{\scriptscriptstyle ZZ,ZZ}$ is described in~\cref{subsec:app:rydberg-chizz-computation}.

Numerical optimization is carried out via gradient-ascent pulse engineering (GRAPE)~\cite{khaneja2005, garon2013, anderson2015, riaz2019}, in which the time-dependent control function $\tilde \Omega(t)$ is discretised as a piecewise function over a time-interval $T$.
The optimization is then described by $N\gg1$ parameters $\phi_1, ..., \phi_N$, such that $\tilde \Omega(t_i) \equiv \Omega e^{-i \phi_i}$, where $t_i = (i+1/2)dt$, and $dt=T/N$.
We choose $N=600$.
Two other parameters are optimized over.
The first is a residual single-qubit phase, as usual in the context of Rydberg gates~\cite{jandura2022, jandura2023, giudice2025}.
The second is the total gate time $T$, optimized together with the other parameters to reduce the number of iterations.
This is uncommon, as usually the optimization is repeated for increasing gate times, until a time $T$ is found for which the cost function drops significantly. 
We observe that the piecewise approximation of the control function, and the combined optimization of parameters and gate time, can lead to numerical instabilities, suppressed by appropriately designed penalties in the cost function, as further commented on in~\cref{subsec:app:rydberg-T-optimization-details}.

\subsection{Gates Performance}
\label{subsec:rydberg-performance}

We computed the performance of the time-optimal and noise-optimised gates, by simulating the Lindbladian dynamics described by the Hamiltonian and jump operators in~\cref{subsec:rydberg-noise}.
The performance was evaluated at experimentally realistic amplitude $\Omega=7.5\times 2\pi\,\mathrm{MHz}$, atomic distance $R=2.5\;\mu\mathrm{m}$, and van der Waals potential $V=63.7 \Omega$.
The simulation was discretised in $3000$ time-steps; 8 vibrational levels were used per atom; shot-to-shot noise in the Rabi frequency was sampled according to the Gauss-Hermite quadrature over 15 realizations.
It was verified that these parameters were sufficient to obtain converged results.
Residual population in the Rydberg state at the end of the pulse was mapped to computational states and leakage according to the branching ratios in~\cref{fig:sub:rydberg-noise-decay}.

In~\cref{fig:rydberg-gate-and-noise}, we compare the time-optimal and optimized pulses for arbitrary angle rotations $R_{ZZ}(\theta)$.
We note that, with the adopted parameters, the error rate of the $CZ$ gate amounts to 0.14\%, thus only slightly better than best reported experimental values.
\Cref{fig:sub:rydberg-optim-pulses} compares their shape, where we observe that the optimized pulses exhibit a smooth profile, and are implemented in approximately twice the time of the time-optimal counterparts.
\Cref{fig:sub:rydberg-optim-chi} compares the probability of undetectable errors under realistic noise, with the optimized pulses consistently providing a two- to three-fold reduction.
Finally, \Cref{fig:sub:rydberg-optim-inf,fig:sub:rydberg-optim-leak} show that the optimized gates' infidelity---post-selected upon leakage events---and probability of leakage are only mildly increased with respect to the time-optimal pulse.
As it will be shown in the next section, these results directly translate to a twofold reduction in the logical error rate of the protocol, with negligible increase in post-selection.

However, two more remarks are in order.
First, we note that previous work has shown that no pulse can suppress leading-order sensitivity of Rydberg gates to atomic motion, unless combined with fine-tuned experimental tricks~\cite{jandura2023}.
These are in turn exposed to other error sources and experimental challenges, and are not assumed in our work.
This vulnerability could limit the suppression of the undetectable error in our optimized pulses.
Secondly, we note that the reduction of the cost function during the optimization process, as opposed to the value associated with time-optimal pulses, was by nearly an order of magnitude, hence more pronounced than observed in~\cref{fig:sub:rydberg-optim-chi}. 
This discrepancy is likely due the difference between the approximate noise model adopted in the optimization and the detailed noise model used in the final benchmarking. 
In a real experimental scenario, the method's improvements may be maximized by spending additional resources in a more detailed optimization process.
We defer this investigation to future work.

\section{Logical Performance}
\label{sec:performance}

In this section, we first outline the underlying assumptions in the simulation of the logical TMR protocol, specifically the adopted noise model and decoding.
Then we show and comment on the numerical performance of its repeat-until-success implementation.

\subsection{Assumptions}
\label{subsec:performance-assumptions}

The protocol is assumed to exploit the all-to-all connectivity enabled by atomic shuttling~\cite{bluvstein2024, zhang2025}, in turn enabling the realization of (fold-)transversal logical gates.
Owing to their deterministic nature and the use of correlated decoding, these require only $\sim1$ round of syndrome extraction~\cite{zhou2024-algorithmicFT}, as opposed to traditional lattice surgery logic.
Despite increasing the total volume of classical decoding, efficient decoding has been shown to remain viable~\cite{cain2024, cain2025, ismail2025}.
As such, our simulations incorporate a single noisy round of stabilizer extraction on the logical qubit following the encoding of the resource state and gate teleportation, as shown in~\cref{fig:sub:teleportation}.
Classical decoding is performed with the recently developed Minimum Weight Parity Factor (MWPF) decoder~\cite{wu2025-mwpf}, with cluster node limit set to 300.
More details on the decoder, and its comparison against popular alternatives, can be found in Ref.~\cite{ismail2025}.

The protocol is simulated against circuit-level noise of varying strength, modeled by the error rate $p_2\equiv p$ of $CZ$ gates---the only natively available Clifford two-qubit gates.
Single qubit gates have depolarizing rate $p_1=p_2/5$, while measurements and resets---only assumed to be available in the computational basis---have bit-flip rate $p_m=p_2/2$ and $p_i=p_2/5$, respectively.
These values roughly reproduce typical state-of-the-art devices using alkali-earth atoms \cite{ma2023, muniz2025, peper2024, finkelstein2024}.
The error channel $\chi(\theta)$ of the non-Clifford transversal rotations, numerically computed as per section~\cref{sec:gate-optim} with parameters corresponding to $p_2=0.14\%$, was linearly rescaled to match the circuit-level error rate.

A key component of the model is the biased erasure noise of the $CZ$ gates, the native two-qubit gates in atomic platforms, exhibited by alkali-earth atoms \cite{ma2023, sahay2023}.
As mentioned in~\cref{subsec:rydberg-noise}, Rydberg decay is one of the main sources of infidelity, and mainly results in \textit{state-dependent} leakage from the $\ket{1}$ state.
As such, re-initialization in $\ket{1}$, or replacement of the atom by a qubit in $\ket1$, does not flip its computational state, but only affects the phase, resulting in a $Z$-biased noise channel.
While the details of the $CZ$ noise modeling, including the adopted Kraus channel, are referred to~\cref{app:logical-simul-noise}, we note that previous work has shown that this form of bias can significantly increase the QEC threshold \cite{wu2022, ma2023, sahay2023, veroni2024}.
Moreover, due to the sensitivity of the TMR protocol to phase-flips, erasures in $CZ$ gates acting on the first two rows of the logical patch---thus in proximity of the transversal physical rotations---are post-selected upon.

\subsection{Simulation and Results}
\label{subsec:simulation-results}

We numerically simulated the logical performance of the RUS procedure for the realization of arbitrary-angle logical rotations $R_{Z,L}(\theta_L)$ in the rotated surface code with distance $d=5,7, \text{or}\; 9$.
The resource state is initially encoded by one round of syndrome extraction at distance $d_s=d-1$ respectively, then $r=d_s/2$ transversal physical rotations are applied, and finally the state is grown to the target distance $d$.
The resource state is injected upon measurement of trivial syndromes, and discarded otherwise.

The key figure of merit used to assess the protocol's robustness is the diamond-norm error $\varepsilon_\diamond$, also known as the worst-case error rate. 
This is more sensitive than infidelity to coherent errors~\cite{sanders2016_diamondnorm, huang2019_diamondnorm}, determines the maximum probability of distinguishing the noisy and ideal outputs~\cite{kueng2016_diamondnorm, kliesch2021_diamondnorm}, and plays a central role in rigorous proofs of fault tolerance thresholds~\cite{aharonov1998, aliferis2005-exRec}. 
In this scenario, due to the strong suppression of coherent noise in the RUS protocol, $\varepsilon_\diamond \simeq p_L$.
More details on the diamond-norm error can be found in~\cref{subsec:app:error-budget:error-metrics}.
Another important metric to assess the scheme's viability is the average number $A(\theta_L)$ of attempts required to successfully encode a resource state. 
Although omitted for compactness, this also depends on the physical error model and details of the TMR implementation.

\begin{figure}[!t]
    \centering
    \includegraphics[width=0.47\textwidth]{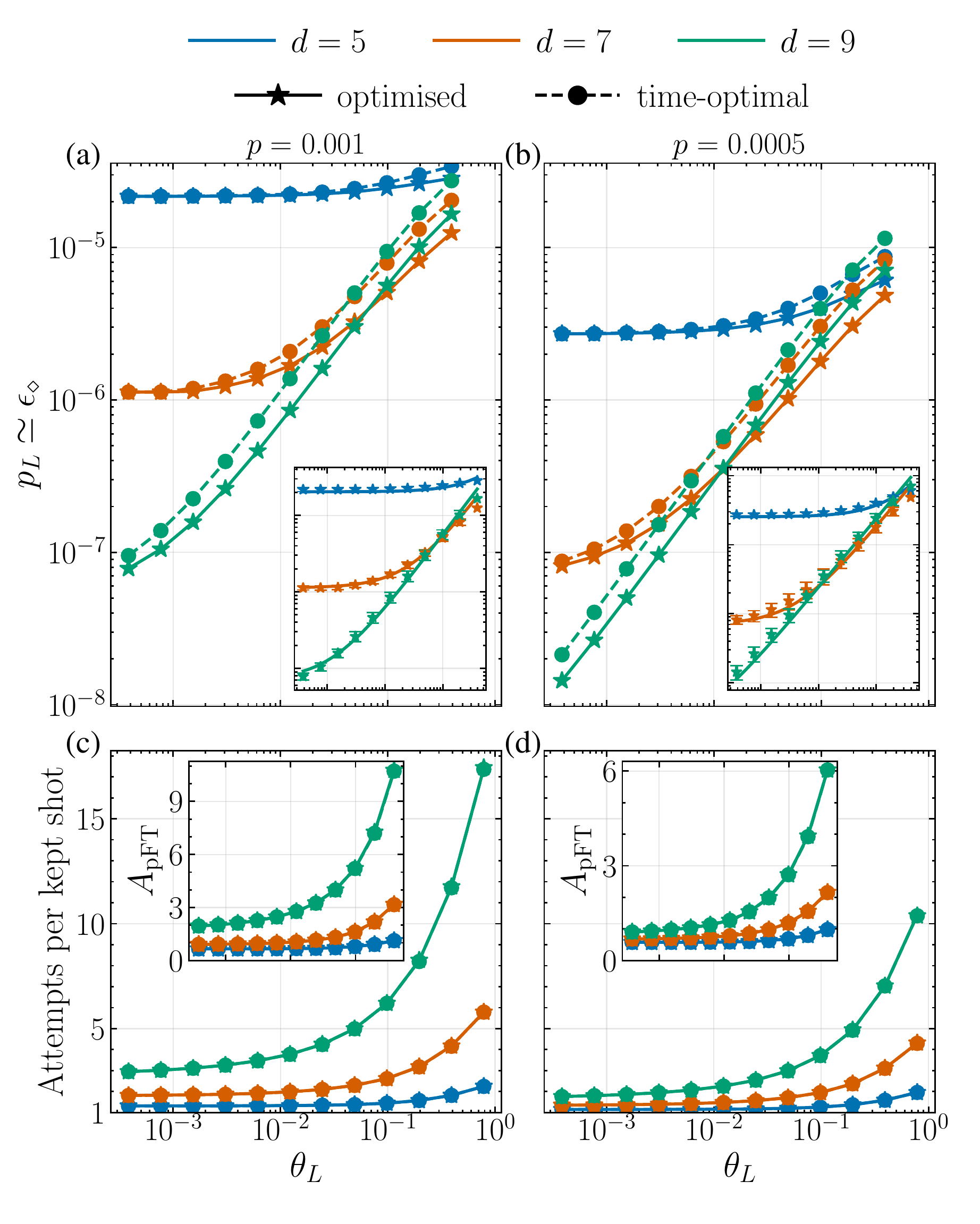}
    \phantomsubfloat{\label{fig:pFT-results-pL0.001}}
    \phantomsubfloat{\label{fig:pFT-results-pL0.0005}}
    \phantomsubfloat{\label{fig:pFT-results-ApFT0.001}}
    \phantomsubfloat{\label{fig:pFT-results-ApFT0.0005}}
    \vspace{-2em}
    \caption{
        Logical performance of the RUS execution of the TMR protocol. 
        The rows show the logical error rate, and the average number of encoding attempts per resource state.
        Columns correspond to two-qubit error rates $p = 0.1\%$ and $0.05\%$ respectively, based on which other error rates are rescaled.
        Code distances $d=5,7,9$ are considered, with corresponding distance $d_s=d-1$ at which the physical transversal rotations are applied.
        The use of time-optimal (dashed) and noise-optimized (solid) Rydberg gates is compared.
        (a-b) Logical error rate of an arbitrary-angle $R_Z(\theta_L)$ rotation.
        Markers are presents for both $p_L$ and $\epsilon_\diamond$, indistinguishable due to their closeness.
        The inset compares numerical simulations (markers) against the analytic fit in~\cref{eq:pL-pFT-ansatz} (lines) upon use of the optimized pulses, showing excellent agreement.
        Errorbars represent the standard error on the numerical estimates, and are shown in the inset only for visual clarity.
        (c-d) Average number of attempts per encoded resource state, i.e. inverse of the probability the encoding procedure passes post-selection.
        Numerical estimates (markers) are compared against the analytical ansatz in \cref{eq:attempts-per-thetaL(step)-definition-main-text} (lines) showing excellent agreement.
        The inset shows the behavior of the $A_{\rm pFT}(\theta_L)$ metric, which can be interpreted as an \textit{effective} number of attempts per encoded state in the RUS process, over a batch of parallel rotations (see~\cref{app:subsec:resources-pft}).
    }
    \label{fig:pFT-results-pL-ApFT}
\end{figure}

We reconstruct these metrics by combining Clifford-circuit simulations, performed with \texttt{stim}~\cite{stim}, with the physical chi matrix of the transversal rotations.
Our method is thoroughly described in~\cref{app:logical-TMR-simul-method}, and we stress that it enables the evaluation of the logical channel without further approximation in the treatment of the non-Clifford operations.
Details on the number of shots simulated at different distances and error rates, and on the estimation of uncertainty, can be found in~\cref{subsec:app:TMR-noise-channel-numerical-hyperparams-approxims}. 
The results are shown in~\cref{fig:pFT-results-pL-ApFT}, for two-qubit error rates $p\equiv p_2=10^{-3}$ and $5\times 10^{-4}$.
The simulations compare the use in the transversal rotations of time-optimal Rydberg gates against the optimized pulses found in \cref{sec:gate-optim}.

The logical error rate is displayed in \cref{fig:pFT-results-pL0.001,fig:pFT-results-pL0.0005}, and is well approximated by the following ansatz
\begin{equation}
    p_L(\theta_L, p, d, r) \simeq \alpha_{\rm RUS}(p,r)\,p\,|\theta_L| + 2 p_{L}^{({\rm inj})} (p,d),
    \label{eq:pL-pFT-ansatz}
\end{equation}
as highlighted in the insets.
The first term is specific to the TMR protocol, depends linearly on $|\theta_L|$, and dominates for larger $|\theta_L|$ and relatively small code distance and error rate. 
As expected, the use of optimized pulses yields approximately a twofold reduction in $\alpha_{\rm RUS}$, notably without additional overhead.
We estimate $\alpha_{\rm RUS}(r,p)\simeq r(0.01 + 4p)$ upon use of the optimized pulses. 
This is between one and two orders of magnitude lower than previous estimates~\cite{toshio2025, ismail2025}, which assumed an angle-independent noise model for the transversal rotations.
Such a reduction stems from the use of post-selection upon leakage, the favorable noise bias in Rydberg gates observed in~\cref{sec:gate-optim}, and the pulse-level optimization.
This reduction significantly improves the application-level resource estimates considered in the next section.
The second error term comprises errors in the Clifford operations involved in state injection, and is suppressed exponentially in the code distance as
\begin{equation}
    p_{L}^{({\rm inj})} (p,d) = \frac{a_{\rm inj} }{d-1} \left(\frac{p}{p_{\rm th}}\right)^{\lceil d/2 \rceil},
    \label{eq:pL-inj-ansatz}
\end{equation}
with fitted $a_{\rm inj}\simeq 0.070$ and $p_{\rm th}\simeq0.012$ under the considered noise model.
%clf data here
Its contribution is twice as much because the RUS protocol requires approximately two repetitions on average.
At near term code distances and error rates, it represents the main limitation to the performance of the TMR protocol at small $|\theta_L|\lesssim 10^{-3}$.
This suggests that the recently proposed combination of the TMR protocol and magic state cultivation~\cite{toshio2026}, whose advantages manifest at smaller angles,  will only be attained in neutral-atom devices at lower error rates.

The average number $A(\theta_L)$ of encoding attempts required per kept resource state is displayed in~\cref{fig:pFT-results-ApFT0.001,fig:pFT-results-ApFT0.0005}.
This is inverse of the probability of accepting the encoded state, thus scaling as
\begin{equation}
    A^{-1}(\theta_L) \approx p_{\theta_L} \cdot (1-p_{\rm erasure}) \cdot p_{\rm pass}.
    \label{eq:attempts-per-thetaL(step)-definition-main-text}
\end{equation}
All factors depend on the details of the TMR implementation and error model, and are, respectively, the probability \cref{eq:pthetal} of collapsing on the correct state, \cref{eq:app:p-erasure} of observing no erasure error, and that of observing no other faults.
With the empirical fit
$p_{\rm pass} \approx (1 -4.5p)^{[2(d^2-d)+(d_s^2-d_s)]}$, excellent agreement is achieved with the numerical simulations.

As further motivated in the next section and in~\cref{app:subsec:resources-pft}, $A(\theta_L)$ does not by itself capture the overhead of parallel executions of multiple RUS rotations.
For application-level scheduling, the overhead is better described by the metric \(A_{\rm pFT}(\theta_L)\).
This can be interpreted, in analogy with standard gate synthesis by Clifford$+T$ decomposition, as the \textit{effective} number of attempts per injection over the whole RUS protocol.
Specifically, for the RUS realization of a rotation by $\theta_L$, with each $s^{\rm th}$ step targeting $\theta_{L,s}$, it is defined as
\begin{align}
    A_{\rm pFT}(\theta_L) &= \max_{s\geq 1}\;  q_s(\theta_L) -1,
    \nonumber \\
    q_s(\theta_L) &= 2^{-s}
        \max\!\left(A(\theta_{L,s}, \frac{A(\theta_{L,s+1})}{2}+1\right)
    .
    \label{eq:ApFT-definition-main-text}
\end{align}
Its dependence on $\theta_L$, $d$ and $p$ is shown in the insets of \cref{fig:pFT-results-ApFT0.001,fig:pFT-results-ApFT0.0005}.
Remarkably, in the regime of main interest, $A_{\rm pFT}\simeq 1$ at moderate code distances, hence smaller than the equivalent metric in magic state cultivation~\cite{gidney2024-cultiv, sahay2025-cultiv, vaknin2026_Tsynth}.
The relevance of the metric to quantifying the overhead of partial fault tolerance in realistic applications is further discussed in~\cref{subsec:applications-limitations-circuit-depth}.

\section{Resource analysis for Near-Term Applications}
\label{sec:applications}

This section quantifies the resource requirements of the transversal STAR architecture and compares this with two fully FT approaches: standard Clifford$+T$ synthesis~\cite{kliuchnikov2023_Tsynth} and Hamming Weight Phasing (HWP)~\cite{gidney2018_Tsynth, kivlichan2020_FH}.  Throughout we quantify the limitations and  trade-offs between low qubit counts,  high accuracy, and short circuit depth, broken down as follows.

\begin{figure*}[t]
    \centering
    \includegraphics[width=\textwidth]{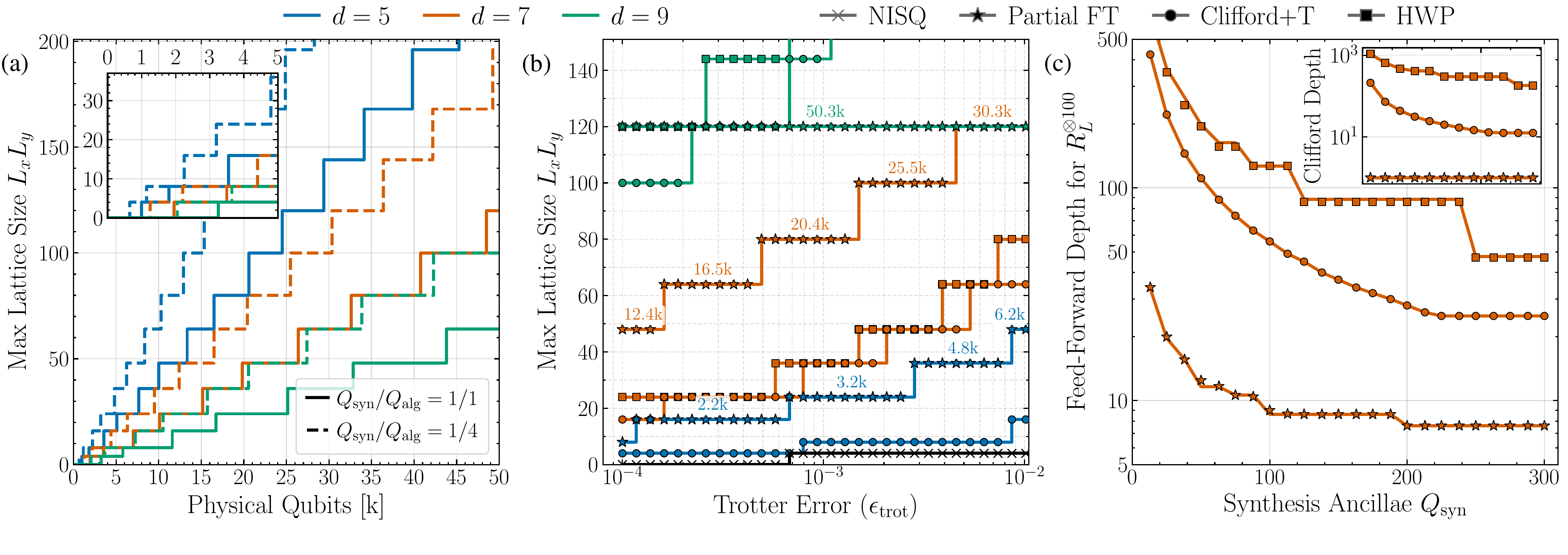}
    \phantomsubfloat{\label{fig:sub:pFTvsFT-physical-qubits}}
    \phantomsubfloat{\label{fig:sub:pFTvsFT-maxM-per-PL}}
    \phantomsubfloat{\label{fig:sub:pFTvsFT-depth}}
    \vspace{-3em}
    \caption{
        Overview of resource requirements for partially and fully FT simulations. 
        For concreteness, the problem considered is the time evolution of the Fermi-Hubbard model on a square lattice with periodic boundaries.
        We consider rotated surface encoding at moderate code distances $d=5,7,9$, and physical error rate $p=10^{-3}$.
        The partially FT approach (stars) is compared against standard Clifford+$T$ synthesis (circles) and Hamming weight phasing HWP (squares), as well as a NISQ implementation (crosses) when applicable.
        (a) Maximum lattice size supported by growing hardware sizes.
        This is independent of the approach, and determined by code distance $d$ and the fraction $Q_{\rm syn}/Q_{\rm alg}$ of logical patches dedicated to the synthesis of non-Clifford operations (solid and dashed lines).
        The inset provides a closer look to smaller hardware sizes.
        (b) Maximum lattice size per allowed Trotter error on the final observable, with evolution-time proportional to the lattice's side.
        At smaller distances, the partially FT approach enables simulation of larger systems, whereas the advantage is lost at distance $d\geq 9$.
        Labels over partially FT datapoints indicate the number of physical qubits required for simulations at given code distance and lattice size, with $Q_{\rm syn}/Q_{\rm alg}=1/4$.
        %clf data here or consequence of
        (c) Circuit depth for the synthesis of one layer of $100$ parallel $R_Z(\theta_L)$ rotations, at code distance $d=7$, given $Q_{\rm syn}$ ancillary logical patches dedicated to the realization of non-Clifford operations. 
        Depth is shown both in terms of the injection of resource states (main) and the associated Clifford depth (inset).
        Datapoints are numerical estimates, whereas lines are heuristic analytic estimates, showing excellent agreement.
        %clifford depth modeling assumptions here
    }
    \vspace{-0.5em}
    \label{fig:pFTvsFT-resources}
\end{figure*}

\Subsecref{subsec:applications-limitations-hardwarep-size} addresses the limitations from hardware size on the availability of logical qubits. 
Given an implemented QEC code and code distance, independent of the approach, hardware size poses stringent requirements on the availability of logical qubits.
See also~\cref{fig:sub:pFTvsFT-physical-qubits}.

\Subsecref{subsec:applications-limitations-logical-error} addresses the limitations from logical error rates on the achievable simulation size.
Focusing on Trotterised dynamics, gate counts are defined by its digital implementation and the target Trotter error; code distance and physical noise define a logical error model.
Together, these determine the feasibility of quantum simulations at different system sizes---assuming the combination of QEC and quantum error mitigation.
See also~\cref{fig:sub:pFTvsFT-maxM-per-PL}.

\Subsecref{subsec:applications-limitations-circuit-depth} addresses the limitations from circuit depth. 
In particular it quantifies the trade-off between depth and the availability of non-Clifford resources, depending on the computation approach.
See also~\cref{fig:sub:pFTvsFT-depth}.

We consider implementations with the rotated surface code.
Altogether, we argue that at moderate code distances $d=5\,\mathrm{or}\,7$, partial fault tolerance can support the simulation of many-body physics at system sizes beyond the reach of its fully FT counterpart, as well as of classical methods, thus providing a pathway to early demonstrations of quantum advantage.
In particular, simulations at distance $5$ beyond the reach of classical methods are expected to require less than $5{,}000$ qubits.
Moreover, we show that partial fault tolerance reduces "feed-forward" and Clifford depths by roughly one and two orders of magnitude, respectively, relative to fully FT implementations.
The reduction in depth is accentuated when only a small fraction of the hardware is allocated to non-Clifford resource-state production, as expected in early FT devices.

In quantifying these trade-offs, we focus on the specific case of time evolution in the (Fermi--)Hubbard model, a paradigmatic model of strongly correlated fermions. 
The model describes spin-$1/2$ fermions hopping between neighboring lattice sites and interacting through an on-site interaction, with Hamiltonian
\begin{align}
  H_{FH}/\hbar = -J\!\sum_{\langle i,j\rangle,\sigma}\!\left(c^{\dagger}_{i\sigma}c_{j\sigma} + \mathrm{\hc}\right) + U\sum_{i=1} n_{i\uparrow}n_{i\downarrow},
\end{align}
where $c_{j\sigma}^\dagger(c_{j\sigma})$ is the fermionic creation(annihilation) operator, $\langle i,j\rangle$ denotes nearest-neighbors on the lattice, and $n_{i\sigma}=c_{i\sigma}^\dagger c_{i\sigma}$.
Despite its simple structure, it exhibits rich many-body physics such as antiferromagnetism, Mott insulating behavior, and a simplified picture of high-temperature superconductivity~\cite{arovasetal2022_fh}. 
Its relative simplicity compared with full electronic-structure Hamiltonians, together with its display of strongly correlated dynamics, has made the Hubbard model a long-standing benchmark for analogue and digital quantum simulation~\cite{ flannigan2022_FH, kivlichan2020_FH, arovasetal2022_fh, campbell2022_FH, wecker2015_fh, jiang2018_fh, stanisic2022_fh, baySmidt2025_FH, kanSymons2025_FH, akahoshi2024a, toshio2025}.
For its digital implementation, the fermionic Hamiltonian is mapped to Pauli operators using the Jordan–Wigner (JW) transformation, requiring no additional ancillary qubits.
Our protocol is tailored to early FT architectures, where Clifford operations still have a non-negligible cost, by making frequent use of the Clifford-count reduction technique introduced in Ref.~\cite{hastings2015_FH}, and adopted in Ref.~\cite{flannigan2022_FH}, of which we provide an extension to periodic boundary conditions.
Details of the Hamiltonian, the adopted Trotter decomposition, and our circuit implementation are given in~\cref{sec:app:fermi-hubbard-fh}.
Results provided in the remainder of the section assume $U=4J$ and half-filling, as this represents a challenging parameter regime for classical simulation~\cite{arovasetal2022_fh, kivlichan2020_FH, campbell2022_FH}.

\subsection{Limitations from hardware size}
\label{subsec:applications-limitations-hardwarep-size}

The physical-qubit budget imposes the first bound on the simulated lattice size.
The spatial footprint of the error-corrected simulation is determined by the following factors.
\begin{enumerate}[label=(\roman*)]
    \item The size of each logical qubit, defined by the encoding and code distance.
    We consider encoding in the rotated surface code, requiring $2d^2-1$ physical qubits per logical patch.
    \item The number $Q_{\rm alg}$ of \textit{algorithmic} logical qubits, i.e. those required for the algorithm implementation, neglecting the cost of non-Clifford operations. 
    This is determined by the selected application and related system size. 
    For the simulation of the Hubbard model on a square $L_x \times L_y$ lattice with periodic boundary conditions, our implementation requires $Q_{\rm alg}=2L_xL_y + L_y$ algorithmic qubits (see~\cref{sec:app:fermi-hubbard-fh} for more details).
    \item The number $Q_{\rm syn}$ of ancillary patches dedicated to the synthesis of non-Clifford operations.
\end{enumerate}
Altogether, the number of physical qubits required amounts to
\begin{equation}
    Q_{\rm phys} \simeq (2d^2-1)(Q_{\rm alg} + Q_{\rm syn}).
\end{equation}

\Cref{fig:sub:pFTvsFT-physical-qubits} quantifies the maximum lattice size in the Hubbard model that can be simulated at different hardware sizes, depending on the adopted code distance and the fraction of resources dedicated to non-Clifford synthesis.
It is not surprising that larger system sizes are achieved at smaller code distance and ancillary footprint.
However, these result in higher error rates and larger circuit depth, respectively, as discussed in the next subsections.

\subsection{Limitations from logical errors}
\label{subsec:applications-limitations-logical-error}

Early FT quantum computation is likely to be combined with quantum error mitigation (QEM) to extend its applicability.
QEM aims to return the noiseless expectation value $\langle O\rangle$ of a chosen observable, at the price of an increased variance in its estimator, hence sampling overhead.
For typical QEM techniques, the overhead increases exponentially with the accumulated error rate
\begin{equation}
    \Sigma_L = \sum_{g\,\in\,\rm circuit}  p_{L,g},
\end{equation}
where each gate $g$ has a logical error rate $p_{L,g}$, independently of the circuit's noise propagation.
For more details, we refer the reader to the review in Ref.~\cite{cai2023_qem}.
Therefore, $\Sigma_L\lesssim 1$ is demanded, as a general rule of thumb, to keep the sampling overhead within reasonable limits.
In this subsection, we quantify the maximum system size that can be simulated while ensuring that $\Sigma_L\lesssim 1$.

The accumulated logical error depends, in the first place, on the number of simulated Trotter steps, which is fixed by the target Trotter error.
This is in turn a function of the evolution time $T$ and the order of the Trotter decomposition.
The Trotter error is taken to scale as 
\begin{align}
    \epsilon_{\rm trot}^{(\rm ord)} \propto 
     N_{\rm trot} \, \left( \frac{TJ}{N_{\rm trot}} \right)^{\rm ord+1},
    \label{eq:trotter-error}
\end{align}
where $\rm ord$ denotes the order of the decomposition.
Rather than relying on theoretical bounds on the Trotter error, which are often loose~\cite{kivlichan2020_FH, campbell2022_FH, baySmidt2025_FH}, the proportionality constants were inferred from simulations performed with the \texttt{ffsim} package~\cite{ffsim}, as described in~\cref{sec:app:fermi-hubbard-fh}.
For typical observables---such as average density, occupancy, spin, double occupancy, density correlations, and spin correlations---representative proportionalities $10^{-2}$ and $3\times10^{-3}$ were inferred for the second- and fourth-order decompositions of the Hubbard Hamiltonian considered in the manuscript.
Setting the evolution time to grow with size as $TJ=(L_x+L_y)/2$ (so that information approximately propagates across the whole 2D system as we scale the size), the number of Trotter steps required to bound the Trotter error below a given target is readily estimated.
Gate counts are then obtained, given the adopted implementation of the Trotter steps.
The implementation we devised is detailed in~\cref{sec:app:fermi-hubbard-fh}.

The accumulated logical error then depends on the error model of the adopted gates.
We assume a single round of syndrome extraction is interleaved with each operation, as in the framework of algorithmic fault tolerance~\cite{zhou2024-algorithmicFT,cain2024,cain2025,serraperalta2025-decodetransversal}.
The available Clifford operations are  $H, S, S_X, CX, CZ$, with error rates given in~\cref{tab:gates-logical-error-rates} as derived in~\cref{subsec:app:logical-cliffords-simulation}. 
Partially FT computation realizes arbitrary-angle $\{R_Z, R_X\}$ rotations by means of the TMR protocol, with error rate~\cref{eq:rus-error-ansatz-main-text}.
Fully FT computation decomposes them into $\{T, T_X\}$ and Clifford gates.
With moderate code distances, the error budget of $T$ gates is assumed to be dominated by the injection process, such that $p_L^{(T)}\simeq p_L^{\rm (inj)}$, defined in~\cref{eq:pL-inj-ansatz}.
Given near-term constraints, such as limited ancillary resources, rotations in the fully-FT scenario are assumed to be realized by the most efficient \textit{ancilla-free} synthesis protocol, requiring $N_{T}^{\rm Tdec}(\epsilon) \simeq 1.52 \log_2 (\epsilon^{-1}) \; - 0.01$
T gates per rotation to achieve precision $\epsilon$~\cite{kliuchnikov2023_Tsynth}.
The optimal precision is chosen by minimizing the total error, i.e. the sum of $\epsilon$ and the error rates of the $T+$Clifford gates involved---further details are provided in~\cref{app:subsec:resources-tsynthesis}.
Finally, a digital NISQ approach is included in the comparison, with a simplified and optimistic error model: error rate $p$ for two-qubit gates and noiseless single qubit operations.

\begin{table}[t]
    \centering
    \caption{FT gates, followed by a single syndrome extraction round, have logical error rate
    \mbox{$p_L=a_{\bullet} (p/p_{\rm th, \bullet})^{\lceil d/2 \rceil}/(d-1)$}, with gate-dependent thresholds $p_{\rm th, \bullet}$ and prefactors $a_{\bullet}$ provided below, as derived in~\cref{subsec:app:logical-cliffords-simulation}.
    $T$ gates' noise is assumed to be dominated by Clifford contributions, given the moderate code distances considered in the manuscript.  
    %clf data here
    }
    \label{tab:gates-logical-error-rates}
    \renewcommand{\arraystretch}{1.5}
    \begin{tabular*}{\columnwidth}{
        @{\extracolsep{\fill}} l c c c c @{}
    }
        \hline\hline
        & $H$ & $S$ \& $S_X$ & $CX$ \& $CZ$ & $T$ \& $T_X$ \\
        \hline
        Threshold $p_{\rm th,\bullet}$
        & $0.015$
        & $0.009$
        & $0.008$
        & $0.012$ \\
        Prefactor $a_{\bullet}$
        & $0.032$
        & $0.033$
        & $0.025$
        & $0.070$ \\
        \hline\hline
    \end{tabular*}
\end{table}

Once the $\epsilon_{\rm trot}$-dependent gate counts and the $(d,p)$-dependent logical error rates are set, the accumulated error rate $\Sigma_L$, governing the cost of error mitigation, can be estimated at given system size.
\Cref{fig:sub:pFTvsFT-maxM-per-PL} shows the maximum system size that can be achieved in a square layout before the accumulated error rate exceeds 1, for different target trotter errors.
We recall that evolution time is chosen to scale proportional to the lattice's side.
The figure compares partially and fully FT approaches at a physical error rate $p=0.001$ and code distances $d=5,7,9$.
For each data point, we evaluate both second- and fourth-order Trotter formulas and retain the best result.
Often, the second-order decomposition was preferable for the partially FT approach, whereas full fault tolerance showed better performance under the fourth-order decomposition.

\begin{table*}[!ht]
    \centering
    \caption{Comparison of different synthesis protocols for the realization of a batch of $b$ parallel $R_Z(\theta_L)$ rotations.
    Estimates are derived in~\cref{sec:app:resources-arbitrary-rotations}, and represent well-approximating ansatze.
    The rotation angle is assumed to be a dyadic fraction of $\pi$, $\theta_L=(2k+1)\pi/2^{C+1}$ with $k$ and $C$ integers.
    $A_{\rm FT}$ is the average number of attempts required to encode the relevant fully FT resource state; $A_{\rm pFT}$ is the effective analogous for the partially FT RUS protocol.
    These depend on code distance and physical noise: typical values for neutral-atom implementations at moderate distance and error rate $p=0.001$ are $A_{\rm pFT} \gtrsim 1$ (see \cref{fig:pFT-results-ApFT0.001}) and $A_{\rm FT} \simeq 1.2$~\cite{vaknin2026_Tsynth}.
    $N_T^{\rm Tdec}(\epsilon)$ is the average number of T gates required to synthesize $R_Z$ rotations with precision $\epsilon$.
    The reduction factor $\mathfrak{r}\geq1$ governs the spacetime tradeoff between the availability of ancillary logical patches and circuit depth.
    %clifford depth modeling assumptions here
    }
    \label{tab:synthesis_comparison}

    \renewcommand{\arraystretch}{1.8}

    \begin{tabular*}{\textwidth}{
        @{\extracolsep{\fill}}
        l
        c
        c
        c
        @{}
    }
        \hline\hline

        &
        Ancillae for
        &
        Feed-forward depth
        &
        Clifford depth
        \\[-8pt]

        &
        non-Clifford synthesis
        &
        per batch
        &
        per batch
        \\[2pt]

        \hline\hline

        Partial FT
        &
        $\displaystyle
        \left\lceil
            (1 + A_{\text{pFT}})
            \frac{b}{\mathfrak{r}}
        \right\rceil$
        &
        $\displaystyle
        \min(\log_2 b + 1.33,C)
        + 2\mathfrak{r}
        \left(
            1-2^{-\min\{C,\lceil\log_2\mathfrak{r}\rceil\}}
        \right)$
        &
        $\displaystyle \lesssim 1$
        \\[6pt]

        Clifford+T
        &
        $\displaystyle
        \left\lceil
            (1 + A_{\text{FT}})
            \frac{b}{\mathfrak{r}}
        \right\rceil$
        &
        $\displaystyle
        \left\lceil
            \mathfrak{r}N_T^{\rm Tdec}(\epsilon)
        \right\rceil$
        &
        $\displaystyle
        \left\lceil
            \mathfrak{r}N_T^{\rm Tdec}(\epsilon)/2
        \right\rceil$
        \\[6pt]

        Reduced HWP
        &
        $\displaystyle
        \left\lceil
            \frac{4}{3}
            \left(1+\frac{2}{3}A_{\text{FT}}\right)
            \frac{b}{\mathfrak{r}}
        \right\rceil$
        &
        $\displaystyle
        \mathfrak{r}
        \left[
            2\log_{3/2}\left(\frac{b}{\mathfrak{r}}\right)
            +N_T^{\rm Tdec}(\epsilon)
        \right]$
        &
        $\displaystyle
        \mathfrak{r}
        \left[
            16\log_{3/2}\left(\frac{b}{\mathfrak{r}}\right)
            +\frac{N_T^{\rm Tdec}(\epsilon)}{2}
        \right]$
        \\[6pt]
        \hline\hline
    \end{tabular*}
\end{table*}

The NISQ approach included in the comparison shows poor performance compared to that of error-corrected computation.
At moderate distances $5$ and $7$, partial fault tolerance yields the lowest error rates.
This is due to the relatively large contribution of errors from Clifford operations, whose usage is significantly reduced in the TMR protocol.
At distances $5$ and $7$, the partially FT protocol targeting Trotter error $\sim0.5\%$ accommodates lattice sizes $36$ and $120$ respectively, beyond the reach of current classical approaches and NISQ digital computation for equivalent evolution times~\cite{danna2025_fermionic, thompson2025_fermionic, papic2026_fermionic, alam2025_fermionicgoogle,alam2025_fermionicquantinuum}, as well as much larger than the fully FT counterparts. 
%clf data here (or consequence of)
This identifies the STAR architecture as a more promising candidate for near-term demonstrations of quantum advantage.
At higher distance, noise from Clifford operations is strongly reduced, and fully FT approaches achieve lower error rates, enabling the simulation of larger system sizes.

This comparison, here outlined for the representative error rate $p=0.001$, was repeated over the range $p\in[10^{-4}, 5\times 10^{-3}]$.
By comparing the maximum lattice size attainable by error-corrected approaches with target Trotter error $0.5\%$, regimes of advantage of partial and full fault-tolerance are identified.
For the purposes of comparing against current classical methods, the limit of classical simulations of the 2D Hubbard model is assumed to be just below lattice size $36$~\cite{papic2026_fermionic, alam2025_fermionicquantinuum} in the equivalent setting, i.e. on a square lattice with periodic boundaries and with evolution times comparable to the length of the lattice.
Moreover, we repeated the comparison among methods under the assumption that $\alpha_{\rm RUS}=1.5$ in~\cref{eq:pL-pFT-ansatz}, as previously estimated for the transversal STAR architecture~\cite{ismail2025}.
\Cref{fig:sub:pFT-in-pd-diagram-sketch} shows that the reduction of the effective RUS prefactor $\alpha_{\rm RUS}$ obtained in our framework establishes a near-term region in which the transversal STAR architecture is not only a lower-overhead alternative to full fault-tolerance, but yields lower errors than the other approaches.
In the long-term, instead, full fault tolerance still appears as a more robust solution, albeit with a significantly larger overhead, as quantified in the next subsection.

\subsection{Limitations from circuit depth}
\label{subsec:applications-limitations-circuit-depth}

In this subsection, we analyze the circuit depth of different approaches in the realization of parallel arbitrary-angle rotations.
We show that the TMR protocol not only requires lower depth, but also that it has a much more advantageous spacetime tradeoff upon reduction of physical resources, independently of the target application.

\Cref{sec:app:resources-arbitrary-rotations} provides a detailed derivation of the resources required for a maximally parallel implementation of the methods---i.e. with the shortest possible depth---and of the increase in depth following a reduction of these resources.
\Cref{tab:synthesis_comparison} summarizes the findings, and compares the different approaches to gate synthesis.
Specifically, it assumes that a batch of $b$ parallel $R_Z(\theta_L)$ rotations is to be implemented, with $\theta_L$ a dyadic fraction of $\pi$.
The table provides reliable analytic approximations of the circuit depth as the availability of ancillae is reduced by a factor $\mathfrak{r}$ with respect to the maximally parallel scenario.
We note that while contributions to depth from feedforward and Clifford operations are provided separately for generality, the total circuit depth is likely less than their direct sum, due to potential concurrent scheduling: this relies on additional practical assumptions and its quantitative characterization is left for future work.
%clifford depth modeling assumptions here
The modeling assumes comparable times for the encoding and the injection of resource states, justified in the context of neutral atom implementations, where measurements and shuttling represent the main time bottleneck.

The feed-forward depth of the TMR protocol has a small logarithmic dependence on the batch; rather remarkably, upon reduction of its ancillary reservoir it only receives an additive increase.
The only Clifford contribution to its circuit depth is a potential $S$ or $S^\dagger$ correction to the RUS process.
In contrast, for Clifford+$T$ synthesis, the depth is set by the target approximation error, with typical T-count $N_T^{\rm Tdec}(\epsilon)>20$, and approximately half as many $S$-type gates.
Reducing the ancillary reservoir increases both depths multiplicatively.
Regarding HWP, its requirements and resources are determined by its specific implementation.
As motivated in~\cref{app:subsec:resources-hwp}, we considered a ``reduced'' version of the protocol, which is particularly suited for near-term implementations with moderate code distance and an ancillary reservoir of limited size.
While HWP reduces the number of rotations to be synthesized, it increases both ancillae requirements and circuit depth.

The significant difference in circuit depth is evident in~\cref{fig:sub:pFTvsFT-depth}.
The figure shows the circuit depth required to implement one batch of $b=100$ parallel $R_Z$ rotations against the number $Q_{\rm syn}$ of logical patches available for the synthesis of non-Clifford resources.
Typical values for implementations at code distance $d=7$ and physical error rate $p=0.001$ are considered, with effective average number of encoding attempts per resource state $A_{\rm pFT}=1.05,\;A_{\rm FT}=1.2$, and $N_T^{\rm Tdec}=25$ $T$ gates per synthesized rotation.
As the size of the non-Clifford resources is comparable to or smaller than the batch size, matching near-term constraints, the feed-forward depth of the TMR protocol is roughly an order of magnitude shorter than the FT counterparts.
Even once the maximally parallel regime is attained, requiring much larger ancillae reservoirs, the partially FT depth remains $\sim 5$ times shorter.

In terms of Clifford operations, partial fault-tolerance yields at least an order of magnitude reduction, growing up to three orders of magnitude once the availability of ancillae is reduced.
However, we note that in near-term regimes with fewer ancillary patches, non-Clifford operations cannot be fully parallelized, leading where possible to concurrent scheduling of Clifford and non-Clifford gates.
This implies that the total depth is not simply the sum of the separate contributions, but is subject to further assumptions---the details of the Clifford+$T$ decomposition, hardware routing, and scheduling optimization---beyond the scope of our work.
%clifford depth modeling assumptions here

\section{Summary and Outlook}
\label{sec:summary}

Partial fault tolerance significantly reduces the resources required for robust quantum computation.
In the STAR architecture, Clifford gates remain fault-tolerant (FT), while arbitrary-angle rotations are natively implemented, avoiding long Clifford+$T$ synthesis sequences~\cite{akahoshi2024, akahoshi2024a, choi2023a, toshio2025, ismail2025, zeng2025_pftbias, ismail2026, toshio2026}.
Specifically, arbitrary-angle rotations are realised via the Transversal Multi-Rotation (TMR) protocol, which combines the use of $r$ physical $R_{ZZ}(\theta)$ rotations with syndrome extraction and post-selection, to prepare angle-dependent resource states.
These are then injected in a repeat-until-success (RUS) process, as described in~\cref{sec:partial-ft}.
This approach reduces the space-time cost of arbitrary rotations, at the price of exposing the protocol to a small set of hardware-dependent faults that must be carefully modeled.
In neutral-atom hardware, reconfigurable connectivity enables the execution of transversal operations, further reducing space-time overhead and logical error rates~\cite{ismail2025}.

In this work, we built on these advances by investigating the co-design of partially FT computation and neutral-atom hardware, ranging from the underlying atomic physics to noise propagation in the logical protocol, and the final target application.
With a combination of detailed modeling and optimization, partial fault tolerance emerges not only as a lower-overhead option, but also as a more robust route for near-term demonstrations of quantum advantage.

Concretely, we first developed a refined model of noise propagation from the physical operations to the logical protocol (cf.~\cref{sec:partial-ft}).
In the first place, we show that the RUS nature of the protocol suppresses the leading coherent contribution to the logical channel, better informing future error mitigation approaches.
Secondly, while it is known that dominant undetectable faults arise in the aforementioned $r$ physical $R_{ZZ}(\theta)$ rotations~\cite{akahoshi2024,choi2023a,toshio2025}, our work revealed an intrinsic resilience to their coherent noise, guiding subsequent optimization of its performance.
Therefore we tailored the execution of the $R_{ZZ}(\theta)$ Rydberg gates in neutral-atom hardware to the logical protocol, exploiting accurate knowledge of the underlying atomic physics (see~\cref{sec:gate-optim}).
In the first place, our modeling reveals a much more favorable scaling in these rotations than assumed in previous analysis~\cite{toshio2025, ismail2025}, as well as a beneficial noise bias amplified by the possibility of post-selecting on leakage events.
Moreover, we further suppressed undetectable physical faults via pulse-level error-selective gate optimization, yielding a two- to three-fold reduction of the critical faults, without additional overhead.
We specifically focused on implementations in ${}^{171}\rm Yb$, a popular candidate for neutral-atom quantum computation~\cite{senoo2025, peper2024, muniz2025}, but our findings can be extended to other atomic species.
Our refined analysis and optimized implementation predict logical error rates approximately an order of magnitude below previous STAR estimates~\cite{toshio2025, ismail2025} (cf.~\cref{sec:performance}).
Specifically, the error rate of $\theta_L$ rotations in the rotated surface code, at distance $d$ and physical error rate $p$, is well modeled by
\begin{equation}
p_L(\theta_L,p,d,r) \simeq
\alpha_{\rm RUS}(p,r)\,p\,|\theta_L| + 2p_L^{(\rm inj)}(p,d),
\end{equation}
with $\alpha_{\rm RUS}(p,r)\simeq r(0.01+4p)$ for the optimized Rydberg pulses and circuit-level noise model considered here.
The first term captures the angle-dependent contribution specific to the TMR encoding---depending on the number $r$ of physical transversal $R_{ZZ}(\theta)$ rotations---while the second term captures the Clifford and injection errors accumulated over the RUS attempts, two on average.

Then, we benchmarked the applicability and scalability of our framework against fully FT alternatives, namely standard Clifford+$T$ synthesis and Hamming Weight Phasing (see~\cref{sec:applications}).
As a representative case-study, we considered time evolution in the Hubbard model with periodic boundaries, for which we devised a simulation circuit tailored to early FT scenarios.
The aforementioned improvements make partial fault tolerance most compelling at moderate code distances $d\leq7$, where the asymptotic advantages of full fault tolerance are not yet dominant.
At physical error rate $p\sim10^{-3}$, partially FT computation, in combination with error mitigation, is expected to surpass the reach of classical methods and NISQ simulations, as well as that of fully FT alternatives at comparable physical resources.
Notably, we predict simulations of $36$ sites at $d=5$ and $100$ sites at $d=7$ to be achievable, with physical footprints of approximately $5$ and $25$ thousand physical qubits respectively.
% clf data here or consequence of

Finally, we analyzed the circuit depth and spacetime tradeoffs of the techniques considered (cf~\cref{subsec:applications-limitations-circuit-depth}).
Even in regimes where fully FT synthesis is expected to achieve lower error rates, the STAR architecture retains a substantially better spacetime tradeoff, with order-of-magnitude reductions in circuit depth.
This leads to order-of-magnitude reductions in circuit depth, particularly pronounced when the number of ancillary logical patches is limited.
For example, for a batch of $b=100$ parallel rotations at $d=7$ and $p=10^{-3}$, repeat-until-success TMR injection depth with realistic ancillary reservoirs is approximately one order of magnitude shorter than fully FT alternatives, and remains five times shorter even with extensive availability of resource states.
The Clifford depth of STAR rotations is lower than the FT counterparts by one to three orders of magnitude, depending on the number of ancillary logical patches available.
%clifford depth modeling assumptions here
Therefore, while partial fault tolerance asymptotically yields higher error rates, its depth and hardware efficiency make it a compelling alternative even in the longer term.

Several directions follow from these results.
In the first place, partially FT protocols can be extended to alternative error-correcting codes, hardware and applications.
The rotated surface code provides a versatile baseline for neutral-atom implementations, but other code families may offer better tradeoffs.
In particular, high-rate QLDPC codes are expected to reduce the physical overhead, although their applicability is likely to depend on the target application, as explored in recent works~\cite{ismail2025,ismail2026}.
Due to generally lower thresholds, their advantage is likely to become more pronounced in longer-term implementations.
Secondly, the different noise structure of partially FT operations calls for tailored quantum error mitigation techniques, as well as suggesting potential connections to quantum metrology~\cite{ott2026_partialFTmetrology, huang2026_partialFTmetrology}.
Our end-to-end analysis and improvements provide a natural starting point for these investigations.

Altogether, our results open new prospects for the resilience and scalability of partially FT computation, and establish a near-term regime in which it is expected to surpass the reach of both classical methods and fully FT approaches.
This makes partial fault-tolerance a compelling route to early demonstrations of practical quantum advantage.
\vspace{-1em}

\section{Acknowledgments}
\label{sec:acknow}

We thank Giacomo Giudice, Martin Kiffner, Sebastian W\"uster, and Jonathan Pritchard for stimulating discussions. The authors acknowledge the use of the University of Oxford Advanced Research Computing (ARC) facility in carrying out this work (http://dx.doi.org/10.5281/zenodo.22558).
This work was supported by the EPSRC through the QQQS programme grant (EP/Y01510X/1) and by the QCi3 hub (EP/Z53318X/1). 
Work by SV was supported by PlanQC GmbH.

AI tools were used in the following form in this work: ChatGPT 5.6 Sol was used to  suggest improvements to syntax and readability during the writing process.
Different versions, from ChatGPT 5.0 to 5.6 Sol, assisted with debugging and with code generation, proposing adjustments to the simulation code, producing Mathematica and SymPy scripts to verify mathematical derivations, and implementing requested changes in the layout of figures (which were produced separately).
AI suggestions were in each case reviewed and edited by the authors and further verified against independent calculations and numerical tests.

\appendix

\section{Detailed Derivation of the Error Budget of the TMR protocol}
\label{sec:app:error-bugget-deriv}

In this appendix, analytic estimates for the error budget of the TMR protocol are derived.
As it will be quantified, three main error mechanisms limit the performance of the TMR protocol: noise from stabilizer extraction and transversal operations, undetected correlated phase-flips, and undetected repeating bit-flips.
The effect of other faults is included in numerical simulations, but not dealt with analytically in this appendix.
This appendix is organized as follows:
\begin{enumerate}[label=(\roman*)]
    \item \Cref{subsec:app:error-budget:error-metrics} provides a few remarks on the adopted error metrics, i.e. average infidelity and diamond-norm error.
    \item \Cref{subsec:app:error-budget:noise-from-qec} briefly comments on the error contributions from noisy Clifford operations.
    \item \Cref{subsec:app:error-budget:noise-from-repeat-flips,subsec:app:error-budget:noise-from-correl-phase-flips} introduce the two sources of noise specific to the TMR protocol---which \textit{partial} fault tolerance owes its name to---and models their effects on the resource state.
    \item \Cref{subsec:app:error-budget:noise-whole-teleport} quantifies error propagation onto the target logical qubits upon gate-teleportation.
    \item \Cref{subsec:app:error-budget:noise-whole-RUS} quantifies error propagation in the whole repeat-until-success scheme, showing error accumulation of incoherent noise, but also a remarkable cancellation of leading coherent noise.
\end{enumerate}

For the purposes of clarifying the notation and procedure, it should be recalled from the main text that the TMR protocol aims to encode the resource state $\ket{m_{\theta_L}}=R_{Z,L}(\theta_L)\ket+$, where $R_{Z,L}(\theta_L)=\mathrm{exp}(-i \theta_L Z_L/2)$.

It does so by applying $r$ transversal rotations, each of weight $w$, followed by stabilizer measurements and post-selection.
In this appendix, $Z^{\otimes w}_q$ denotes the tensor product of the $Z$ operator over $w$ consecutive qubits, starting at index $wq$.
Then, the transversal rotations have form $\bigotimes_{q=0}^{r-1} R_{Z^{\otimes w}_q}(\theta)$, for some angle $\theta$.
The noise $\mathcal{E}$ acting on each transversal rotation is modeled by the chi-matrix $\chi_{\scriptscriptstyle P,P'}$, such that 
\begin{equation}
    \mathcal{E}(\rho) = \sum_{P,P'\in \mathcal{P}^{\otimes w}} \chi_{\scriptscriptstyle P,P'} P \rho P',
\end{equation}
with $\mathcal{P}^{\otimes w}$ the set of $w$-qubit Pauli operators.

\subsection{Error Metrics}
\label{subsec:app:error-budget:error-metrics}

As mentioned in the main text, the adopted figures of merit are the average infidelity $1-F$, and the diamond-norm error $\varepsilon_\diamond$. 
The latter, also known as the worst-case error rate, is defined as the diamond-norm distance between the noise channel $\mathcal{E}$ that we wish to characterize and the identity channel $\mathcal{I}$,
\begin{equation}
    \varepsilon_\diamond(\mathcal{E}) = \frac{1}{2} \norm{\mathcal{E} - \mathcal{I}}_\diamond 
    = \frac{1}{2} \sup_\rho \norm{\left( \mathcal{E}\otimes \mathcal{I} - \mathcal{I}^{\otimes2}\right)(\rho)}_1,
\end{equation}
where $\norm{A}_1 = \tr{\sqrt{A^\dagger A}}$, and the supremum is over all density matrices of dimension $D^2$, with $D$ the Hilbert-space dimension of the system acted upon by $\mathcal{E}$.
The diamond-norm error is intuitively understood as the maximum probability of distinguishing the outputs of the noisy and ideal operation~\cite{kliesch2021_diamondnorm, kueng2016_diamondnorm}.
It is sub-additive, and it plays a central role in rigorous proofs of fault tolerance thresholds~\cite{aharonov1998, aliferis2005-exRec, kliesch2021_diamondnorm, kueng2016_diamondnorm}.
Notably, the diamond-norm error is more sensitive than the average infidelity to coherent errors~\cite{sanders2016_diamondnorm, huang2019_diamondnorm}.

As an example---which will be of great interest in the remainder of the appendix---given a general completely-positive trace-preserving $Z$-error channel of the form
\begin{equation}
    \mathcal{N}(\rho; p,q) = (1-p)\rho + p Z\rho Z - (iqZ\rho + \hc),
    \label{eq:general-z-channel}
\end{equation}
where $p,q$ are real, then $1-F=2p/3$, whereas $\varepsilon_\diamond = \sqrt{p^2 + q^2}=p\sqrt{1+(q/p)^2}$~\cite{huang2019_diamondnorm}.
This implies that the average infidelity and diamond-norm error between rotations by a target angle $\theta_0$ and an incorrect angle $\theta_1$ are
\begin{align}
    1-F(\theta_0,\theta_1)  &= 
    \frac{2}{3} \sin^2 \left(\frac{\theta_0-\theta_1}{2} \right), \nonumber \\
    \varepsilon_\diamond(\theta_0, \theta_1) &= 
    \left |\sin \left(\frac{\theta_0 - \theta_1}{2} \right) \right|,
    \label{eq:errors-rotations}
\end{align}
such that potentially $\varepsilon_\diamond \gg 1-F$.
Finally, we note that, allowing for a non-CPTP channel of the form in~\cref{eq:general-z-channel} with complex $q$, as it is possible upon post-selection, the diamond distance generalizes to $\varepsilon_\diamond = \sqrt{p^2 + |q|^2}$, as tested numerically.

\subsection{Noise from QEC}
\label{subsec:app:error-budget:noise-from-qec}

The first contribution to logical faults is given by noise in the syndrome extraction and state injection.
This is common to FT quantum computation in general, and it is exponentially suppressed with the code distance $d$, provided that the physical error rate $p$ is below threshold $p_\mathrm{\rm th}$. 
It is often well captured by an effective Pauli-noise model, with 
\begin{align}
    p_L^{\rm (inj)}\simeq a_{\rm inj} \left(\frac{p}{p_\mathrm{\rm th}} \right)^{\ceil{\frac{d+1}{2}}},
    \label{eq:qec-error-contrib}
\end{align}
where $a_{\rm inj}$ and $p_\mathrm{inj}$ depend on the physical noise model and details of the specific implementation.
For stochastic Pauli noise, $1-F^{(\rm inj)}=2p_L^{\rm inj}/3$ and $\varepsilon_\diamond^{(\rm inj)}=p_L^{(\rm inj)}$.

\subsection{Noisy resource state under correlated phase-flips}
\label{subsec:app:error-budget:noise-from-correl-phase-flips}
The first error mechanism specific to the TMR protocol originates from undetected correlated phase flips that occur shortly before or shortly after the transversal $R_{Z^{\otimes w}}$ rotations, and is typically the dominant error.
To understand how these flips affect the protocol, recall~\cref{eq:big-tmr-result}: the transversal rotations amount to the sum of the desired $\theta_L$ rotation, plus other $Z$-error terms that map the logical state outside the logical subspace.
Assuming that, for even $r$, the $S_X^\dagger$ correction is applied as specified in the previous section, the error terms that are closest to the logical subspace---hence dominating the error budget---are
\begin{equation}
    -i
    \sum_{q=0}^{r-1} \sqrt{p_{\theta_1}}\,Z_q^{\otimes w}\, \ket{m_{(-1)^{\lfloor r/2 \rfloor} \theta_1}}_L,
    \label{eq:z-error-terms-explicit}
\end{equation}
where the effective rotation angle $\theta_1$ satisfies
\begin{equation}
    \tan {\frac{\theta_1}{2}} = - \tan^{r-2} \frac{\theta}{2} = - \tan^{(1-\frac{2}{r})} \frac{\theta_L}{2}, 
    \label{eq:theta1}
\end{equation}
and the probability of collapsing onto each one of these terms is
\begin{equation}
    p_{\theta_1}(\theta,r) 
    = \cos^{2(r-1)} \frac{\theta}{2} \sin^{2} \frac{\theta}{2}
    + \sin^{2(r-1)} \frac{\theta}{2} \cos^2 \frac{\theta}{2}.
    \label{eq:p1}
\end{equation}
Therefore, if the logical state collapses onto any of these error terms, and, in addition, the corresponding $Z^{\otimes w}_q$ error occurs shortly before or after
\footnote{A \(Z_q^{\otimes w}\) fault occurring sufficiently far before or after the transversal rotations would be detected by the intervening stabilizer measurements.}, then the net effect is the production of the logical state $\ket{m_{\theta_1}}_L$: the state is rotated incorrectly, yet undetectably so.

Denoting by $p$ the circuit-level error rate, and by $\chi$ the chi-matrix of the error channel acting on each of the transversal rotations, the incoherent probability of correlated-phase flips on any of the physical rotations is given by
\begin{equation}
    p_{Z^{\otimes w}} \simeq r\;[a_Z\,p^w + \chi_{\scriptscriptstyle Z^{\otimes w},Z^{\otimes w}}(\theta)].
\end{equation}
Moreover, in contrast to previous works, we provide a complete modeling that does not neglect the contribution of coherent (i.e. off-diagonal) errors in the physical rotations, which amounts to
\begin{equation}
    q_{Z^{\otimes w}} \simeq -i\;r\;\chi_{\scriptscriptstyle Z^{\otimes w}, I}(\theta).
\end{equation}

Then, after the transversal rotations and stabilizers measurements, the resource state is in leading order
\begin{align}
    m^{(Z^{\otimes w})}_\mathrm{un} 
    =\;& 
    (1-p_{Z^{\otimes w}})\,p_{\theta_L} \,
    \ket{m_{\theta_L}} \bra{m_{\theta_L}}_L 
    \nonumber \\
    &\quad\;+\,\,p_{Z^{\otimes w}}\,p_{\theta_1}\,
    \,\ket{m_{\theta_1}} \bra{m_{\theta_1}}_L
    \nonumber \\
    &\quad\;+\qquad\;\,\,q_{\mathrm{off}}\,
    \,\ket{m_{\theta_1}} \bra{m_{\theta_L}}_L + \hc,
\end{align}
where $q_\mathrm{off}=q_{Z^{\otimes w}} \sqrt{p_{\theta_L} p_{\theta_1}}$.
Note, the resource state is currently unnormalised due to post-selection.
Denoting $\Delta = \theta_1 -\theta_L$, this can then be re-written as
\begin{align}
    m^{(Z^{\otimes w})}_\mathrm{un}(\theta_L) = 
    (1-\,p_{Z^{\otimes w}})\,p_{\theta_L} \, &\ket{m_{\theta_L}} \bra{m_{\theta_L}}_L 
    \nonumber \\
    + \,p_{Z^{\otimes w}}\,p_{\theta_1} \, R_{Z,L}(\Delta) &\ket{m_{\theta_L}} \bra{m_{\theta_L}}_L R_{Z,L}^\dagger(\Delta)  
    \nonumber \\
    + \,q_{\mathrm{off}}\,R_{Z,L}(\Delta) &\ket{m_{\theta_L}} \bra{m_{\theta_L}}_L + \hc.
\end{align} 

\subsection{Noisy resource state under repeating flips}
\label{subsec:app:error-budget:noise-from-repeat-flips}

The second error contribution specific to the TMR protocol originates from undetected flips that anticommute with the $Z_L$ operator.
Practically, such faults require two events:
\begin{enumerate}
    \item an $X$- or $Y$-flip shortly before the rotations, which effectively reverses the rotation direction, as $R_{Z_1^{\otimes w}} X_1 = X_1 R^\dagger_{Z_1^{\otimes w}}$, and occurs with probability $O(p)$;
    \item a second flip shortly after, either during the rotation itself with probability $\chi_{\scriptscriptstyle X,X}$ or $\chi_{\scriptscriptstyle Y,Y}$, or in one of the subsequent gates with probability $O(p)$.
\end{enumerate}
In summary, the first flip inverts the sign of rotation, while the second makes the fault go undetected. 
The probability of this combination of errors scales linearly with the number of qubits in the support of $Z_L$, which is the smaller code distance $d_s$ at the time of the transversal rotations.

Therefore, the probability $p_{\rm flip}$ for this  error reads 
\begin{equation}
    p_{\rm flip} =
    d_s[a_X\,p^2 + a_{X,r}\,p\,(\chi_{\scriptscriptstyle X,X}+\chi_{\scriptscriptstyle Y,Y})(\theta)],
\end{equation}
where $a_X, a_{X,r}>0$ re circuit- and noise-dependent prefactors arising from the combinatorics of the contributing error processes; $(\chi_{\scriptscriptstyle X,X}+\chi_{\scriptscriptstyle Y,Y})$ depend on the physical angle of rotation $\theta$.
In this scenario, off-diagonal errors in the transversal rotations, e.g. $\chi_{\scriptscriptstyle X,Y}$, are projected out by stabilizer measurements.
As this error anticommutes with $Z_L$, it realizes a flipped rotation $R_{Z,L}(-\theta_L)$.
Then, to leading order, the noisy resource state is
\begin{align}
    m^{(\rm flip)}_\mathrm{un}(\theta_L) \simeq\,
    (1-p_{\rm flip}) p_{\theta_L} &\ket{m_{\theta_L}} \bra{m_{\theta_L}}_L \nonumber \\
    +\;p_{\rm flip}\,p_{\theta_L} \,&\ket{m_{-\theta_L}} \bra{m_{-\theta_L}}_L.
\end{align}
Note, the state is currently unnormalised due to post-selection.

In previous studies~\cite{toshio2025, ismail2025}, this contribution was neglected because their noise models assumed $\chi_{\scriptscriptstyle Z^{\otimes w},Z^{\otimes w}} \sim O(p)$, making the correlated phase-flips (see ~\cref{subsec:app:error-budget:noise-from-correl-phase-flips}) overwhelmingly dominant.
As it is shown in later sections, this is not necessarily the case.
Moreover, due to its quadratic dependence on the error rate, this noise source can contribute to the ultimate limits of the TMR protocol, independently of the code distance and of the implementation of the transversal rotations.

\subsection{Gate-teleportation Noise}
\label{subsec:app:error-budget:noise-whole-teleport}

In this subsection, we first derive the noisy state $\mathfrak{M}(\theta_L)$ of the resource qubit. 
Then, we quantify noise propagation onto the input logical qubit $\Psi$ through the teleportation channel $\mathcal{T}^{(\mu)}(\Psi;\theta_L)$, conditioned on the teleportation measurement $\mu=0$ or 1.

In the previous subsections, we defined $m^{(Z^{\otimes w})}_\mathrm{un}(\theta_L)$ and $m^{(\rm flip)}_\mathrm{un}(\theta_L)$ as the unnormalised resource states under the most relevant error sources: correlated phase-flips and repeating flips, respectively.
The resulting state under their combined action is
\begin{align}
    \mathfrak{M}(\theta_L) =\;
     ( 1 - x_{\theta_L} ) \, &\ket{m_{\theta_L}} \bra{m_{\theta_L}}_L \nonumber \\
    +\, x_{\theta_L}\,Z_L &\ket{m_{\theta_L}} \bra{m_{\theta_L}}_L Z_L
    \nonumber \\
    -\, i y_{\theta_L} Z_L &\ket{m_{\theta_L}} \bra{m_{\theta_L}}_L + \hc,
    \label{eq:resource-state-channel}
\end{align}
where, in leading order, the noise coefficients are
\begin{align}
    x_{\theta_L} \simeq\;&
    \frac{p_{\theta_L}}{n_{\theta_L}} 
    \left[ 
    p_{Z^{\otimes w}}\,p_{\theta_1} \sin^2 \frac{\Delta}{2} 
    + p_{\rm flip}\,p_{\theta_L} \sin^2 \theta_L
    \right], \nonumber \\[5pt]
    y_{\theta_L} \simeq\;&
    \frac{p_{\theta_L}}{n_{\theta_L}} 
    \Biggl[  
    \left(
        p_{Z^{\otimes w}}\,p_{\theta_1} \cos \frac{\Delta}{2} + q_{\mathrm{off}}
    \right) \sin \frac{\Delta}{2} 
    \nonumber \\
    &\hspace{1cm}- p_{\rm flip}\,p_{\theta_L} \sin \theta_L
    \Biggr],
    \label{eq:teleport-resource-state-channel-coefficients}
\end{align}
with normalisation factor $n_{\theta_L}$---close to one in the small angle scenario---defined by
\begin{align}
    \frac{n_{\theta_L}}{p_{\theta_L}} = 
    \left[ 
    (1-p_{Z^{\otimes w}})p_{\theta_L} + p_{Z^{\otimes w}}\,p_{\theta_1} + 2\,\mathrm{Re}(q_{\mathrm{off}}) \cos \frac{\Delta}{2}
    \right],
\end{align}
where terms of order $O(p_{\rm flip}\,p_{Z^{\otimes w}} + p_{\rm flip}\,q_\mathrm{off})$ have been neglected.
To model gate teleportation, note that the resource state has matrix elements
\begin{align}
    \mathfrak{M}_{0,0}(\theta_L)
    &= \left[ 1 + 2\,\mathrm{Im}(y_{\theta_L}) \right]/2 \nonumber \\
    \mathfrak{M}_{0,1}(\theta_L) 
    &= e^{-i\theta_L} \left[ 1 - 2x_{\theta_L} - 2i\,\mathrm{Re}(y_{\theta_L}) \right]/2 \nonumber \\
    \mathfrak{M}_{1,0}(\theta_L)
    &= e^{+i\theta_L} \left[ 1 - 2x_{\theta_L} + 2i\,\mathrm{Re}(y_{\theta_L}) \right]/2 \nonumber \\
    \mathfrak{M}_{1,1}(\theta_L)
    &= \left[ 1 - 2\,\mathrm{Im}(y_{\theta_L}) \right]/2.
\end{align}
Upon gate teleportation, the $Z_L$-type errors on the noisy resource state $\mathfrak{M}({\theta_L})$ propagate on the target logical qubit, originally in state $\Psi$.
To perform teleportation, a $\text{CX}=\ket{0}\bra{0} \otimes \hat{\mathbb{I}} + \ket{1}\bra{1} \otimes \hat{X}$ gate controlled on $\Psi$ targeting $\mathfrak{M}$ is applied, yielding
\begin{align}
    \text{CX}\,(\Psi \otimes \mathfrak{M})\,\text{CX}^\dagger 
    &=\quad\! \Psi_{00} \ket{0}\bra{0} \otimes \mathfrak{M} \nonumber \\
    &\quad + \Psi_{01} \ket{0}\bra{1} \otimes \mathfrak{M}X \nonumber \\
    &\quad + \Psi_{10} \ket{1}\bra{0} \otimes X\mathfrak{M} \nonumber \\
    &\quad + \Psi_{11} \ket{1}\bra{1} \otimes X\mathfrak{M}X.
\end{align}
This is followed by the measurement of the resource state with outcome $\mu$, yielding the unnormalised output
\begin{align}
    & \begin{pmatrix}
          \Psi_{00} \, \mathfrak{M}(\theta_L)_{0\oplus\mu,0\oplus\mu} 
        & \Psi_{01} \, \mathfrak{M}(\theta_L)_{0\oplus\mu,1\oplus\mu} \\
          \Psi_{10} \, \mathfrak{M}(\theta_L)_{1\oplus\mu,0\oplus\mu} 
        & \Psi_{11} \, \mathfrak{M}(\theta_L)_{1\oplus\mu,1\oplus\mu}
    \end{pmatrix}.
\end{align}
We denote the teleportation outcome $\mathcal{T}^{(\mu)}(\Psi; \theta_L)$ as twice this post-measurement state, as normalized for real $y_{\theta_L}$.
This can be re-written as
\begin{align}
    \mathcal{T}^{(\mu)}(\Psi; \theta_L) =\,
     ( 1 - x_{\theta_L} ) \, &\Psi_{(-1)^{\mu}\theta_L}
     \nonumber \\
    +\, x_{\theta_L}\,Z_L &\Psi_{(-1)^{\mu}\theta_L} Z_L
    \nonumber \\
    -\, i y_{\theta_L}^{(\mu)} Z_L &\Psi_{(-1)^{\mu}\theta_L} + \hc,
    \label{eq:teleport-channel-complicated}
\end{align}
where $\Psi_{\bullet} = R_{Z,L}\left(\bullet\right) \, \Psi \, R^\dagger_{Z,L}\left(\bullet\right)$, such that $\Psi_{(-1)^{\mu}\theta_L}$ is the noiseless output upon the recorded measurement, whereas $y_{\theta_L}^{(0)}=-y_{\theta_L}^{(1)}=y_{\theta_L}$.
In fact, upon opposite measurements, both the rotation angle $\theta_L$ and the coherence $y_{\theta_L}$ are flipped, as $\mathfrak{M}_{i,j}$ transforms into $\mathfrak{M}_{i\oplus 1,j\oplus 1}$ under sign-flip of both $\theta_L$ and $y_{\theta_L}$.

The normalized output is
\begin{equation}
    \tilde{\mathcal{T}}^{(\mu)}(\Psi; \theta_L) =
    \frac{\mathcal{T}^{(\mu)}(\Psi; \theta_L)}{2P^{(\mu)}(\Psi; \theta_L)},
\end{equation}
where
\begin{equation}
    P^{(\mu)}(\Psi; \theta_L) = \frac{1}{2} + (-1)^{\mu} \mathrm{Im}(y_{\theta_L}) \mathrm{Tr}(Z_L \Psi )
    \label{eq:teleport-probability-Pm}
\end{equation}
is the probability the $\mu$ measurement is observed, conditioned on the input $\Psi$.

We note that $\tilde{\mathcal{T}}(\Psi;\theta_L)$ is also protocol- and noise-dependent, as most quantities depend on the number $r$ of transversal rotations and the noise strength $p$.
The dependence is kept implicit for compactness.
The resulting infidelity and diamond-norm error are $2x_{\theta_L}/3$ and $\sqrt{x_{\theta_L}^2+|y_{\theta_L}|^2}$ respectively. 
However, these values only describe an individual instance of gate teleportation.
As shown in the next subsection, complex error accumulation and cancellation occurs within the whole repeat-until-success protocol.

\subsection{Noise propagation in the RUS process}
\label{subsec:app:error-budget:noise-whole-RUS}

Gate teleportation is probabilistic, as the gate is correctly injected with probability $\sim1/2$ upon successful measurement, whereas the inverse gate is injected upon opposite measurement.
A flipped gate can be corrected simply by implementing gate-teleportation again, with double the original angle.
Therefore, a complete gate injection protocol is a repeat-until-success (RUS) scheme, in which gate teleportation is repeated until the desired measurement is observed.
At every $s^{\rm th}$ step, with $s=1,2,\ldots$, a different angle $\theta_{L,s}$ is injected (see main text).

The previous subsection modeled a single teleportation attempt. 
In this subsection, error propagation in the whole RUS protocol is examined.
It will be useful to define the following ansatz describing the action of a $R_{Z,L}(\theta)$ rotation on an input $\Psi$, followed by noise, whose incoherent and coherent error contributions are $x,y$ :
\begin{align}
    \mathcal{N}(\Psi;\,\theta,x,y) 
    =\;& 
    \,(1-x)\; \Psi_{\theta} + x\, Z_L\, \Psi_{\theta}\, Z_L \nonumber \\
    &- i y\, Z_L\, \Psi_{\theta} \; + \hc,
    \label{eq:ansatz-noise-process-firstdef}
\end{align}
recalling that $\Psi_{\theta} $ denotes $R_{Z,L}\left(\theta\right) \, \Psi \, R^\dagger_{Z,L}\left(\theta\right)$.
Coefficients $x,y$ generally depend on the angle of rotation $\theta$.

As it will become clear at the end of the subsection, it is now convenient to work with generally unnormalised states.
Extending the notation adopted in previous subsections, the action of 
gate-teleportation at the $s^{\rm th}$ step of the RUS process on an input $\Psi$ can be modeled as
\begin{align}
    \mathcal{T}_s^{(\mu)}(\Psi;\theta_L) = 
    \mathcal{N}\left(\Psi; \,(-1)^{\mu} \theta_{L,s},x_s,y_s^{(\mu)}\right)
    \label{eq:z-channel-rus-each-teleport} 
\end{align}
where $\mu$ is the difference between the observed and target measurement.
The $x_s,y_s$ error weights denote the error with respect to the state associated to the observed measurement $\mu$, and, owing to their angle-dependence, they are defined by $s$.
As commented in the previous subsection, and critical to the error modeling of the RUS process, the diagonal error $x$ is independent of the measurement outcome, whereas the coherence $y$ is inverted upon inverted measurement, i.e. $y^{(1)}_s = -y^{(0)}_s$, due to the one-sided action of the $Z$ operator propagating onto the measurement outcome.

While $\mathcal{T}_s^{(\mu)}$ describes the action of each individual instance of gate-teleportation, let $\rho^{(\mu)}_s(\rho;\theta_L)$ be the cumulative result after the $s^\mathrm{\rm th}$ step.
Thus, it is defined as
\begin{align}
    \rho^{(\mu)}_s(\Psi;\theta_L) =&\; \mathcal{T}_s^{(\mu)} \circ \mathcal{T}_{s-1}^{(1)} \circ ... \circ \mathcal{T}_{1}^{(1)} (\Psi; \theta_L),
\end{align}
and can be modeled as
\begin{align}
    \rho^{(\mu)}_s(\Psi;\theta_L) = 
    \; \mathcal{N}(\Psi, \Theta_{s}^{(\mu)};\,p_s^{(\mu)},q_s^{(\mu)}) + \delta_s^{(\mu)} \Psi_{\Theta_{s}^{(\mu)}},
    \label{eq:z-channel-rus-accumulated}
\end{align}
where $\Theta_{s}^{(\mu)} = [(-1)^{\mu}2^{s-1}-(2^{s-1}-1)]\theta_L$, such that $\Theta_{s}^{(0)}=\theta_L$, whereas $\Theta_{s}^{(1)}$ is the accumulation of wrongful rotations;
$\delta_s^{(\mu)}$, defined shortly, tracks a variation in the trace stemming from the use of unnormalised states and complex $y_s^{(\mu)}$.
These relations allow to study the propagation of errors in between attempts of gate-teleportation.
As $\rho^{(\mu)}_s(\Psi;\theta_L) = \mathcal{T}_s^{(\mu)} (\rho_{s-1}^{(1)}; \theta_L)$, we observe the recursive relation
\begin{align}
    p_s^{(\mu)} &= 
    p_{s-1}^{(1)} + x_s 
    \;- 2 x_s\,p_{s-1}^{(1)} 
    + 2 \mathrm{Re}(q_{s-1}^{(1)*}\,y_s^{(\mu)}) 
    \nonumber \\
    &\quad+\, x_s \delta_{s-1}^{(1)}, 
    \nonumber \\
    q_s^{(\mu)} &= 
    q_{s-1}^{(1)} + 
    y_s^{(\mu)}
    - 2x_s\mathrm{Re}(q_{s-1}^{(1)}) 
    - 2p_{s-1}^{(1)} \mathrm{Re}(y_s^{(\mu)})
    \nonumber \\
    &\quad+\, y_s^{(\mu)} \delta_{s-1}^{(1)},
    \nonumber \\
    \delta_{s}^{(\mu)} &= 
    \delta_{s-1}^{(1)} 
    + 4\,\mathrm{Im}(y_s^{(\mu)}) \mathrm{Im}(q_{s-1}^{(1)}).
    \label{eq:rus-error-propagation}
\end{align}
It can be noted that the form of the update of $\delta_{s}^{(\mu)}$ recalls that of the non-hermitian term in~\cref{eq:teleport-probability-Pm}.
To leading order, unravelling the recursive relations yields
\begin{align}
    p_s^{(\mu)} &\simeq \sum_{t=1}^{s}
    x_t + 2 \mathrm{Re}\left(
    y_s^{(\mu)}\sum_{t=1}^{s-1} y_t^{(1)*} 
    + \sum_{t=2}^{s-1}\sum_{u=1}^{t-1} y_t^{(1)*} y_u^{(1)} 
    \right), 
    \nonumber \\
    q_s^{(\mu)} &\simeq 
    y_s^{(\mu)}+\sum_{t=1}^{s-1}
    y_t^{(1)},
    \nonumber \\
    \delta_s^{(\mu)}
    &\simeq
    4\,\mathrm{Im}\!\left(y_s^{(\mu)}\right)
    \sum_{t=1}^{s-1}
    \mathrm{Im}\!\left(y_t^{(1)}\right)
    \nonumber \\
    &\quad
    +
    4
    \sum_{t=2}^{s-1}
    \mathrm{Im}\!\left(y_t^{(1)}\right)
    \sum_{u=1}^{t-1}
    \mathrm{Im}\!\left(y_u^{(1)}\right),
    \label{eq:rus-error-propagation-approx}
\end{align}
up to, respectively, $O(x_s^2 + x_s y_s^2 + y_s^4)$, $O(x_sy_s + y_s^3)$, and $O(x_s y_s^2 + y_s^4)$.

The RUS process terminates once a measurement $\mu=0$ is obtained.
Thus, its final result is the average of the $\mu=0$ normalized cumulative outputs $\tilde \rho_s^{(0)}(\Psi, \theta_L)$, weighted by the probability of their occurrence, which we denote $\Pi^{(0)}_s(\Psi; \theta_L)$.
The normalized states are described by
\begin{align}
    \tilde{\rho}^{(0)}_s(\Psi;\theta_L) &=\; \tilde{\mathcal{T}}_s^{(0)} \circ \tilde{\mathcal{T}}_{s-1}^{(1)} \circ ... \circ \tilde{\mathcal{T}}_{1}^{(1)} (\Psi; \theta_L)
    \nonumber \\
    &=\frac{1}{2^{s}\Pi^{(0)}_s}\;\mathcal{T}_s^{(0)} \circ \mathcal{T}_{s-1}^{(1)} \circ ... \circ \mathcal{T}_{1}^{(1)} (\Psi; \theta_L)
    \nonumber \\
    &=\frac{1}{2^{s}\Pi^{(0)}_s}\;\rho_s^{(0)} (\Psi; \theta_L).
\end{align}
Therefore, the final result of the RUS process is best described by the unnormalised outputs of the protocol, as
\begin{align}
    \rho_\mathrm{RUS}(\Psi; \theta_L) 
    =\;& \sum_{s=1}^{\infty} \Pi^{(0)}_s(\Psi; \theta_L) \;\tilde{\rho}^{(0)}_s(\Psi; \theta_L) 
    \nonumber \\
    =\;& \sum_{s=1}^{\infty} \frac{1}{2^{s}} \rho^{(0)}_s(\Psi; \theta_L) 
    \nonumber \\
    =\;& 
    \mathcal{N}(\Psi;\, \theta_L, p_\mathrm{RUS}, q_\mathrm{RUS})
    + \delta_{\rm RUS}(\theta_L) \Psi_{\theta_L},
    \label{eq:rus-error-ansatz}
\end{align}
recalling that $\Psi_{\theta_L} = R_{Z,L}\left(\theta_L\right) \, \Psi \, R^\dagger_{Z,L}\left(\theta_L\right)$ denotes the correctly rotated state, and we where have
\begin{align}
    p_\mathrm{RUS}(\theta_L)
    &= \sum_{s=1}^{\infty} \frac{1}{2^{s}}  p^{(0)}_s(\theta_L),
    \nonumber \\
    q_\mathrm{RUS}(\theta_L)
    &= \sum_{s=1}^{\infty} \frac{1}{2^{s}}  q^{(0)}_s(\theta_L),
    \nonumber \\
    \delta_{\rm RUS}(\theta_L)
    &= \sum_{s=1}^{\infty} \frac{1}{2^{s}}  \delta^{(0)}_s(\theta_L).
    \label{eq:rus-error-coefficients}
\end{align}
As it will be argued in the following subsection, $\delta_{\rm RUS}(\theta_L)\simeq0$. 
\vspace{1em}

\subsubsection{Noise Cancellation in the Clifford-terminated RUS}

Previous work~\cite{toshio2025, ismail2025} has observed that, by choosing $\theta_L = \pi/2^{C+1}$, the RUS process is terminated deterministically after $C$ unsuccessful steps by the fully-FT application of an $S$ gate, as the ${(C+1)}^\mathrm{\rm th}$ step requires the injection of $\theta_{L,{C+1}}=\pi/2$.
In the first place, we highlight that this Clifford termination is applicable to any dyadic fraction of $\pi/2$, i.e.
\begin{equation}
        \theta_L = \frac{\pi}{2} \frac{2k+1}{2^C},
        \label{eq:app:dyadic-thetaL}
\end{equation}
for any integer $k$.
The form of~\cref{eq:app:dyadic-thetaL} drastically expands the set of convenient choices for $\theta_L$ for algorithmic purposes, as it provides a dense set of angles, capable of approximating any other up to arbitrary precision.

Remarkably, we observe that the Clifford-terminated RUS (ctRUS) protocol has \textit{near-vanishing} coherent error.
For the moment, we assume that the terminating Clifford $S$ correction is noiseless.
The output of the Clifford-terminated RUS protocol is
\begin{align}
    \rho_\mathrm{ctRUS}(\Psi;\theta_L,C)
    =\;&
    \sum_{s=1}^{C}
    \frac{1}{2^{s}}
    \rho_s^{(0)}(\Psi;\theta_L)
    \nonumber \\
    &\quad
    +
    \frac{1}{2^C}
    S\;\rho_{C}^{(1)}(\Psi;\theta_L)\;S^\dagger .
    \label{eq:ctRUS-output-noiseless-S}
\end{align}
The perfect Clifford correction maps the final failed branch back to the target logical rotation.
Therefore,
\begin{equation}
    S\rho_{C}^{(1)}(\Psi;\theta_L)S^\dagger
    =
    \mathcal{N} \left(
        \Psi;\,\theta_L,p_{C}^{(1)},q_{C}^{(1)}
    \right)
    +
    \delta_{C}^{(1)}\Psi_{\theta_L}.
\end{equation}
It follows that the full Clifford-terminated protocol is described by a single effective noise channel,
\begin{align}
    \rho_\mathrm{ctRUS}(\Psi;\theta_L,C)
    =&\;
    \mathcal{N}
    \left( \rho;\,\theta_L, p_\mathrm{ctRUS},q_\mathrm{ctRUS}
    \right)
    \nonumber \\
    &\;+\; \delta_\mathrm{ctRUS}(\theta_L,C)\Psi_{\theta_L},
    \label{eq:ctRUS-ansatz-error}
\end{align}
with
\begin{align}
    p_\mathrm{ctRUS}(\theta_L, C)
    &=
    p_\mathrm{RUS}^{(C)}
    +
    \frac{1}{2^C}p_{C}^{(1)},
    \nonumber \\
    q_\mathrm{ctRUS}(\theta_L, C)
    &=
    q_\mathrm{RUS}^{(C)}
    +
    \frac{1}{2^C}q_{C}^{(1)},
    \nonumber \\
    \delta_\mathrm{ctRUS}(\theta_L,C)
    &=
    \delta_\mathrm{RUS}^{(C)}
    +
    \frac{1}{2^C}\delta_{C}^{(1)},
    \label{eq:ctRUS-error-coefficients}
\end{align}
as $p_\mathrm{RUS}^{(C)}$, $q_\mathrm{RUS}^{(C)}$, and $\delta_\mathrm{RUS}^{(C)}$ denote the weighted contribution of the successful RUS branches appearing before Clifford termination.
These are, to leading order,
\begin{align}
    p_{\mathrm{RUS}}^{(C)} &\simeq
    \sum_{s=1}^{C}\!\left[
    \left(\frac{1}{2^{s-1}}\!-\!\frac{1}{2^C}\right)x_s
    -\frac{2}{2^C}\mathrm{Re}\!\left(
    y_s^{(0)}
    \sum_{t=1}^{s-1} y_t^{(0)*}
    \right)
    \right],
    \nonumber \\
    q_\mathrm{RUS}^{(C)} &\simeq
    \frac{1}{2^C}
    \sum_{s=1}^{C}
    y_s^{(0)},
    \nonumber \\
    \delta_\mathrm{RUS}^{(C)}
    &\simeq
    -\frac{4}{2^C}
    \sum_{s=2}^{C}
    \mathrm{Im}\!\left(y_s^{(0)}\right)
    \sum_{t=1}^{s-1}
    \mathrm{Im}\!\left(y_t^{(0)}\right).
    \label{eq:rus-truncated-error-propagation-approx}
\end{align}
Before proceeding further, we can note that even in absence of cut-off, for $C\rightarrow \infty$, both $\delta_{\rm RUS}$ and $q_{\rm RUS}$ approach zero.
The final failed branch has, to leading order,
\begin{align}
    p_{C}^{(1)} &\simeq
    \sum_{s=1}^{C} x_s
    + 2 \mathrm{Re}\left[
    y_{C}^{(0)}
    \sum_{s=1}^{C-1} y_s^{(0)*}
    +
    \sum_{s=1}^{C-1} y_s^{(0)}
    \sum_{t=1}^{s-1} y_t^{(0)*}
    \right],
    \nonumber \\
    q_{C}^{(1)} &\simeq
    -\sum_{s=1}^{C} y_s^{(0)},
    \nonumber \\
    \delta_{C}^{(1)}
    &\simeq
    4
    \sum_{s=2}^{C}
    \mathrm{Im}\!\left(y_s^{(0)}\right)
    \sum_{t=1}^{s-1}
    \mathrm{Im}\!\left(y_t^{(0)}\right).
    \label{eq:ctRUS-term-error-propagation-approx}
\end{align}
Substituting~\cref{eq:rus-truncated-error-propagation-approx,eq:ctRUS-term-error-propagation-approx} into~\cref{eq:ctRUS-error-coefficients}, we observe that the leading coherent contribution cancels.
The final state in~\cref{eq:ctRUS-ansatz-error} has noise coefficients
\begin{align}
    p_{\mathrm{ctRUS}}(\theta_L, C)
    &=
    \sum_{s=1}^{C} \frac{1}{2^{s-1}}x_s
    + O(x_s^2+x_s y_s^2+y_s^4),
    \nonumber \\
    q_\mathrm{ctRUS}(\theta_L, C)
    &= O(x_s y_s+y_s^3),
    \nonumber \\
    \delta_\mathrm{ctRUS}(\theta_L,C)
    &= O(x_s y_s^2+y_s^4).
    \label{eq:ctRUS-error-propagation-approx}
\end{align}

In practice, the terminating Clifford gate is generally noisy.
If its stochastic error probability is denoted by $p_L^{(S)}$, then the above expressions neglect contributions of order $2^{-C}p_L^{(S)}$, together with higher-order products between the noise of the $S$ gate and the RUS steps.
The Clifford error rate $p_L^{(S)}$ is suppressed exponentially in code distance, and these terms are further reduced by the probability $2^{-C}$ of reaching the terminating branch, hence are typically negligible.
Moreover, since the terminating correction is a Clifford operation, its noise can be Pauli-twirled~\cite{dur2005-twirling,emerson2007-twirling}, hence it does not contribute a coherent-noise term to $q_\mathrm{ctRUS}$.

Altogether, the termination cancels the leading contribution to the coherent-noise coefficient $q_\mathrm{ctRUS}$, as well as the $O(y_s^2)$ coherent-noise contribution to $p_\mathrm{ctRUS}$.
The corresponding branch-normalisation contribution $\delta_\mathrm{ctRUS}$ cancels as well.
This provides an intrinsic error-suppression mechanism for dyadic choices of $\theta_L$, in addition to giving a natural deterministic cutoff for the RUS process.
Moreover, this result extends to other Clifford-terminated probabilistic processes, such as the fully FT injection of $T$ gates.

\section{Details on Rydberg Gates Optimization}
\label{sec:appendix-rydberg}

\subsection{On the negligible impact of axial motion}
\label{subsec:app:rydberg-axial-fluctuations}

Rydberg gates in alkali atoms (Rb, Cs) are significantly impacted not only by radial fluctuations, but axial too.
In this subsection, we estimate the contribution of axial position fluctuations to the variance of the Rydberg interaction for the ytterbium parameters used in this work.
Specifically, we show that, within a semiclassical picture of atomic motion, their impact is an increase in interaction fluctuations by just a few percentage points.

Under the assumption of harmonic traps and thermal equilibrium, atoms experience shot-to-shot fluctuations in position and velocity with variance
\begin{align}
    \sigma_x^2 &\simeq 
    \frac{\hbar}{2m\omega_x} \coth{\left( \frac{\beta \hbar\omega_x}{2}\right)}, 
    \nonumber \\
    \sigma_{v_x}^2 &\simeq
    \frac{\hbar\omega_x}{2m} \coth{\left( \frac{\beta \hbar\omega_x}{2}\right)},
    \label{eq:app:rydberg-variance-pos-and-vel}
\end{align}
where $\omega_x$ is the trap frequency along the axis considered. 

In our setup, we take the interatomic separation to be $\vec{R} = (R_0+\Delta x,\,\Delta y,\,\Delta z)$, with the radial trap frequency inducing fluctuations in the $x,y$ plane, whereas $z$ is identified as the axial direction.
The Rydberg potential is $V=V_0 (R/R_0)^{-6}$, thus
\begin{equation}
    V=V_0 \left[
        \left(1 + \frac{\Delta x}{R_0}\right)^2 + \frac{\Delta y^2}{R_0^2} + \frac{\Delta z^2}{R_0^2}
        \right]^{-3}.
\end{equation}
The standard deviations of the relevant terms are
\begin{alignat}{2}
\sigma_{\Delta x} = \sigma_{\Delta y} &= \sqrt{2} \sigma_x &&\,\simeq\, 0.019 R_0, 
\nonumber \\
\sigma_{\Delta x^2} = \sigma_{\Delta y^2} &= 2\sqrt{2} \sigma^2_x &&\,\simeq\, 0.0005 R_0^2, 
\nonumber \\
\sigma_{\Delta z^2} &= 2\sqrt{2} \sigma^2_z &&\,\simeq\, 0.011 R_0^2,
\end{alignat}
where numerical values are obtained under the assumption of interatomic separation $R_0 = 2.5 \mu \rm m$, temperature $2 \,\mu\rm K$, radial trap frequency $\omega_{x}/2\pi = 50 \,\rm kHz$, and axial frequency $\omega_{z}/2\pi = 10 \,\rm kHz$, roughly reproducing state-of-the-art values~\cite{peper2024, muniz2025-ancillae}.
These values justify the expansion
\begin{equation}
V \approx V_0 \left[
1 \;-\; \frac{6\Delta x}{R_0}
\;-\; \frac{3\,\Delta z^{2}}{R_0^{2}}
\right],
\end{equation}
where second-order fluctuations in $x$ and $y$ were neglected due to their much smaller deviation. 
The isolated contributions from $x$ and $z$ displacements to fluctuations in interaction strength are $\sigma_{V,x} = 6\sqrt{2} V_0 \,\sigma_x/R_0$ and $\sigma_{V,z} = 6\sqrt{2} V_0 \,\sigma_z^2/R_0^2$, respectively.  
As they add in quadrature, their combination yields
\begin{equation}
    \sigma_V = \sigma_{V,x} \sqrt{1 + \frac{\sigma_{V,z}^2}{\sigma^2_{V,x}}}
    \simeq 1.04 \,\sigma_{V,x}.
\end{equation}
That is, including axial fluctuations leads to only a $\sim 4\%$ increase in the deviation of the interaction strength.
Given the computational cost of its inclusion in simulations, they were neglected.

\subsection{Noise Approximation for Gate Optimization}
\label{subsec:app:rydberg-noise-approximation}

In the optimization process, a semiclassical approximation of the noise was considered, reducing the computational cost of the procedure. 
Specifically, the Hamiltonian in~\cref{eq:rydberg-hamiltonian} is assumed, with shot-to-shot gaussian fluctuations on its parameters, determined by the semiclassical variance of position and velocity in~\cref{eq:app:rydberg-variance-pos-and-vel}: 
\begin{enumerate}[label=(\roman*)]
    \item Fluctuations $\delta \Delta_q$ in detuning---different for each qubit---with standard deviation $\sigma_\Delta = 2 \pi \sigma_v / \lambda$. 
    We considered $\sigma_\Delta = 0.005\,\Omega$.
    \item Fluctuations $\delta V$ in the interaction strength, with standard deviation $\sigma_V = 6\sqrt2 V\,\sigma_x/R$. 
    We considered $\sigma_V = 7\,\Omega$. 
    \item Fluctuations $\delta \Omega$ in amplitude, which also give rise to additional shifts in detuning, due to self-light shifts.
    We considered $\sigma_\Omega = 5 \times 10^{-3}\Omega$, and $\delta \Delta = 0.2\,\delta\Omega$.
\end{enumerate}
The driving amplitude was assumed to be constant throughout the protocol, ignoring finite ramping times.

Despite the possibility of detecting and post-selecting upon most decay errors (see~\cref{subsec:rydberg-noise}), inclusion of Rydberg decay in the optimization process was found to be beneficial.
Details of inclusion in the computation of the cost function are provided in the next subsection.

\subsection{Cost Function Computation}
\label{subsec:app:rydberg-chizz-computation}

We wish to implement the target gate $U_0$ on $N_q$ qubits.
In practice, noisy versions $U_n$ of the gate are implemented, yielding the channel $\mathcal{U}(\rho) = \overline{U_n \rho U_n^\dagger}$, where the overline denotes the average over noisy realizations.
To make the role of noise explicit, note that \(\mathcal{U}(\rho)\) is equivalent to an "afterwards" noise channel \(\mathcal{E}(\rho)\) following the application of the target gate, such that
\begin{align}
    \mathcal{U}(\rho) &= \mathcal{E}(U_0 \rho U_0^\dagger),
    \nonumber \\
    \mathcal{E}(\rho) &= \mathcal{U}(U_0^\dagger\rho U_0).
\end{align}
Equivalently, this expands in terms of its chi-matrix as
\begin{equation}
    \mathcal{E}(\rho) = \sum_{Q,Q'} \chi_{\scriptscriptstyle Q,Q'} Q \rho Q',
\end{equation}
Its diagonal entries are computed as
\begin{equation}
    \chi_{\scriptscriptstyle Q,Q} = \frac{1}{2^{3N_q}} \sum_{P \in \mathcal{P}^{\otimes N_q}} (-1)^{c_{P,Q}} \, \mathrm{tr} \left\{ P \, \mathcal{E}(P) \right\},
    \label{eq:chi-def}
\end{equation}
where $\mathcal{P}^{\otimes N_q}$ denotes the set of $N_q$-qubits Pauli operators, and $c_{P,Q} = 0\,(1)$ if $P$ and $Q$ (anti)commute.
This expression can then be used to compute the relevant chi-matrix element in numerical simulations.

Combining the previous expressions, again denoting with the overline average over noisy realizations,
\begin{align}
    \chi_{\scriptscriptstyle Q,Q} &=
    \frac{1}{2^{3N_q}} \sum_{P \in \mathcal{P}^{\otimes N_q}} (-1)^{c_{Q,P}} \, \mathrm{Tr} \left\{ P \, \overline{U_n U_0^\dagger P U_0 U_n^\dagger} \right\} 
    \nonumber \\
    &\simeq 
    \frac{1}{2^{3N_q}} \sum_{P \in \mathcal{P}^{\otimes N_q}} (-1)^{c_{Q,P}} \sum_{i \in \mathcal{C}} \langle i | \overline{P U_n U_0^\dagger P U_0 U_n^\dagger}| i \rangle,
\end{align}
where $\mathcal{C}$ denotes a set of orthonormal basis states in the computational subspace of $N_q$ qubits.
Then, with \( P = \sum_{s,t\in\mathcal{C}} P_{st} |s\rangle \langle t| \), where \( P_{st} = \langle s | P | t \rangle \), it reads
\begin{align}
    \chi_{\scriptscriptstyle Q,Q} = \frac{1}{2^{3N_q}} &\sum_{P \in \mathcal{P}^{\otimes N_q}} (-1)^{c_{Q,P}}
    \nonumber \\
    &\;\sum_{s,t,p,q \in \mathcal{C}} P_{st} P_{pq} \overline{\langle t | U_n U_0^\dagger \ket{p}\bra{q} U_0 U_n^\dagger| s \rangle}.
\end{align}

Conveniently rearranging, this yields
\begin{align}
    \chi_{\scriptscriptstyle Q,Q} &= \frac{1}{2^{3N_q}} 
    \sum_{s,t,p,q \in \mathcal{C}}  \sigma_{st,pq}(Q)\; \overline{
    \langle t | U_n U_0^\dagger \ket{p}\bra{q} U_0 U_n^\dagger| s \rangle,
    }
\end{align}
where 
\begin{equation}
    \sigma_{st,pq}(Q) = \sum_{P \in \mathcal{P}^{\otimes N_q}} (-1)^{c_{Q,P}} P_{st} P_{pq},
\end{equation}
are gate-independent coefficients.

For simplicity, the average over noises was taken to be the sum of separate contributions from fluctuations in Hamiltonian parameters and Rydberg decay, each with different form of the noisy evolution operator $U_n$ and weighting of the average.
The noisy evolution depends on the optimized control parameters and total duration. i.e. $U_n \equiv U_n(\vec{x},T)$: $x_0$ denotes a residual single-qubit phase after the application of the Rydberg Hamiltonian, whereas each $x_{k>0}$ parameter represents the laser phase at the $k^\mathrm{\rm th}$ time-step.
Computation of the time evolution operator occurs in the limit of a large enough number of time-steps, thus is well approximated by a simple first-order decomposition.

Shot-to-shot fluctuations were incorporated in the cost function by averaging over multiple realizations
\begin{equation}
    U_n^{(H)}(\vec{x},T) = 
    R_Z^{\otimes N_q}(x_0)\prod_{k=N}^{1} e^{-i \frac{H_n({x_k})}{\hbar} \frac{T}{N} },
\end{equation}
by means of Gauss-Hermite quadrature. 
That is, rather than using probabilistic sampling, the noisy realizations were selected at fixed values, and averaged with fixed weights over their mesh.
We averaged over the 2,000 most probable noisy realizations in the mesh, with the mesh given by seven points per noisy variable, i.e. $7^4$ realizations for two qubits.

Rydberg decay is incorporated by replacing the time-evolution operator at time step $k'$ with a jump operator $L_q^{(i)}=\sqrt{\gamma_i} \ket{i}\bra{r}$, representing Rydberg decay of the $q^{\rm th}$ qubit to the computational state $\ket{i}$.
Thus, the average was performed, with equal weight $dt$, on all possible operators
\begin{align}
    U_{q, i, k'}^{(\gamma)}(\vec{x},T) =&\; R_Z^{\otimes N_q}(x_0)
    \left( \prod_{k=N}^{k'+1} e^{-i \frac{H_0({x_k})}{\hbar} \frac{T}{N}} \right)
    \nonumber \\
    &\; L_q^{(i)}
    \left(\prod_{k=k'-1}^{1} e^{-i \frac{H_0({x_k})}{\hbar} \frac{T}{N}}\right),
\end{align}
where $H_0$ is the noiseless Hamiltonian, similar to the approach used in~\cite{jandura2024}.

Finally, we point out that the gradient can be analytically computed using the product rule, both for the gate parameters $\vec{x}$ and gate duration $T$.

\subsection{Numerical Instabilities}
\label{subsec:app:rydberg-T-optimization-details}

In this subsection, we comment on some undesired behavior in the numerical optimization of the pulse shape $\phi(t)$, and on the adopted solutions.
Concretely, we added penalty-terms to the cost-function in~\cref{eq:cost-function}, to (i) ensure a smooth pulse profile, hence its experimental feasibility, and (ii) to avoid erroneous local solutions stemming from the combined optimization of the pulse shape and its duration $T$.

The pulse was smoothened to match experimental constraints, by the addition of the regularizer 
\begin{equation}
    \mathrm{cost_{sm}} = \eta_\mathrm{sm} N \sum_{i=1}^{N-1} (\phi_{i+1} - \phi_i)^2,
\end{equation}
already used in the optimization of Rydberg interactions \cite{giudice2025, zeng2025}. 
Our results were found by setting $\eta_\mathrm{sm}=3\times 10^{-9}$.

The optimization of the gate duration $T$ if unconstrained, was observed to occasionally lead to undesired behavior such as null gate time, or, oppositely, overly long gate durations.
These were prevented by the regularizer
\begin{equation}
    \mathrm{cost}_T = \frac{1 - \tanh{[50 \Omega (T - T_\mathrm{tmop})]}}{2} + \eta_T \Omega(T - T_\mathrm{tmop}),
\end{equation}
conditioned on the duration $T_\mathrm{tmop}$ of the corresponding time-optimal pulse, where the first term imposes a duration longer than that of the time-optimal solution, as necessarily the case, and the second term provides a small penalty against longer pulses.
We set $\eta_T = 10^{-9}$, to ensure that the latter penalty only affects the optimization process once a robust solution is found, driving the procedure to shorter equally-robust solutions.

\section{Circuit-level Noise}
\label{app:logical-simul-noise} 

As mentioned in~\cref{subsec:performance-assumptions}, the logical protocol is simulated against circuit-level noise of varying strength, modeled by the error rate $p\equiv p_2$ of two-qubit gates.
This section provides the exact noise acting on $CZ$ gates in our simulations, accounting for both Rydberg decay errors and other noise sources.
In particular, we assume that a representative fraction $f_d=0.5$ of the error rate comes from Rydberg decay, allowing the impact of different noise sources to be captured by the model.
Then, we assume that
\begin{enumerate}
    \item a fraction $R_e = 0.98$ of the errors due to decay results in biased erasures, in line with the estimates from~\cite{ma2023, sahay2023};
    \item the remaining noise due to decay leads to depolarizing noise, but with no contribution from $XX$, $XY$, $YX$, and $YY$ errors, which would require unlikely double-decay events, as excitations to the Rydberg state are only possible from $\ket{1}$;
    \item the remaining fraction $1-f_d=0.5$ of the error rate comes from sources other than decay; this contributes to $Z$-biased noise, owing to the fact that, in Rydberg gates, $\ket{0}$ and $\ket1$ are uncoupled in the absence of decay.
\end{enumerate}
Therefore, for the majority of $CZ$ gates, we assume the noise channel to be Pauli noise where the probabilities for each noise term
\begin{equation}
\begin{pmatrix}
 & p_{IX} & p_{IY} & p_{IZ} \\
p_{XI} & p_{XX} & p_{XY} & p_{XZ} \\
p_{YI} & p_{YX} & p_{YY} & p_{YZ} \\
p_{ZI} & p_{ZX} & p_{ZY} & p_{ZZ}
\end{pmatrix}
\nonumber
\end{equation}
are the following fraction of $p_2$
\begin{equation}
\left(\frac{1-f_d}{3} + f_d\frac{R_e}{4}\right)
\begin{pmatrix} &0&0&1\\ 0&0&0&0\\ 0&0&0&0\\ 1&0&0&1 \end{pmatrix}
+
f_d \frac{1-R_e}{11}
\begin{pmatrix} &1&1&1\\ 1&0&0&1\\ 1&0&0&1\\ 1&1&1&1 \end{pmatrix}
\label{eq:app:cz-biased-noise}
\end{equation}
and $p_{II}=1-\sum_{P\neq II} p_P$. 
Note that the weight $f_d R_e/4$ follows from the fact that one of the Pauli operators originating from biased erasures is the identity.

Since post-selection is used in the encoding of the resource state, erasure in the CZ gates acting on the first two rows of the logical patch was assumed to be post-selected upon, instead of replaced by biased errors. 
These locations were chosen due to their proximity to the support of the transversal rotations, such that phase errors on these CZ gates can more easily increase the failure rate of the TMR protocol.
Therefore, their error model does not include the $f_d R_e /4$ term in~\cref{eq:app:cz-biased-noise}.
Moreover, the rate of acceptance of the resource state was reduced by the probability of observing no erasure,
\begin{equation}
    1-p_{\rm erasure} = (1-p_2 f_d R_e)^{n^{(2)}_{CZ}}
    \label{eq:app:p-erasure}
\end{equation}
where $n^{(2)}_{CZ}$ is the distance-dependent number of relevant CZ gates.

For completeness, we also recall that single qubit gates have depolarizing rate $p_2/5$, while measurements and resets, only available in the computational basis, have bit-flip rate $p_m=p_2/2$ and $p_i=p_2/5$, respectively.

\section{Numerical Methods for Simulation of Logical Gadgets}
\label{app:logical-TMR-simul-method} 

In this appendix, we outline the numerical modeling we implemented for the simulation of logical gadgets used in the manuscript : the TMR protocol and other Clifford operations.
This appendix is organised as follows
\begin{enumerate}[label=(\roman*)]
    \item In~\cref{subsec:app:TMR-noise-channel-deriv-for-numerics}, we derive an explicit modeling of the noisy TMR channel, accounting for both the noise in the non-Clifford transversal rotations and standard circuit-level noise.
    \item In~\cref{subsec:app:TMR-noise-channel-numerical-estimatation}, we outline how Clifford simulators, such as \texttt{stim}~\cite{stim}, can be used to accurately estimate the noisy TMR channel.
    We note that the exactness of the method will be obtained at the price of a loss in efficiency.
    \item In~\cref{subsec:app:TMR-noise-channel-deep-circuit-estimation} we show how to compute the performance of the TMR protocol within a deeper circuit, to better reproduce its performance in a realistic scenario.
    For the purpose, we also compute the logical flip rate of quantum memory experiments and state injection.
    \item In~\cref{subsec:app:logical-cliffords-simulation} we benchmark the performance of transversal Clifford operations under the considered noise.
    \item In~\cref{subsec:app:TMR-noise-channel-numerical-hyperparams-approxims} we comment on the sampling statistics and uncertainty estimation.
\end{enumerate}

\subsection{Full working modeling of the TMR protocol}
\label{subsec:app:TMR-noise-channel-deriv-for-numerics}

This subsection derives a closed form describing the full TMR channel under noise.
This includes accounting for noise in (i) the initial error detection $\mathrm{QED}_\mathrm{a}$, (ii) the transversal noisy rotations $\mathcal{R}^{\otimes r}$, (iii) the following error detection $\mathrm{QED}_\mathrm{b}$, and finally (iv) gate teleportation $\mathcal{T}$.
We stress that our method allows for the noisy non-Clifford $\mathcal{R}^{\otimes r} (\rho) =\sum_{\alpha\beta} \mathcal{R}_{\alpha\beta} P_\alpha \rho P_\beta$ rotations to be accounted for without approximations, and to address all noise sources outlined in~\cref{subsec:partial-ft-error-budget,subsec:partial-ft-rus}, and thoroughly described in~\cref{sec:app:error-bugget-deriv}.
The modeling exploits the Clifford nature of the syndrome extraction circuit, and assumes Pauli noise on all gates other than the transversal rotations, as customary in QEC and legitimated by the practical implementation of Pauli Twirling for Clifford operations.

The resource state is, before any encoding, $\ket{0}\bra{0}^{\otimes n}$, over $n$ physical qubits.
The logical resource state is initialized by the suitable syndrome extraction circuit $\mathrm{QED}_\mathrm{a}$.
As this is a Clifford circuit, any Pauli error can be propagated to its end, up to a global phase.
Therefore, the resulting state after $\mathrm{QED}_\mathrm{a}$ can be written as
\begin{align}
    m_{L}^{(1)} = \mathrm{QED_1}(\ket{0}\bra{0}^{\otimes n})
    = \sum_\mathbf{a} p_\mathbf{a} P_\mathbf{a} \ket{+}\bra{+}_L P_\mathbf{a},
\end{align}
where $\ket{+}\bra{+}_L$ is the result of the ideal circuit, $P_\mathbf{a}$ denotes a physical Pauli operator, and any global phase has been canceled.

After the transversal rotations of the TMR, here encoded by the chi-matrix $\mathcal{R}_{\alpha\beta}$, the state is in
\begin{align}
    m_{L}^{(TMR)} &= \mathcal{R}^{\otimes r} (m_{L}^{(1)})
    \nonumber \\
    &= \sum_{\alpha,\beta} \mathcal{R}_{\alpha\beta} P_\alpha m_{L}^{(1)} P_\beta
    \nonumber \\
    &=\sum_{\mathbf{a},\alpha,\beta} \mathcal{R}_{\alpha\beta}\,p_\mathbf{a}\,  P_\alpha P_\mathbf{a} \ket{+}\bra{+}_L P_\mathbf{a} P_\beta.
\end{align}
Note that $\mathcal{R}_{\alpha\beta}$ represents the overall action of the noisy gates, not an error channel following their ideal implementation. 

The resource state then goes through more rounds of error syndrome extraction, denoted $\mathrm{QED_b}$.
Its final state, which we denote $\mathfrak{M}_L$ as in~\cref{subsec:app:error-budget:noise-whole-teleport}, reads $\mathfrak{M}_L = \mathrm{QED_2}(m_{L}^{(TMR)})$. 
Breaking down the action of the second error detection cycle, it can be written as
\begin{widetext}
    \begin{equation}
    \mathfrak{M}_L = \mathrm{Tr}_{\mathbf{z}} \left(
        \mathcal{E} \left(
            \sum_{\mathbf{a},\alpha,\beta} 
            \mathcal{R}_{\alpha\beta} \,p_\mathbf{a}\,  
            P_\alpha 
            P_\mathbf{a} \ket{+}\bra{+}_L P_\mathbf{a} 
            P_\beta 
            \; \otimes \; \ket{\mathbf{z}_{\mathbf{a},\alpha}}\bra{\mathbf{z}_{\mathbf{a},\beta}} 
    \right)
    \right),
\end{equation}
\end{widetext}
where $\mathcal{E}$ represents noise occurring during the error detection process, $\mathrm{Tr}_{\mathbf{z}}$ represents the ancillae measurement, and $\mathbf{z}_{\mathbf{a},l}$ is the ancillae bit-string measured after the stabilizers' extraction.
Note that Pauli noise after the transversal rotation can be considered to be applied at the end of a noiseless extraction owing to the Clifford nature of the circuits, as previously explained for $\mathrm{QED_a}$.
Upon the ancillae measurement, only terms with identical left and right syndromes $\mathbf{z}_{\mathbf{a},l} = \mathbf{z}_{\mathbf{a},r}$ are preserved, in turn requiring $P_\alpha \equiv_L P_\beta$, i.e. $P_\alpha = P_\beta$ up to logical operators and stabilizers. 
Concretely, as the TMR is applied on the support of $Z_L$, the relation $P_\alpha \equiv_L P_\beta$ imposes
\begin{align}
    P_\beta = i^{v_p(\alpha,\beta)} Z_L^{v_0(\alpha,\beta)} \prod_k S_k^{v_k(\alpha,\beta)} P_\alpha
    \label{eq:app:logically-equivalent-Pl-Pr}
\end{align}
where $S_k$ denotes the stabilizers with full support on the qubits acted upon by the transversal rotations, $v_p\in\{0,1,2,3\}$, and $v_{0}, v_k \in \{0,1\}$, .
Therefore, after measurement and post-selection, we can write
\begin{align}
    \mathfrak{M}_L
    &= \mathcal{E'} \left(
    \sum_{(\alpha,\beta)_L } \mathcal{R}_{\alpha\beta} 
    P_\alpha  \rho_L^{(1)} P_\beta
    \right) \nonumber \\
    &= 
    \sum_{(\alpha,\beta)_L } \mathcal{R}_{\alpha\beta} 
    \sum_{\mathbf{a}, \mathbf{b}} 
    \,p_{\mathbf{a}, \mathbf{b}} \, 
            P_\mathbf{b} P_\alpha P_\mathbf{a} 
            \ket{+}\bra{+}_L P_\mathbf{a} P_\beta 
            P_\mathbf{b}
            \;
    ,
\end{align}
where $(\alpha,\beta)_L $ denotes all pairs of Pauli operators satisfying~\cref{eq:app:logically-equivalent-Pl-Pr}, and $\mathbf{b}$ identifies the average over noisy trajectories in $\mathrm{QED}_b$.

The same errors have the same logical correction.
Therefore, we wish to replace the right-applied operator $P_\beta$ with $P_\alpha$.
Using~\cref{eq:app:logically-equivalent-Pl-Pr} and rearranging yields
\begin{align}
    \mathfrak{M}_L = 
    &\sum_{(\alpha,\beta)_L } \mathcal{R}_{\alpha\beta} 
    \nonumber \\
    &\;\sum_{\mathbf{a,b}} 
    p_{\mathbf{a,b}} \, 
    s_{\mathbf{a},\alpha,\beta} \, 
            P_\mathbf{b} P_\alpha P_\mathbf{a}
            \ket{+}\bra{+}_L
            Z_L^{v_0(\alpha,\beta)}
            P_\mathbf{a} P_\alpha P_\mathbf{b}
    ,
    \label{eq:app:rho2-tmr-modeling}
\end{align}
where $s_{\mathbf{a},\alpha,\beta} = \pm i^{v_p(\alpha,\beta)}$, with the sign determined by the commutation relation between $P_\mathbf{a}$ and $Z_L^{v_0(\alpha,\beta)} \prod_k S_k^{v_k(\alpha,\beta)}$.

The resource state $\mathfrak{M}_L$ is then injected via gate-teleportation onto a target logical qubit, initially in $\Psi$.
The outcome of the teleportation measurement is denoted $\mu$.
Under a Pauli noise model, the action of the protocol is again modeled as the ideal scheme followed by residual noise $Q_\mathbf{c}$, while previously accumulated Pauli errors $P_\mathbf{b} P_\alpha P_\mathbf{a}$ propagate through the protocol to residual errors $Q_\mathbf{b} Q_l Q_\mathbf{a}$.
Denoting $\mathbf{t}$ as the combination of each $\mathbf{a}, \mathbf{b}, \mathbf{c}$ trajectories, and $Q_{\mathbf{t},l} = Q_\mathbf{c} Q_\mathbf{b} Q_l Q_\mathbf{a}$, the result reads
\begin{align}
    \rho_L\,(\Psi) = 
    &\sum_{(\alpha,\beta)_L } \mathcal{R}_{\alpha\beta} (-1)^{\mu \cdot v_0(\alpha,\beta)}
    \nonumber \\
    &\quad\sum_{\mathbf{t}} 
    p_{\mathbf{t}} \, 
    s_{\mathbf{t},\alpha,\beta} \,
            Q_{\mathbf{t},l}\,
            \Psi Z_L^{v_0(\alpha,\beta)}
            Q_{\mathbf{t},l}^{\dagger}
            .
\end{align}
where the phase $(-1)^{\mu \cdot v_0(\alpha,\beta)}$, arises from the one-sided action of $Z_L^{v_0(\alpha,\beta)}$ on the teleportation measurement outcome $\mu$. 
It is determined by observing that
\begin{enumerate}
    \item if $v_0(\alpha,\beta) = 0$, then $Z_L^{v_0(\alpha,\beta)} \equiv I$, and no additional phase is ever acquired;
    \item if $v_0(\alpha,\beta) = 1$, then the adopted resource state is effectively $\ket{+}\bra{-}_L$, such that the phase between measurements in $\mu=0$ and $1$ differs by $(-1)^{\mu}$.
\end{enumerate}
The result $\mu$ of the measurement can be determined and used in post-processing.
With \(r\) physical rotations, the target outcome is \(\mu=0\) when \(\lfloor r/2\rfloor\) is even and \(\mu=1\) when it is odd.

Upon decoding, the state is mapped back onto the logical computational state, yielding the final state
\begin{align}
    \rho_L^{(\mu)} (\Psi) = 
    &\sum_{(\alpha,\beta)_L }
    \mathcal{R}_{\alpha\beta} 
    (-1)^{\mu \cdot v_0(\alpha,\beta)}
    \nonumber \\
    &\sum_{\mathbf{t}} 
    p_{\mathbf{t}} \, 
    s_{\mathbf{t},\alpha,\beta} \,
            Q_{L,(\mathbf{t},l)}
            \Psi
            Z_L^{v_0(\alpha,\beta)}
            Q_{L,(\mathbf{t},l)}
            .
\end{align}
While in most QEC experiments it suffices to determine whether the logical state of interest was flipped, here, due to the one-sided action of the $Z_L^{v_0(\alpha,\beta)}$ operator, the exact resulting logical operator $Q_L$ must be determined.
Fortunately, this can be done by simply applying each trajectory to both initial states $\ket+_L, \ket0_L$.
Any other global phase in $Q_{L,(\mathbf{t},i)}$ cancels out.

Finally, it will be useful for its numerical evaluation, to express the final state in terms of probabilities for different outcomes, rather than a sum over trajectories.
First, we introduce $p_{\eta}(\alpha,\beta)$ as the fraction of accepted trajectories with commutation parity $\eta$ over all attempted trajectories.
More precisely, $p_{\eta}(\alpha,\beta)$ for $\eta=0\,(1)$ is the unnormalised probability that a trajectory passed post-selection given the error $P_a$ before the transversal rotations, such that $P_a$ (anti)commutes with the difference between $P_\beta$ and $P_\alpha$, i.e. $Z_L^{v_0(\alpha,\beta)} \prod_k S_k^{v_k(\alpha,\beta)}$.
As such, $p_{\eta}(\alpha,\beta)$ encodes the probability distribution over trajectories of the $\pm1$ sign of $s_{\mathrm{\mathbf{t}},\alpha,\beta}$.
These probabilities are unnormalised to track the loss of norm due to post-selection.
Second, we define $p_{Q,\alpha,\eta}$ as the probability of having a logical flip $Q_L$ conditioned on $P_\alpha$ and $\eta$.
This yields
\begin{align}
    \rho_L^{(\mu)} (\Psi) = 
    &\sum_{(\alpha,\beta)_L }
    \mathcal{R}_{\alpha\beta} i^{v_p(\alpha,\beta)}
    (-1)^{\mu \cdot v_0(\alpha,\beta)}
    \nonumber \\
    &\;\sum_{\eta\in {0,1}}
    {(-1)^\eta\,p_{\eta}(\alpha,\beta)}
    \nonumber \\
    &\;\;\sum_{Q} 
    p_{Q,\alpha,\eta} \,
            Q_L
            \Psi
            Z_L^{v_0(\alpha,\beta)}
            Q_L
            .
\end{align}

Finally, the range of the outermost sum can be further reduced by considering a subset of Pauli operators $\mathcal{G}$, such that all $P_\alpha$ can be generated from them, the logical $Z_L$ operator and $S_k$ stabilizers. 
Then, following similar steps as before, 
\begin{align}
    \rho_L^{(\mu)} (\Psi) = 
    &\sum_{g\in\mathcal{G}}
    \;\sum_{\alpha,\beta \,\equiv_L g}
    \mathcal{R}_{\alpha\beta} i^{v_p(\alpha,\beta)}
    (-1)^{\mu \cdot v_0(\alpha,\beta)} 
    \nonumber \\
    &\;\sum_{\eta\in {0,1}}
    {(-1)^\eta\,p_{\eta}(\alpha,\beta)}
    \nonumber \\
    &\;\;\sum_{Q} 
    p_{Q,g,\eta} \,
            Q_L
            Z_L^{v_0(\alpha,g)}
            \Psi
            Z_L^{v_0(\beta,g)}
            Q_L
            .
    \label{eq:app:tmr-performance-over-generators-computation}
\end{align}

By computing this expressions for a suitable selection of input states, one obtains not only the logical infidelity, but an accurate description of the whole logical channel.
By repeated application of the TMR channel, with, at each step, suitable measurement outcome $\mu$ and angle-dependent matrix $\mathcal{R}_{\alpha\beta}(\theta_L)$, the channel for the whole RUS procedure can be estimated.

\subsubsection{Clifford Correction for even \texorpdfstring{$r$}{r}}
\label{subsec:app:TMR-noise-channel-clifford-correction}

As described in~\cref{subsec:partial-ft-design} and depicted in~\cref{fig:sub:teleportation}, if the TMR involves an even number of transversal physical rotations, then the gate $S_{X_L}^\dagger = \exp(i \pi X_L / 4)$ is to be applied on the resource state before teleportation measurement.
In this brief subsection, we highlight how to incorporate its action on the final state, as well as to propagate its noise.

To incorporate its action on the resource state $\mathfrak{M}_L$ described in~\cref{eq:app:rho2-tmr-modeling}, note that it imparts a phase $e^{i\pi/4}$ on the left side, while on the right side its action depends on $v_0(\alpha,\beta)$, as $X_L$ and $Z_L$ anti-commute: it imparts a $(\alpha,\beta)$-dependent phase $e^{-i\pi/4} (e^{+i\pi/4})$ for $v_0(\alpha,\beta)=0\,(1)$.
This phase then propagates onto the final logical state.

To incorporate its infidelity, we first note that it is Clifford and has a transversal implementation up to a qubit permutation~\cite{chen2024_transversalS}, hence its infidelity falls exponentially with the code distance.  
Nevertheless, errors can occur and propagate : specifically, $X_L$ and $Y_L$ flips modify the teleportation-measurement outcome.
Denoting their probability $p_{{\rm flip}_L}^{(S_X)}$, the final result $\rho_L^{(\mu)}(\ket{\sigma_L})$ of a gate teleportation attempt with measurement $\mu$ on the target logical qubit $\Psi$, effectively is
\begin{align}
    \rho_L^{(\mu)}\left(\Psi;p_{{\rm flip}_L}^{(S_X)}\right) =
    \left(1-p_{{\rm flip}_L}^{(S_X)}\right) &\,\rho_L^{(\mu)}(\Psi;0) 
    \nonumber \\
    +\;p_{{\rm flip}_L}^{(S_X)} &\,\rho_L^{(\mu \oplus 1)}(\Psi; 0).
\end{align}
We numerically computed under the assumed noise model \mbox{$p_{{\rm flip}_L}^{(S_X)} \simeq 0.026 (p/0.009)^{\lceil d/2 \rceil }/(d-1)$}, as part of the Clifford simulations described in~\cref{subsec:app:logical-cliffords-simulation}.
%clf data here

\subsection{Numerical estimation of TMR performance}
\label{subsec:app:TMR-noise-channel-numerical-estimatation}

The previous subsections derived the quantities required for numerical reconstruction.
In this subsection, we explicitly explain how the relevant error parameters are accessed by means of Clifford simulations performed in \texttt{stim}~\cite{stim}.
As evident from~\cref{eq:app:tmr-performance-over-generators-computation}, an accurate estimate of the logical errors in the TMR protocol requires more information than usually extracted from Clifford simulations.
This can be obtained by running the same noise realisation for multiple initial states, in turn made possible by manually inserting errors in the circuit according to their probability distribution---at the price of slowing down significantly the simulations' speed.

The outermost sum is dealt with by simulating multiple noisy trajectories for each generating Pauli $g\in\mathcal{G}$, in place of the transversal rotation. 
To compute $p_{\eta}(\alpha,\beta)$, the initialization of the resource state is simulated with the same errors for both $\ket{0}_L$ and $\ket{+}_L$---differing by a layer of $H$ gates---allowing the measurement of $Z_L$ and $X$-type stabilizers respectively.
Their flip indicates anti-commutation of $P_\mathbf{a}$ and said operators, allowing to compute $s_{\mathrm{\mathbf{t}},\alpha,\beta}$ for each trajectory, hence $p_{\eta}(\alpha,\beta)$.
Then, the same trajectory is extended with the following operations, and possible additional errors, to compute the probability distribution $p_{Q,g,\eta}$ of logical flips, conditioned on $P_g$ and $\eta$.
To exactly individuate what logical error occurred in each trajectory, this was simulated for the target logical qubit in both $\ket{0}_L$ and $\ket{+}_L$.

Finally, we would like to point out how each noise source discussed in~\cref{sec:app:error-bugget-deriv} is reflected in~\cref{eq:app:tmr-performance-over-generators-computation}: (i) logical flips from syndrome extraction are represented by $p_{Q,g,\eta}$; (ii) noise from repeating flips is encoded in $p_{\eta}(\alpha,\beta)$, and (iii) $\mathcal{R}_{\alpha\beta}$ encodes noise in the transversal rotations themselves, with dominant error contributions being variation in the correlated phase-flips terms with respect to the noiseless scenario.

\begin{figure*}[t]
    \centering
    \includegraphics[width=\textwidth]{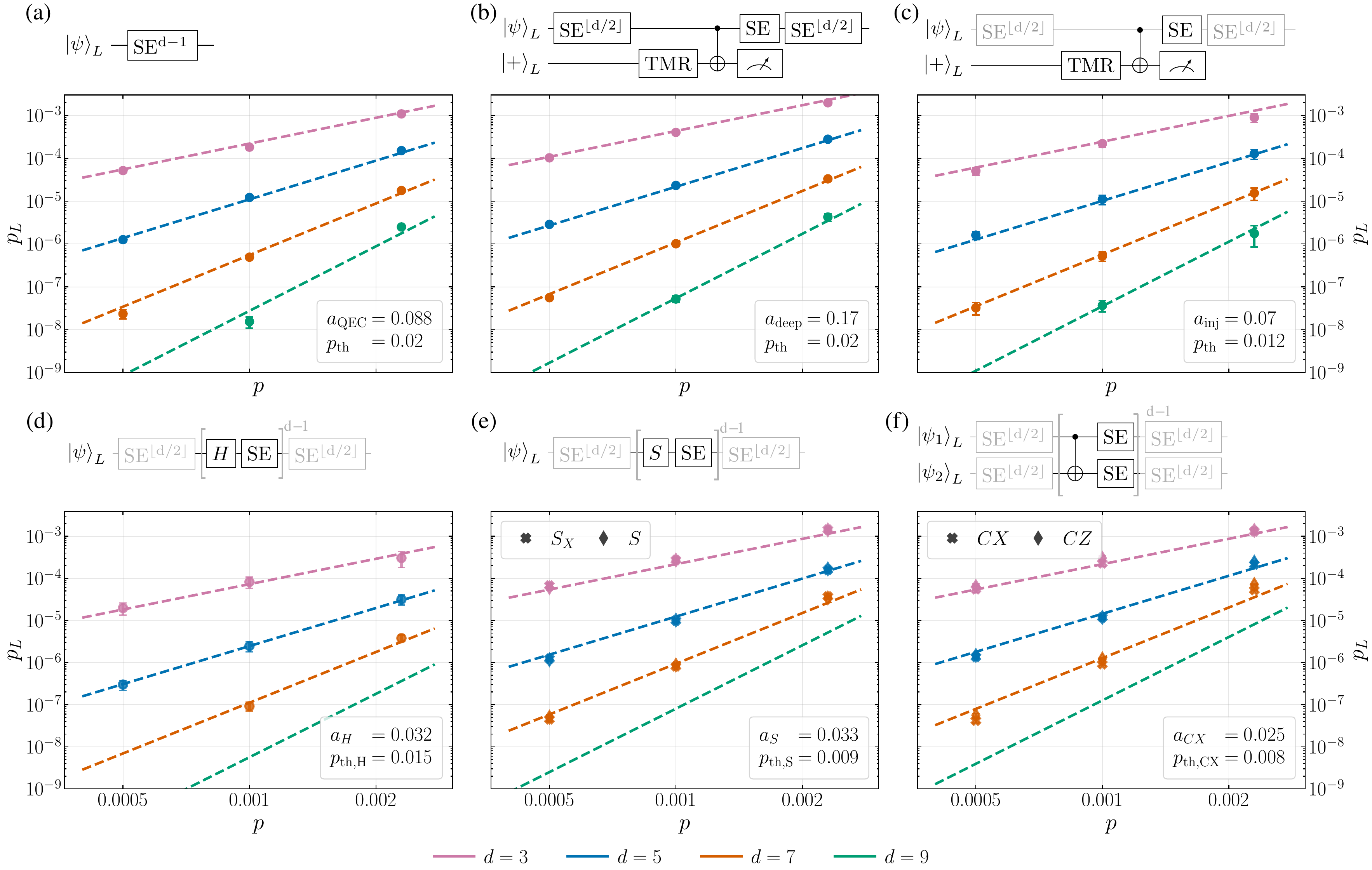}
    \phantomsubfloat{\label{fig:app:pL-qec-dmin1-rounds}}
    \phantomsubfloat{\label{fig:app:pL-deep}}
    \phantomsubfloat{\label{fig:app:pL-reconstructed-tmr}}
    \phantomsubfloat{\label{fig:app:pL-H}}
    \phantomsubfloat{\label{fig:app:pL-S}}
    \phantomsubfloat{\label{fig:app:pL-CX}}
    \vspace{-2em}
    \caption{
        Logical error rate of Clifford operations against physical error rate $p$, for code distances $3,5,7$ and $9$.
        In each panel, the diagram above the plot displays the simulated circuit, including both the components whose error rate is measured (black), and those whose contribution is deconvolved away (light grey).
        Fitted thresholds and noise prefactors are shown, with the logical error rate of isolated gadgets (c-f) further normalised by a factor $d-1$.
        Errorbars denote sampling standard error.
        (a) $d-1$ rounds of syndrome extraction.
        (b) TMR gadget in between $d-1$ rounds of syndrome extraction.
        (c) Isolated TMR gadget, i.e. TMR encoding, teleportation, and one round of extraction.
        (d) Isolated logical $H$ gadget.
        (e) Isolated logical $S$ or $S_X$ gadget.
        (f) Isolated logical $CX$ or $CZ$ gadget.
        %clf data here
    }
    \label{fig:app:pLs-clifford}
\end{figure*}

\subsection{TMR logical performance in deep circuit}
\label{subsec:app:TMR-noise-channel-deep-circuit-estimation}

Evaluating the protocol's performance, specifically the $p_{Q,g,\eta}$ rates, in practical scenario requires embedding it within multiple QEC rounds.
First, the logical TMR gadget (comprising gate teleportation and a single syndrome-extraction round) is simulated in a deeper circuit, as preceded and followed by $\lfloor d/2\rfloor$ noisy QEC rounds. 
To isolate the noise contribution from the TMR gadget alone, simulations of the QEC memory are also performed. 

To accurately parameterize~\cref{eq:app:tmr-performance-over-generators-computation}, we must extract the isolated TMR error probabilities, $p_{Q,g,\eta}$, from the total logical probabilities obtained in the deep circuit simulations, which we denote as $p^{(\mathrm{deep})}_{Q,g,\eta}$. 
As the total deep circuit noise is the composition of the TMR gadget channel and the QEC memory channel, their contributions can be separated working with the Pauli Transfer Matrix (PTM) formalism. 
For any stochastic Pauli channel with error distribution $p_Q$, the PTM eigenvalues $\lambda_P$ are
\begin{equation}
    \lambda_P = \sum_{Q \in \mathcal{P}} (-1)^{c_{P,Q}} \, p_Q,
\end{equation}
where $\mathcal{P} = \{I, X, Y, Z\}$ and $c_{P,Q} = 0\,(1)$ if $P,Q$ (anti)commute.
Their composition in the deep circuit corresponds to the multiplication of their respective eigenvalues
$\lambda^{(\mathrm{deep})}_{P,l,\eta} = \lambda^{(\mathrm{TMR})}_{P,l,\eta} \lambda^{(\mathrm{QEC})}_P$. 
We can therefore deconvolve~\cite{mangini2022_noisedeconv} the channels and isolate the eigenvalues of the TMR protocol simply as
\begin{equation}
    \lambda^{(\mathrm{TMR})}_{P,l,\eta} = \frac{\sum_{R \in \mathcal{P}} (-1)^{c_{P,R}} \, p^{(\mathrm{deep})}_{R,l,\eta}}{\sum_{R \in \mathcal{P}} (-1)^{c_{P,R}} \, p^{(\mathrm{QEC})}_R}.
\end{equation}
Finally, the isolated TMR logical error probabilities $p_{Q,\alpha,\eta}$ to be inserted into~\cref{eq:app:tmr-performance-over-generators-computation} are recovered by applying the inverse transform to the isolated eigenvalues
\begin{align}
    p_{Q,\alpha,\eta} 
    &= \frac{1}{4} \sum_{P \in \mathcal{P}} (-1)^{c_{P,Q}} \, \lambda^{(\mathrm{TMR})}_{P,l,\eta}
    \nonumber \\
    &= \frac{1}{4} \sum_{P \in \mathcal{P}} (-1)^{c_{P,Q}} \,  
    \frac{\sum_{R \in \mathcal{P}} (-1)^{c_{P,R}} \, p^{(\mathrm{deep})}_{R,l,\eta}}{\sum_{R \in \mathcal{P}} (-1)^{c_{P,R}} \, p^{(\mathrm{QEC})}_R}.
\end{align}
Due to statistical deviations, the procedure occasionally results in negative, albeit very small, error rates.
We therefore project the reconstructed Pauli distribution onto the probability simplex, i.e. assumed the closest valid Pauli distribution in Euclidean distance.
minimizing its euclidean distance from the unconstrained estimate using the sampled covariance matrix.

Quantum memory error rates were determined by simulations of logical flips on states $\ket{0}_L$ and $\ket{+}_L$, as customary in studies of quantum memories and Clifford gates.
This it motivated by the assumption of a negligible $Y_L$ error rate---as is usually the case for the surface code, and further motivated by our $Q_L$-resolving TMR simulations.
Results are displayed in~\cref{fig:app:pL-qec-dmin1-rounds}. 
The memory error rate over $d-1$ rounds fits the standard ansatz
\begin{equation}
    p_L^{(\rm QEC)}(p,d) = a_{\rm QEC} \left( \frac{p}{p_{\rm th}}\right)^{\ceil{d/2}},
\end{equation}
with $a_{\rm QEC}\simeq0.088$ and $p_{\rm th}\simeq0.02$.
%clf data here
In the regime considered, the error rate per extraction round at code distance $d$ can be computed as simply $p_L^{(\rm QEC)}(p,d)/(d-1)$.

Moreover, additional simulations were performed to probe logical error rates in the TMR protocol at small rotation angle $|\theta_L|$.
In fact, the effect of the noise sources specific to the TMR protocol decreases with the angle $\theta_L$ of rotation---as outlined in~\cref{subsec:partial-ft-error-budget} and derived in~\cref{subsec:app:error-budget:noise-from-repeat-flips,subsec:app:error-budget:noise-from-correl-phase-flips}.
Then, the main error contribution is simply noise from syndrome extraction and transversal gates.
To accurately probe their behavior at low error rates and high code distance, many more trajectories are needed than can be simulated with the resource-intensive procedure outlined in~\cref{subsec:app:TMR-noise-channel-numerical-estimatation}.
Therefore, additional simulations were performed to simply compute the probabilities of logical flips on states $\ket{0}_L$ and $\ket{+}_L$, for the deep circuit comprised of the full TMR gadget within rounds of syndrome extraction.
Results are displayed in~\cref{fig:app:pL-deep}. 
Combining simulations of the deep circuit and the memory experiment, the contribution of the TMR gadget was isolated.
This is shown in~\cref{fig:app:pL-reconstructed-tmr}, and fitted by
\begin{equation}
    p_L^{(\rm inj)}(p,d) = \frac{a_{\rm inj}}{d-1} \left( \frac{p}{p_{\rm th}} \right)^{\ceil{d/2}},
\end{equation}
with $a_{\rm inj}\simeq0.070$ and $p_{\rm th}\simeq0.012$.
%clf data here
We note that, considering the chi-matrix decomposition of the transversal rotations in the TMR protocol, $\mathcal{R}^{\otimes r} (\rho) =\sum_{\alpha, \beta} \mathcal{R}_{\alpha\beta} P_\alpha \rho P_\beta$, those simulations were performed only for the term $P_\alpha=I^{\otimes 2r}$, which is overwhelmingly dominating in the context of small rotation angle and error rates.
Moreover, owing to the use of extensive post-selection in the resource state encoding, the failure rate is dominated by the transversal CNOT and following syndrome extraction, thus providing a good estimate of the error rate of a general resource state injection.

\subsection{Logical Clifford performance in deep circuit}
~\label{subsec:app:logical-cliffords-simulation}

We simulated the logical Clifford gates required by the application benchmarked in this manuscript, i.e. the simulation of the Hubbard model (see~\cref{subsec:applications-limitations-logical-error}).
These are $\{H, S, S_X, CX,CZ\}$.
Working in the framework of algorithmic fault tolerance~\cite{zhou2024-algorithmicFT,cain2024,cain2025,serraperalta2025-decodetransversal}, we identify a Clifford gadget as the corresponding gate followed by a single round of syndrome extraction.
We note that the transversal $S$ gate proposed in Ref.~\cite{chen2024_transversalS} and hereby adopted, includes a round of syndrome extraction in its minimal execution, thus bringing the total number of extraction rounds per gadget to two.

As explained for the TMR protocol in the previous subsection, we simulated logical gadgets preceded and followed by multiple rounds of syndrome extraction, and then deconvolved the relevant channel. 
Each Clifford gadget was repeated $d-1$ times to better capture its performance within a deep circuit.
Simulation ran for initial states $\ket{0}_L$ and $\ket{+}_L$ (and product thereof for controlled gates), again under the assumption of a negligible $Y_L$ error rate.

Results are shown in~\cref{fig:app:pL-H,fig:app:pL-S,fig:app:pL-CX} for, respectively, the $H$ gate, the $S$-like gates, and the controlled gates.
The $S$-like gates had equivalent performance, and so did the controlled gates.
Performance of the individual gadget was fit against the ansatz $p_L^{(\bullet)} = a_{\bullet} (p/p_{\rm th,\bullet})^{\lceil d/2 \rceil}/(d-1)$.
Corresponding thresholds and prefactors are provided in the figure and~\cref{tab:gates-logical-error-rates}.

\subsection{Details of numerical simulations}
~\label{subsec:app:TMR-noise-channel-numerical-hyperparams-approxims}

For the estimation of the TMR channel in~\cref{eq:app:tmr-performance-over-generators-computation}, the number of shots was increased with the code distance to maintain good accuracy as logical errors become more rare.
Specifically, for each generating operator $P_g$, simulations included 4, 16, 64 million trajectories at code distances 5,7,9 respectively, or half as many passing post-selection, whichever was reached first.
We also observe that the (noisy) channel of the non-Clifford transversal rotations $\mathcal{R}_{\alpha\beta}$ enters the numerical computation~\cref{eq:app:tmr-performance-over-generators-computation} by means of post-processing, rather than by sampling its noise.
This reduces the required sampling overhead, and allows to use the same simulations for different rotation angles $\theta_L$.
Finally, we note that owing to the structure of $\mathcal{R}_{\alpha\beta}$ and the extensive use of post-selection, multiple $P_g$ yield negligible contributions: their simulations thus had a reduced number of shots, with the statistical uncertainty reflected in the errorbars in~\cref{fig:pFT-results-pL-ApFT}.

For the estimation of logical flips in the memory, deep TMR circuits, and Clifford gates, the number of shots was increased with both the code distance and the inverse of the physical error rate.
Memory simulations used $\min(2.5\cdot 10^3 \cdot (0.02/p)^{\ceil{d/2}}, 2\times10^9)$ shots per logical basis.
Deep TMR simulations proceeded until that many shots were simulated or half as many accepted after post-selection.
As these simulations have much larger statistics than the ones performed to estimate the full TMR channel, their results were used to provide a more accurate estimate of $p_{Q,I^{\otimes 2r},0}$, i.e. of the logical flip rates upon trivial commutation on the statistically-leading term.
Clifford gates' simulations had half as many shots per logical basis as in the memory experiments, owing to their longer runtime, and to their generally higher error rate which demanded fewer statistics. 
%clf data here

Sampling standard errors were estimated using a Dirichlet smoothing model, followed by linear uncertainty propagation.
For each simulated Pauli term, acceptance is modeled as binomial with Jeffreys $\operatorname{Beta}(1/2,1/2)$ prior.
Conditioned on acceptance, the joint distribution of commutation outcomes and logical flips was modeled as multinomial.
Probabilities are then evaluated at the mean of a Dirichlet posterior, with prior concentration $0.8$ uniformly distributed over the observed categories, following the recommendation from Ref.~\cite{berger2015_multinomial,alvares2018_multinomial}.
Finally, the multinomial covariance evaluated at those estimates determines the uncertainties.
For the TMR protocol, these uncertainties, affecting the $p_{\eta}(\alpha,\beta)$ and $p_{Q,\alpha,\eta}$ coefficients entering the logical channel, were propagated through normalization, RUS procedure, and calculation of the logical error rate, using linear propagation implemented by the Python \texttt{uncertainties} package~\cite{lebigot_uncertainties}.
The resulting error bars are reported in \cref{fig:pFT-results-pL-ApFT}.

\section{Details on the simulation of the Fermi-Hubbard model}
\label{sec:app:fermi-hubbard-fh}

The Hubbard model is one of the central paradigms in the study of strongly correlated quantum many-body systems, and  has long served as a benchmark target for quantum simulations~\cite{wecker2015_fh, jiang2018_fh, stanisic2022_fh, flannigan2022_FH, kivlichan2020_FH, arovasetal2022_fh, campbell2022_FH, baySmidt2025_FH, kanSymons2025_FH, akahoshi2024a, toshio2025}. 
Its Hamiltonian $H = H_J + H_U$ is the sum of a hopping (kinetic) term $H_J$ and an on-site potential term $H_U$, defined as
\begin{align}
  H_J/\hbar &= -J\!\sum_{\langle i,j\rangle,\sigma}\!\left(c^{\dagger}_{i\sigma}c_{j\sigma} + \mathrm{\hc}\right),
  \nonumber \\
  H_U/\hbar &= U\sum_{i=1} n_{i\uparrow}n_{i\downarrow},
\end{align}
Here $\langle i,j\rangle$ denotes nearest-neighbour lattice sites and $\sigma\in\{\uparrow,\downarrow\}$ labels the spin orbital.
The system has $M$ lattice sites and therefore $2M$ fermionic modes.

\begin{table*}[!t]
    \centering
    \def\arraystretch{1.4}
    \begin{tabular*}{\textwidth}{@{\extracolsep{\fill}} p{1.9cm}||c|c||c|c||c|c||c}
         & \multicolumn{2}{c||}{$R_{X/Z}$}
         & \multicolumn{2}{c||}{$S_X$}
         & \multicolumn{2}{c||}{$CX/CZ$}
         & \multirow{2}{*}{\makecell{Algorithmic\\ancillae}}
         \\
         \cline{1-7}
         & \,Count\, & \,Depth\,
         & \,Count\, & \,Depth\,
         & \,Count\, & \,Depth\,
         &
         \\
         \hline
         \hline
         $H_{v, {\rm odd}}$
         & $2 L_x L_y$
         & $1$
         & $4 L_x L_y$
         & $2$
         & $8 L_x L_y - 6 L_y$
         & $6 L_x - 4$
         & \makecell{$L_y$}
         \\
         \hline
         $H_U$
         & $L_x L_y$
         & $1$
         & $0$
         & $0$
         & $2 L_x L_y$
         & $2$
         & $0$
         \\
         \hline
         \hline
         $H_h^{(\mathrm{pBCs})}$
         & $4 L_x L_y$
         & $2$
         & $8 L_x L_y$
         & $4$
         & $6 L_x L_y - 6 L_y$
         & $L_x/2 + 3$
         & $0$
         \\
         \hline
         $H_{v, {\rm even}}^{(\mathrm{pBCs})}$
         & $2 L_x L_y$
         & $1$
         & $4 L_x L_y$
         & $2$
         & $8 L_x L_y - 8$
         & $12 L_x + L_y - 6$
         & $L_y$
         \\
         \hline
         Total$^{(\mathrm{pBCs})}$
         & $(14 N {+} 2) L_x L_y$
         & $8 N {+} 1$
         & $(24 N {+} 4) L_x L_y$
         & $12 N {+} 2$
         & \makecell{$(32 N {+} 8) L_x L_y$ \\ $-\, (18 N {+} 6) L_y - 8 N$}
         & \makecell{$(19 L_x {+} L_y) N$ \\ $+\, 6 L_x - 4$}
         & $L_y$
         \\
         \hline
         \hline
         $H_h^{(\mathrm{oBCs})}$
         & $4 (L_x L_y - L_y)$
         & $2$
         & $8 (L_x L_y - L_y)$
         & $4$
         & $4 (L_x L_y - L_y)$
         & $4$
         & $0$
         \\
         \hline
         $H_{v,\rm even}^{(\mathrm{oBCs})}$
         & $2 (L_x L_y - 2 L_x)$
         & $1$
         & $4 (L_x L_y - 2 L_x)$
         & $2$
         & $8 L_x L_y - 16 L_x - 6 L_y + 12$
         & $6 L_x - 4$
         & $L_y$
         \\
         \hline
         Total$^{(\mathrm{oBCs})}$
         & \makecell{$(14 N {+} 2) L_x L_y$ \\ $-\; 4 N (L_x + 2 L_y)$}
         & $8 N {+} 1$
         & \makecell{$(24 N {+} 4) L_x L_y$ \\ $-\; 8 N (L_x + 2 L_y)$}
         & $12 N {+} 2$
         & \makecell{$(28 N {+} 8) L_x L_y - 16 N L_x$ \\ $-\, (20 N {+} 6) L_y + 12 N$}
         & \makecell{$(12 L_x {+} 4) N$ \\ $+\, 6 L_x - 4$}
         & $L_y$
         \\
         \hline
    \end{tabular*}
    \caption{Gate counts and circuit depths for single sweeps of the Hubbard Hamiltonian terms on an $L_x \times L_y$ lattice, with $L_x,L_y\geq 4$, and their total.
    $H_{v,{\rm odd}}$ and $H_U$ are independent of boundary conditions, whereas the other estimates are separated between periodic (pBCs) and open (oBCs). 
    Hopping counts are summed over both spin species, which execute in parallel.
    Gates are grouped into single-qubit rotations $R_{X/Z}$, $S_X$ gates, and two-qubit Clifford gates $CX/CZ$.
    The last column gives the ancillae required for the optimized computation of JW strings, reused throughout the circuit.
    The total rows give the resources for a second-order Trotterised evolution of $N$ steps with the ordering $H_{v,{\rm odd}}/2 \to H_U/2 \to H_h/2 \to H_{v, {\rm even}} \to H_h/2 \to H_U/2 \to H_{v, {\rm odd}}/2$, with consecutive $H_{v, {\rm odd}}/2$ terms merged in between steps.
    \label{tab:fh_resources}
    }
\end{table*}

For digital simulation, the fermionic Hamiltonian is mapped to Pauli operators using the Jordan--Wigner (JW) transformation, chosen because it requires no additional ancillary qubits, oppositely to other mappings.
The on-site potential term is simple. 
For fixed-particle number simulations, after a convenient shift of the chemical potential, which can be accounted for in classical post-processing~\cite{campbell2022_FH, baySmidt2025_FH}, the on-site interaction becomes
\begin{equation}
  H_U/\hbar = \frac{U}{4}\sum_{i} Z_{i\uparrow}Z_{i\downarrow}.
\end{equation}
By contrast, the hopping terms are non-local.
A hopping term between modes $i$ and $j$ maps to an $XX+YY$ rotation dressed by the JW parity string
\begin{equation}
  Z_{\mathrm{JW}}^{(i,j)} = \prod_{i<q<j} Z_q,
\end{equation}
acting on all qubits lying between $i$ and $j$ in the chosen JW ordering.
Explicitly, the kinetic term becomes
\begin{equation}
  H_J/\hbar = -\frac{J}{2}\!\sum_{\langle i,j\rangle,\sigma}\!(
X_{i\sigma}X_{j\sigma} + Y_{i\sigma}Y_{j\sigma}) Z_{\mathrm{JW}}^{(i,j)}.
\end{equation}

In the following, we provide an explicit scheme for the Trotterised simulation of time evolution in the Fermi-Hubbard model, specific to a two-dimensional $L_x \times L_y$ lattice with periodic boundary conditions (BCs). 
The main difficulty is the implementation of the hopping Hamiltonian $H_J$, as the JW strings make geometrically local hopping terms generally non-local in qubit space.
To reduce gate overhead, following Refs.~\cite{hastings2015_FH, flannigan2022_FH}, we use the fact that each JW string only depends on the parity of the qubits lying between the two hopping endpoints.
These parities can be computed in a nested fashion with a small number of ancillae.
Specifically, the parity of the $(i,i+1)$ qubits is first stored on an ancilla; then re-used when calculating that of $(i-1,i+2)$; this is in turn stored and then re-used when calculating that of $(i-2,i+3)$, and so on.
Compared with evaluating every JW string independently, this nested computation substantially reduces the gate overhead.
Moreover, hopping between two endpoints preserves the parity of the qubits in the intermediate string, so the hopping rotations can be commuted through parity-computation layers and executed in parallel.

\begin{figure*}[!p]
    \includegraphics[width=0.99\linewidth]{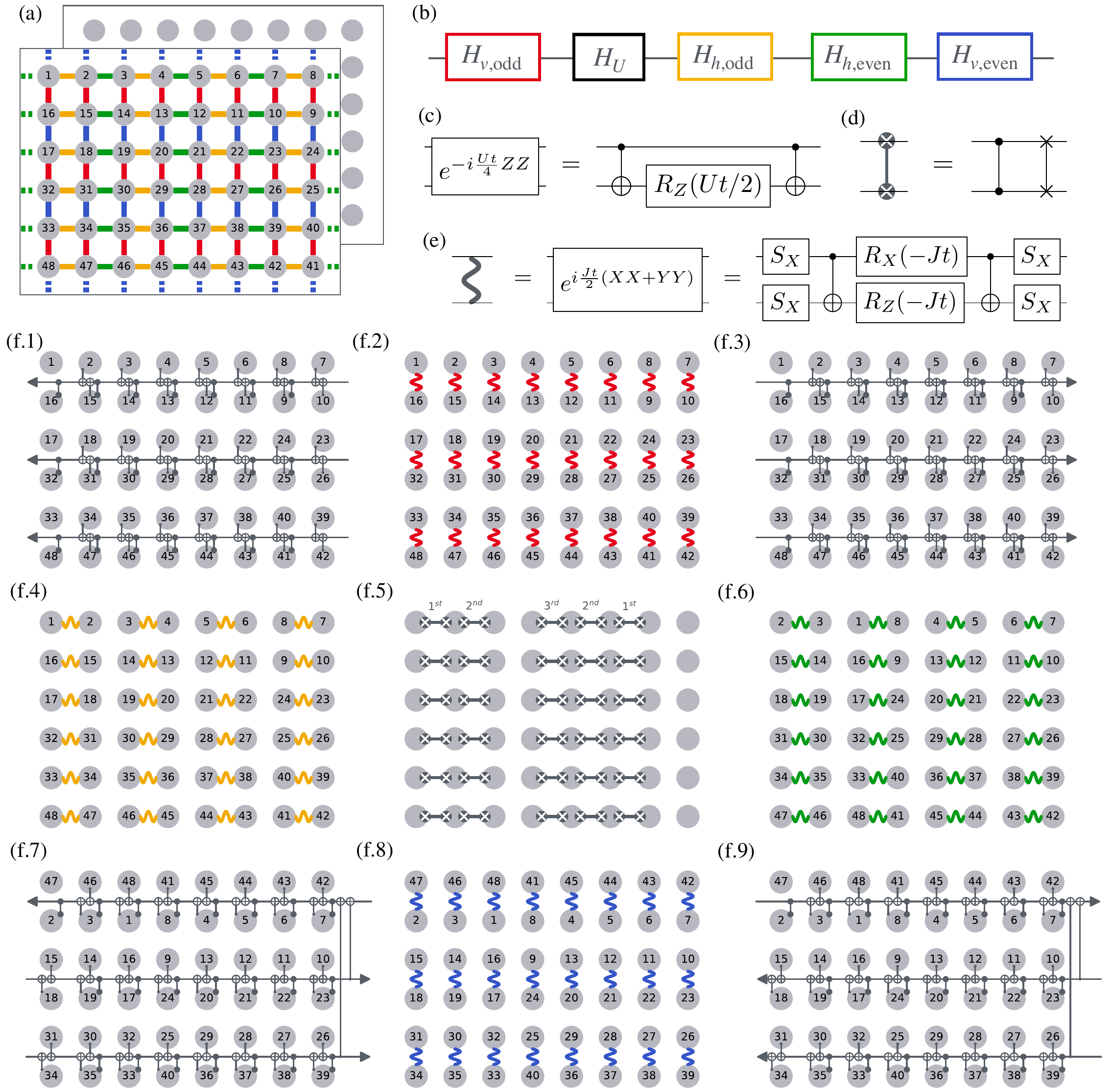}
    \phantomsubfloat{\label{fig:app:fh:model}}
    \phantomsubfloat{\label{fig:app:fh:trotter}}
    \phantomsubfloat{\label{fig:app:fh:Uoperator}}
    \phantomsubfloat{\label{fig:app:fh:fswap-operator}}
    \phantomsubfloat{\label{fig:app:fh:tunnel-operator}}
    \phantomsubfloat{\label{fig:app:fh:f-whole}}
    \vspace*{-1.5em}
    \caption{
        Overview of the digital simulation of the 2D Fermi-Hubbard model with periodic BCs.
        (a) Schematic depiction of the model on a $8\times 6$ lattice: 
        the two planes identify the two spin orbitals;
        the sites' numbering follows the traditional "snake ordering";
        the colors identify sets of mutually commuting hopping terms.
        (b) First-order Trotter decomposition. 
        The same ordering of the terms---with its inverse---is used to construct the higher order decompositions we adopt.
        (c) $ZZ$ rotation.
        (d) Fermionic SWAP.
        (e) Tunneling operator.
        (f) Simulation of the hopping terms. 
        Site labels are those of panel (a).
        For visualization purposes, sites' positions are displayed such that the JW ordering matches an underlying snake-path.
        Initially, this only differs from the sites' indexing by interchanging the two rightmost columns.
        (f.1) Efficient computation of the JW strings for $H_{v,\rm odd}$.
        (f.2) Simulation of $H_{v,\rm odd}$ hopping.
        (f.3) Uncomputation of the JW strings used for $H_{v,\rm odd}$.
        (f.4) Simulation of $H_{h,\rm odd}$, trivially implemented due to a convenient initial choice of JW ordering.
        (f.5) Application of fermionic SWAPs, in the order specified by the labels, mutating the JW ordering.
        (f.6) Simulation of $H_{h,\rm even}$, including hoppings wrapping across the boundaries.
        (f.7) Efficient computation of the JW strings for $H_{v,\rm even}$. 
        The last row is displayed at the top for visualisation purposes, without implying a novel change in the JW ordering.
        Bulk JW parities are computed first.
        Then these are mapped on another ancilla, which is used to further store and update the parity of qubits between the first and last row, to account for the periodic BCs.
        (f.8) Simulation of $H_{v,\rm even}$ hopping.
        (f.9) Uncomputation of the JW strings for $H_{v,\rm even}$.
    }
    \label{fig:app:fh}
\end{figure*}

Using this trick, we extend the scheme proposed in Ref.~\cite{flannigan2022_FH} to periodic BCs, assuming that both $L_x$ and $L_y$ are even.
We decompose the Hamiltonian into five non-commuting parts: the on-site interaction $H_U$ and four hopping Hamiltonians,
\begin{equation}
  H_J = H_{h,\mathrm{even}} + H_{h,\mathrm{odd}} + H_{v,\mathrm{even}} + H_{v,\mathrm{odd}}.
\end{equation}
These four hopping terms are depicted in~\cref{fig:app:fh:model} as, respectively, green, yellow, blue and red bonds.
The even terms include the hopping terms that wrap around the periodic boundaries.

We now assume a second-order Trotter decomposition.
Half the second-order Trotter step---essentially the originating first-order step---is depicted in~\cref{fig:app:fh:trotter}.
Explicitly, the full second-order step is
\begin{widetext}
    \begin{equation}
        V_2(t) = 
        e^{-i H_{v,\mathrm{odd}} \frac{t}{2}}
        e^{-i H_U \frac{t}{2}}\,
        e^{-i H_{h,\mathrm{odd}} \frac{t}{2}}
        e^{-i H_{h,\mathrm{even}} \frac{t}{2}}
        \;e^{-i H_{v,\mathrm{even}} t}\;
        e^{-i H_{h,\mathrm{even}} \frac{t}{2}}
        e^{-i H_{h,\mathrm{odd}} \frac{t}{2}}\,
        e^{-i H_U \frac{t}{2}}
        e^{-i H_{v,\mathrm{odd}} \frac{t}{2}}.
    \end{equation}
\end{widetext}
Over multiple Trotter steps, consecutive half-steps of the $H_{v,\mathrm{odd}}$ term are merged.
This ordering was chosen as it allows for the simulation of the two most costly terms, $H_{v,\mathrm{even}}$ and $H_{v,\mathrm{odd}}$ to occur half as frequently as the others.

\begin{figure*}[!ht]
    \includegraphics[width=\linewidth]{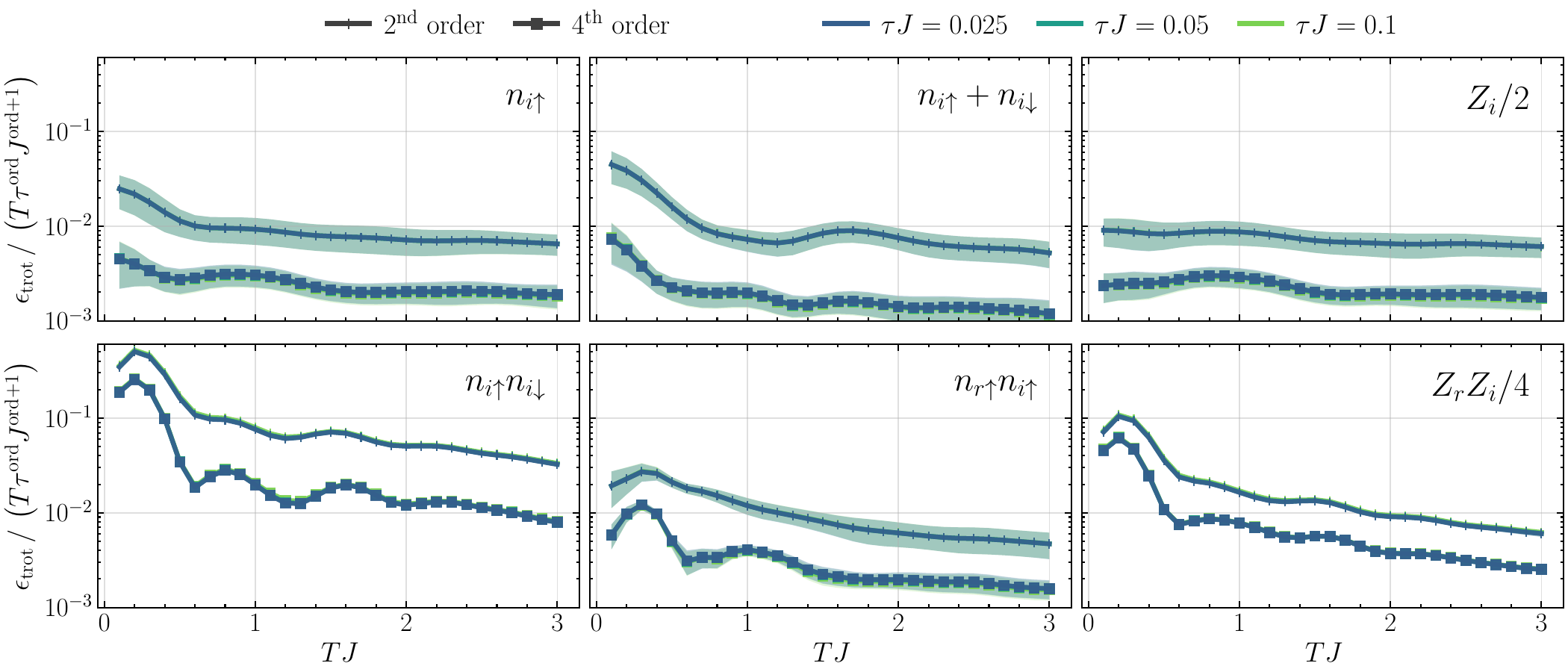}
    \vspace*{-2em}
    \caption{
    Estimates of the Trotter error in the adopted $2^{\rm nd}$ and $4^{\rm th}$ order decompositions, at half-filling and $U=4J$, simulated in a $4\times 4$ lattice.
    Each panel shows for the annotated observable, at different evolution times $T$, the estimated proportionality between the Trotter error $\epsilon_{\rm trot}$ and $T \tau^{\rm ord} J^{\rm ord +1}$, with $\tau$ denoting the step size, and the error defined as per~\cref{app:eq:rms-trotter-error}.
    The curves' collapse confirms the expected scaling $\epsilon_{\rm trot}\propto (\tau J)^{\rm ord}$ at fixed $T$.
    The increasingly flat profile legitimates the scaling $\epsilon_{\rm trot}\propto TJ$ at fixed $\tau$.
    The shaded regions highlight $1\sigma$ standard deviation over 40 initial states, uniformly sampled from the degenerate ground subspace of the hopping Hamiltonian.
    }
    \label{fig:fh-trotter-estimate}
\end{figure*}

The potential term is readily implemented, independently of the ordering, as
\begin{equation}
    e^{-i H_U t} = \bigotimes_{j=1}^M e^{-i \frac{Ut}{4} Z_{j\uparrow} Z_{j \downarrow} },
\end{equation}
with the implementation of each $j^\mathrm{\rm th}$ term shown in~\cref{fig:app:fh:Uoperator}.

The simulation of the hopping terms requires two other building blocks: fermionic SWAPs and tunneling $XX+YY$ rotations, respectively depicted in~\cref{fig:app:fh:fswap-operator,fig:app:fh:tunnel-operator}.
The starting JW ordering follows the pattern shown in~\cref{fig:app:fh:f-whole}, obtained from the typical snake ordering by interchanging the two rightmost sites in each row. 
For each of the two spin orbitals, the simulation of the hopping Hamiltonian proceeds as follows, in one-to-one correspondence with~\cref{fig:app:fh:f-whole}:
\begin{enumerate}[label=(f.\arabic*), leftmargin=*]
  \item The vertical term $H_{v,\mathrm{odd}}$ requires the computation of JW strings' parities, as per the trick used in Refs.~\cite{hastings2015_FH, flannigan2022_FH}.
  These are computed onto $L_y/2$ ancillae---one per pairs of rows---in a nested fashion (CNOTs in the figure).
  Interleaved with the computation of the parities, each $Z_\mathrm{JW}^{(i,j)}$ operator is applied on one of the hopping qubits (CZ in the figure).
  \item As the JW strings have been applied, the corresponding tunneling rotations are simulated.
  \item The JW strings are then uncomputed by reversing the parity-computation circuit of (f.1). 
  This completes the implementation of $H_{v,\mathrm{odd}}$.
  \item $H_{h,\mathrm{odd}}$ currently contains only nearest-neighbor terms. 
  It is therefore straightforwardly simulated by applying the local tunneling rotations.
  \item The JW ordering is then modified by applying fermionic swaps across each row.
  \item In the updated ordering, $H_{h,\mathrm{even}}$ is now local and readily simulated with tunneling rotations.
  \item The nested parity-computation procedure is used for $H_{v,\mathrm{even}}$. 
  Because of the periodic BCs, the JW strings connecting the first and last rows include the bulk string, and are therefore computed after the shorter bulk parities have been obtained.
  \item After these JW strings have been computed, the corresponding tunneling rotations are applied.
  \item Finally, the JW strings are uncomputed by reversing the circuit of (f.7), completing the implementation of $H_{v,\mathrm{even}}$.
\end{enumerate}
We observe that this scheme yields a lower 2-qubit gate count than FSWAP networks, as well as the most recent implementation with JW mapping compiled to an equivalent digital gate set~\cite{alam2025_fermionicgoogle}.
However, it requires $L_y$ ancillae and larger circuit depth.

The resulting per-step gate counts and circuit depths are reported in Table~\ref{tab:fh_resources} for each term in the Hamiltonian as well as the combination over the second-order Trotter decomposition considered. 
The counts and depths are further separated into single-qubit rotations $R_{X/Z}$, $S_X$ gates, and two-qubit Clifford gates $CX/CZ$.
As this scheme adapts straightforwardly to open BCs, estimates are provided for this case as well.
We note that in this scenario the requirement that $L_x$ and $L_y$ be even is relaxed, since it was only needed for an efficient treatment of the hopping terms wrapping around the periodic boundaries.

Finally, we note that the construction extends naturally to higher-order product formulas.
We consider a fourth-order Suzuki decomposition obtained as a symmetric product of five second-order steps,
\begin{equation}
  V_4(t) = V_2(vt)\,V_2(vt)\,V_2\big([1-4v]t\big)\,V_2(vt)\,V_2(vt),
\end{equation}
with $v = 1/(4 - 4^{1/3})$.
As in the second-order case, the half-steps of adjacent second-order steps merge at every interior boundary.
The difference in time step affects only the rotation angles, leaving the gate counts unchanged.
Consequently, the per-step counts and depths are exactly five times those of the second-order scheme reported in Table~\ref{tab:fh_resources}, for both periodic and open BCs.

Finally, we estimate the Trotter error $\epsilon_{\rm trot}^{(\rm ord)}$ of the considered $2^{\rm nd}$ and $4^{\rm th}$ order decompositions.
Specifically, we estimate the proportionality constant for the ansatz $\epsilon^{(\rm ord)}_{\rm trot}\propto N \,(\tau J)^{\rm ord+1}$, where $N$ is the number of Trotter steps and $\tau$ is the step size.
Note, $N \,\tau^{\rm ord+1}=T \,\tau^{\rm ord}$, where $T$ is the total evolution time.
We use the \texttt{ffsim} package~\cite{ffsim} to simulate trotterised and exact time-evolution, and compute the root-mean-square difference 
\begin{equation}
    \epsilon_{\rm trot} = \sqrt{
    \frac{1}{L_x L_y} \sum_{i=1}^{L_xL_y} \left| \langle O_i\rangle_{\rm trot} - \langle O_i\rangle_{\rm exact} \right|^2
    } 
    \label{app:eq:rms-trotter-error}
\end{equation}
on various observables $O$ of typical interest, averaged over the lattice.
These are density $n_{i,\uparrow}$, occupancy $n_{i,\uparrow} + n_{i,\downarrow}$, double occupancy $n_{i,\uparrow} n_{i,\downarrow}$, spin $Z_{i}/2$, density correlation $n_{r,\uparrow} n_{i,\uparrow}$, and spin correlation $Z_{r} Z_{i}/4$ ~\cite{wecker2015_fh,jiang2018_fh,stanisic2022_fh}.
In the latter two, $r$ denotes a reference site in the periodic lattice.
The considered setting is the classically challenging regime characterized by half-filling and $U=4J$.
Simulations were performed at lattice size $4\times4$.
We intend our analysis to be representative of low-temperature behavior, thus we considered initial states uniformly sampled in the span of the degenerate ground subspace of the hopping Hamiltonian $H_J$~\cite{wecker2015_fh,jiang2018_fh,stanisic2022_fh}. 

\Cref{fig:fh-trotter-estimate} displays, separately for each observable, the estimated value of the proportionality constant at different steps of the evolution.
Simulations were performed over evolution times up to $3J^{-1}$, for three different step sizes $\tau J=0.025,\,0.05,\,0.1$, and averaging results over 40 initial states.
On the one hand, the curves' collapse confirms the expected scaling $\epsilon^{\rm (ord)}_{\rm trot} \propto (\tau J)^{\rm ord}$ at fixed $T$.
On the other hand, curves are not perfectly flat, implying a more complex dependence on evolution time than simply linear.
However, an increasingly flat profile is observed with larger times, legitimating the adopted ansatz.
Different observables inevitably provide quantitatively different estimates.
To provide a versatile heuristic description, we set the proportionality constants to $10^{-2}$ and $3\times 10^{-3}$ for $2^{\rm nd}$ and $4^{\rm th}$ order decompositions respectively.

\section{Resources for Execution of Arbitrary Rotations}
\label{sec:app:resources-arbitrary-rotations}

In this appendix, we compare three ways of executing a batch of $b$ logical single-qubit rotations by an arbitrary-angle $\theta_L$: partial fault-tolerance, Clifford+$T$ decomposition with magic state cultivation, and Hamming Weight Phasing (HWP).
Throughout this subsection, resources are quoted for one batch; in an algorithm containing $N_R$ such rotations, the corresponding counts should be multiplied by $N_R/b$, up to integer roundings.
The total number of ancillary patches available is denoted $Q_{\rm syn}$, as in~\cref{subsec:applications-limitations-hardwarep-size}. 
We focus on the cost of both Clifford and non-Clifford operations, as both have a significant impact in near term quantum computation.

We further assume $\theta_L$ to be a dyadic fraction of $\pi/2$ of the form $\theta_L=(2k+1)\pi/2^{C+1}$, where $k$ and $C$ are integers.
This is a convenient ansatz, as it provides a few practical advantages (as explained later in this section) and is capable of approximating any angle to arbitrary precision.

In the remainder of the section, for each method, the following are determined
\begin{enumerate}
    \item the average number of operations (both Clifford and non);
    \item the required number of ancillae for a maximally parallel implementation, i.e. with the smallest possible circuit depth.
    \item the circuit depth in the maximally parallel implementation, i.e. upon availability of a large enough ancilla reservoir to fully parallelize the encoding of the resource states;
    \item the effect of spacetime tradeoffs, in which the number of ancillae is reduced at the expense of a larger depth.
\end{enumerate}
With regard to circuit depth, analytic estimates will often be only approximate.
Nevertheless, they show excellent afreement with numerical simulations, and provide further insight on the depth's scaling in the regime of interest and on the relevant spacetime tradeoffs.
Results presented in the main text use estimates derived from numerical simulations rather than simplified analytic ansätze.
To convert these counts into a common depth unit, we assume comparable times for the encoding and the injection of resource states.
This assumption is justified in the context of neutral atom implementations, in which measurements and shuttling represent the main time bottleneck.

\subsection{Resource Overhead of Partial Fault tolerance}
\label{app:subsec:resources-pft}

In the partially FT implementation, each logical rotation is executed by preparing a resource state and injecting it through a RUS procedure.
No fully fault-tolerant $T$ gate is required in this model.
At each $s^{\rm th}$ step of the RUS scheme, with $s=1,2,\ldots$ the injected angle is doubled, such that, after $C$ unsuccessful steps, the angle to be injected is an odd multiple of $\pi/2$. 
This corresponds to a Clifford gate, hence it can be implemented deterministically, thus ending the RUS injection.

To determine the gate count, it should be noted that a qubit reaches stage $s$ with probability $2^{-(s-1)}$. 
Therefore, the expected number of injections in a batch of size $b$ at the $s^{\rm th}$ step is $b_s=2^{-(s-1)} b$, and the total over the whole RUS process is
\begin{equation}
    N_{\rm inj}^{\rm pFT}(b,C) =
    \sum_{s=1}^{C} b_s =
    2b\left(1 - 2^{-C}\right).
    \label{eq:pft-total-injections}
\end{equation}

To determine depth and ancillae requirements, first we focus on the maximally parallel implementation of the scheme.
We denote $A_{s}\equiv A(\theta_{L,s})$ the average number of attempts required to successfully encode a resource state at the required angle.
We note that this further depends on the code distance and physical error rate, and is described in~\cref{subsec:simulation-results}.
For a maximally parallel implementation, enough ancillae are required to be able to carry out the $s^{\rm th}$ step's injection while simultaneously preparing resource states for the next step.
The injection requires $b_s$ resource states; the preparation step targets the encoding of $b_{s+1}$, and can re-use the $b_s(A_s-1)$ resource states that failed to be encoded in the previous step.
The total number of ancillae required at step $s$ grows with batch size, so we denote it as $b\,q_s$, where from the previous arguments it follows that
\begin{align}
    b\,q_s &= b_s + \max\!\left[\,b_{s+1} A_{s+1},\, b_s(A_s - 1)\,\right]
\end{align}
such that
\begin{equation}
    q_s = 2^{-(s-1)}\left[
        1 +\max\!\left(\frac{A_{s+1}}{2}, A_s - 1\right)
    \right].
\end{equation}
Then, to be maximally parallel we need an ancilla reservoir of size
\begin{equation}
    Q_{\rm syn}^{\rm pFT,prl}(b) = \lceil b (A_{\rm pFT}+1) \rceil,
    \label{eq:pft-max-ancillae}
\end{equation}
where $A_{\rm pFT} = \max_s q_s -1$ is a batch-independent metric, and can be interpreted as an \textit{effective} number of attempts per injection in the RUS protocol.
In fact, its definition makes it directly comparable to the average number of attempts of T states in magic state cultivation (see next subsection).

The average circuit depth is the expectation value of the step $s_{\rm max}$ at which the RUS process over the batch is terminated..
This can be computed using the tail-sum formula for non-negative integer random variables, 
\begin{align}
    D^{\rm pFT, prl}(b,C)
    =& \mathbb{E}[s_{\rm max}]
    \nonumber \\
    =& \sum_{s=1}^{\infty} 1-[1-2^{-(s-1)}]^b,
    \label{eq:pft-depth-tailsum}
\end{align}
where the sum runs over the probabilities that the RUS process over the whole batch extends to or beyond step $s$.
This expectation value, up to negligible corrections, evaluates to $\log_2 b+1/2 +\gamma_{em}/\ln 2$ where $\gamma_{em}\approx0.577$ is the Euler-Mascheroni constant~\cite{eisenberg2008_Tsynth}.
Once the cutoff $C$ is taken into account, the average depth is approximately
\begin{equation}
    D^{\rm pFT, prl}(b,C) \simeq \min(\log_2 b+1.33, C).
    \label{eq:pft-max-depth}
\end{equation}

Finally, we consider the effects of a spacetime tradeoff in which the size of the ancillae reservoir is reduced by a factor $\mathfrak{r}$, i.e.
\begin{equation}
    Q_{\rm syn}^{\rm pFT}(b;\mathfrak{r}) =
    \left\lceil b (A_{\rm pFT}+1)/\mathfrak{r} \right\rceil.
    \label{eq:pft-reduced-budget}
\end{equation}
In this scenario, each step of the RUS protocol contributes on average depth $\lceil b q_s/Q_{\rm syn}^{\rm pFT}(b;\mathfrak{r}) \rceil \approx \lceil \mathfrak{r}\,q_s/(A_{\rm pFT} +1)\rceil$ instead of unity.
The change in depth then is, on average,
\begin{equation}
    \Delta D^{\rm pFT}(b,C;\mathfrak{r})
    \approx
    \sum_{s=1}^{C}
    \left[ 
        \left\lceil
            \frac{ \mathfrak{r}\,q_s}{A_{\rm pFT}+1}
        \right\rceil
    -1 \right].
    \label{eq:pft-depth-increase-exact-sum}
\end{equation}
Under the assumption, satisfied in practice in \cref{subsec:simulation-results}, that the $A_s$ coefficients vary slowly enough to approximate $\lceil q_s/(A_{\rm pFT}+1) \rceil \sim \lceil 2^{-(s-1)} \rceil $, only the first $s_\mathfrak{r}=\lceil\log_2 \mathfrak{r}\rceil$ steps are impacted by the reduction in ancillae, and we further simplify the terms in the above sum to $\sim \mathfrak{r}\,2^{-(s-1)}$.
Taking into account the presence of the cutoff, we denote $m_{\mathfrak{r}C} = \min(s_\mathfrak{r}, C)$, and estimate
\begin{align}
    \Delta D^{\rm pFT}(b,C;\mathfrak{r})
    &\sim \sum_{s=1}^{m_{\mathfrak{r}C}} 2^{-(s-1)}\mathfrak{r}
    \nonumber \\
    &\sim 2\mathfrak{r}(1-2^{-m_{\mathfrak{r}C}})
    \label{eq:pft-depth-increase-approx}
\end{align}
Therefore, the total depth is approximately
\begin{align}
    D^{\rm pFT}(b,C;\mathfrak{r}) 
    \simeq&\; \min(\log_2 b+1.33, C)
    \nonumber \\
    & + 2\mathfrak{r}(1-2^{-m_{\mathfrak{r}C}}),
\end{align}
in good agreement with exact numerical simulations, as shown in~\cref{fig:FTstuff-pFT-depth}.
Remarkably, the depth is only logarithmic in batch size, and, upon spacetime tradeoffs, it only has a small additive increase.

Finally, we also note that the only Clifford operation that must be accounted for at the logical level is the possible $S$ correction to be applied after the $C^{\rm th}$ step of the RUS process, which can always be applied as one parallel Clifford layer.
Therefore, partial fault tolerance has Clifford depth per parallel batch $\lesssim 1$.

\subsection{Resource Overhead of Clifford+\texorpdfstring{$T$}{T} Decomposition}
\label{app:subsec:resources-tsynthesis}

The fully FT baseline synthesizes each arbitrary rotation directly into a Clifford+$T$ sequence.
Near-term quantum hardware is likely to have limited size, thus we adopt the most efficient ancilla-free synthesis protocol, estimated to require
\begin{equation}
    N_{T}^{\rm Tdec}(\epsilon) \simeq 1.52 \log_2 (\epsilon^{-1}) \; - 0.01
\end{equation}
T gates.
This is the mixed protocol provided in Ref.~\cite{kliuchnikov2023_Tsynth}, in turn improving the original proposals in Refs.~\cite{campbell2017_Tsynth, hastings2017_Tsynth}.

\begin{figure*}[t]
    \centering
    \includegraphics[width=\textwidth]{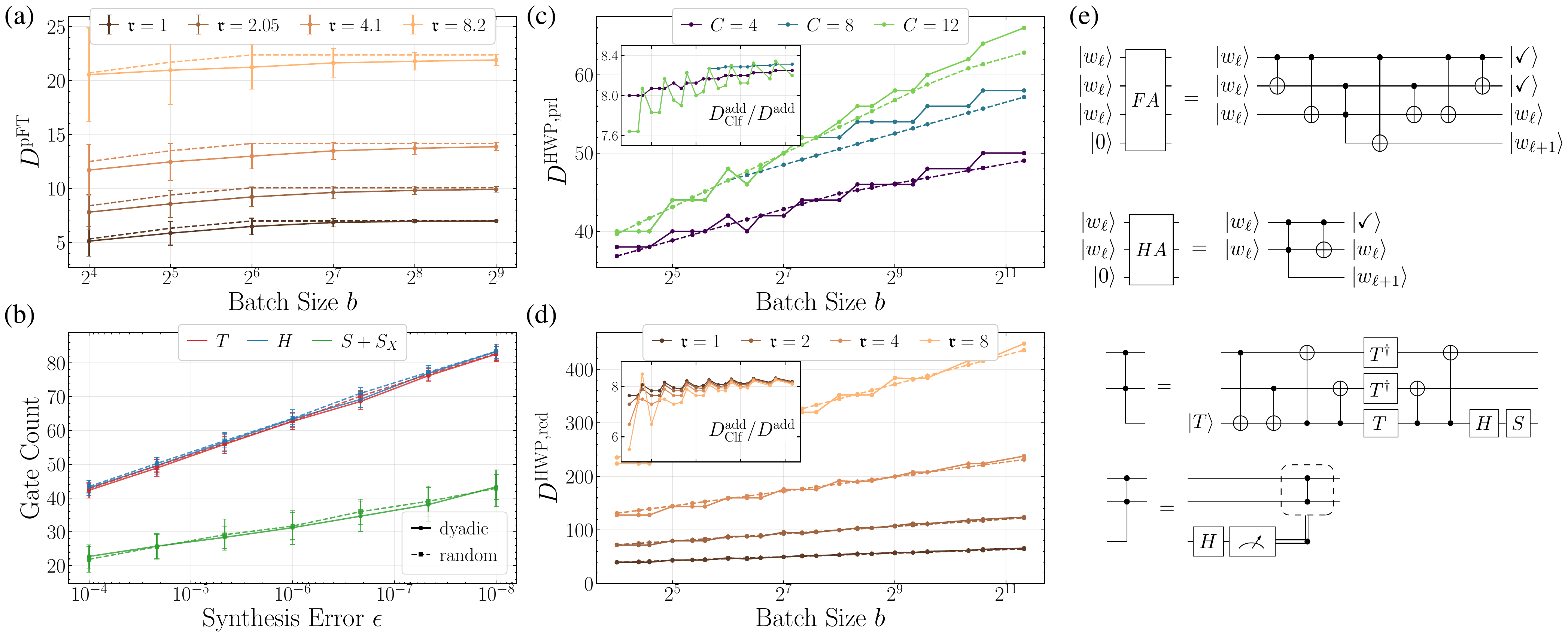}
    \phantomsubfloat{\label{fig:FTstuff-pFT-depth}}
    \phantomsubfloat{\label{fig:FTstuff-Tsynthesis}}
    \phantomsubfloat{\label{fig:FTstuff-HWP-depth}}
    \phantomsubfloat{\label{fig:FTstuff-HWP-reduced-depth}}
    \phantomsubfloat{\label{fig:FTstuff-building-blocks}}
    \vspace{-2em}
    \caption{
        Overview of resource estimation for the synthesis of arbitrary-angle rotations.
        The angles considered are modeled as dyadic fractions of $\pi$, i.e. $\theta_L = (2k+1)\pi/2^{C+1}$ with integer $k$, unless specified otherwise.
        (a) Mean feed-forward depth of the partially FT approach for different batch sizes, at $C=7$. 
        It shows good agreement between exact scheduling (solid) and the analytic approximation in~\cref{eq:pft-depth-increase-approx} (dashed).
        $\mathfrak{r}$ denotes the reduction in the size of the ancilla-reservoir (see~\cref{eq:pft-reduced-budget}), with values chosen to minimize rounding effects given $A_{\rm pFT}\approx1.05$.
        (b) Mean gate counts for Clifford$+T$ synthesis with the Ross-Selinger decomposition.
        Similar estimates are returned whether the target angle is a dyadic fraction of $\pi$ or random.
        Benchmarked for angles in $[0.003, 0.2]$.
        (c) Mean feed-forward depth of maximally-parallel HWP for different batch sizes and cutoffs $C$, assuming $T$-depth 26 for the synthesis of the required rotations.
        It shows good agreement between exact numerical simulations and the heuristic ansatz in~\cref{eq:hwp:depth-heuristic-ansatz}. 
        The inset shows the ratio of the Clifford depth with respect to the feed-forward depth, over the same x-axis range.
        (d) Mean feed-forward depth of reduced HWP for different batch sizes and no cutoff $C\rightarrow \infty$, at different ancillae-reduction factor $\mathfrak{r}$. 
        It shows good agreement between exact numerical simulations (solid) and the analytic ansatz in~\cref{eq:hwp-reduced-batch-depth-approx} (dashed).
        The inset shows the ratio of the Clifford depth with respect to the feed-forward depth, over the same x-axis range.
        (e) Building blocks of HWP. 
        $FA$($HA$) denotes a full(half)-adder, taking three(two) unprocessed input qubits of weight $w_{\ell}$ and returning two(one) processed qubits, one unprocessed qubit of weight $w_{\ell}$ and one of weight $w_{\ell+1}$. 
        The final CNOT of the full-adder can be neglected.
        Below, the logical-AND and the inverse-AND gadgets are depicted, to be used in the adders and in their inverse, respectively.
    }
    \label{fig:FTstuff}
\end{figure*}

The corresponding Clifford overhead is determined by the Matsumoto--Amano normal form of the Clifford+$T$ decomposition~\cite{matsumoto2008_Tsynth,giles2013_Tsynth}, which is independent of the synthesis technique adopted.
In the normal form, up to at most single boundary operators, each $T$ gate appears in one of two elementary patterns, both containing H gates, and only one of which contains an S gate.
Therefore, it is expected that, on average, 
\begin{equation}
    N_H^{\rm Tdec} \simeq N_T^{\rm Tdec}, \qquad 
    N_S^{\rm Tdec} \simeq N_T^{\rm Tdec}/2.
    \label{eq:standard-tsyn-clifford-to-T-ratio}
\end{equation}
This was numerically verified for Ross--Selinger decompositions generated with \texttt{pygridsynth}~\cite{ross2016-selinger,yamamoto2026_Tsynth} over the precision range relevant here. 
The ratio between S and T count remains the same once the probabilistic $S$-gate correction to $T$-gate teleportation is taken into account and merged with surrounding Clifford operators, as shown in~\cref{fig:FTstuff-Tsynthesis}.
The number of $H$ gates is reduced to $N_T^{\rm Tdec} \,\rm mod\,2=0\,\rm or\, 1$ if $T_X$ and $S_X$ gates are available.
Finally, we note that all Clifford gates can be moved to the end of the synthesis and effectively reduced to a single Clifford gate if $T_Y$ gates are also natively available~\cite{litinski2019}, but this scenario extends beyond the assumptions of the manuscript.

Let $A_{\rm FT}$ be the mean number of attempts required to successfully prepare a $T$ resource state.
Recent proposals optimized for a neutral-atom noise mode~\cite{vaknin2026_Tsynth} estimate $A_{\rm FT}\simeq 1.3$ for relevant error rates and code distances.
In the maximally parallel scenario, the ancilla reservoir provides $b$ resource states to be injected, while using $A_{\rm FT} b$ logical patches to encode the next layer's resource states, thus has size
\begin{equation}
    Q_{\rm syn}^{\rm Tdec,prl}(b) = \lceil b(A_{\rm FT}+1) \rceil.
    \label{eq:standard-tsyn-max-ancillae}
\end{equation}
With this footprint, all $b$ rotations are synthesized in parallel, yielding non-Clifford depth
\begin{equation}
    D^{\rm Tdec, prl}(\epsilon) = N_T^{\rm Tdec}(\epsilon).
    \label{eq:standard-tsyn-depth}
\end{equation}

As all qubits in the batch require the same amount of injection, reducing the side-patch budget by a factor $\mathfrak{r}\geq 1$ yields a multiplicative increase in depth.
Thus, the spacetime tradeoff is
\begin{align}
    Q_{\rm syn}^{\rm Tdec}(b;\mathfrak{r}) &=
    \left\lceil \frac{b(A_{\rm FT}+1)}{\mathfrak{r}} \right\rceil, 
    \nonumber \\
    D^{\rm Tdec}(\epsilon;\mathfrak{r}) &=
    \left\lceil \mathfrak{r} N_T^{\rm Tdec}(\epsilon) \right\rceil.
    \label{eq:standard-tsyn-tradeoff-x}
\end{align}
The Clifford depth is proportional to the T-depth as specified by~\cref{eq:standard-tsyn-clifford-to-T-ratio}.

\subsection{Resource Overhead of Hamming Weight Phasing}
\label{app:subsec:resources-hwp}

Hamming Weight Phasing (HWP) is a technique introduced to reduce the number of synthesized arbitrary rotations, applicable when equiangular rotations are applied in parallel~\cite{gidney2018_Tsynth, kivlichan2020_FH}.
Instead of synthesizing $b$ independent rotations, one computes the Hamming weight of the batch, applies rotations to the Hamming-weight register of size $m_{Cb}$, and then uncomputes the register.
For rotation angles that are dyadic fractions of $\pi$, only the first $m_{Cb} = \min\!\left(C - 1,\, \lfloor \log_2 b \rfloor + 1\right)$ digits of the Hamming weight are to be computed, as no advantage is obtained from computing the next ones.
Moreover, it will be useful to define $\sigma_2(b,C)$ as the Hamming weight of the lowest $m_{Cb}$ bits of $b$ in binary form.

The reduction in synthesis cost comes at the expense of additional Clifford and T gates required to compute and uncompute the Hamming weight.
The building blocks are the so called full- and half-adders, in turn relying on the use of logical-AND and inverse-AND gadgets~\cite{gidney2018_Tsynth}.
These are depicted in~\cref{fig:FTstuff-building-blocks} as from Ref.~\cite{kanSymons2025_FH}.
Essentially, each qubit is assigned a weight, and can be "processed" or "unprocessed".
Initially, all qubits have weight 1 and are unprocessed, i.e., they have not yet contributed to the computation of the Hamming weight.
A full(half)-adder introduces an ancillary qubit, consumes four T states, and takes three(two) unprocessed qubits of the same weight; then outputs two(one) processed qubits of same weight, one unprocessed qubit of same weight, and one unprocessed qubit of weight twice as large. 
The computation of the Hamming-weight bits terminates when exactly one unprocessed qubit is left for each weight $w_\ell =2^\ell$, with
$\ell=0,\ldots,m_{Cb}-1$, constituting the Hamming-weight register.
Full adders are mostly used, with half adders only required at most once per weight, if two unprocessed qubits of same weight are left.

To estimate non-Clifford counts, the number of full and half adders need to be determined.
Denote $b_{\ell}$, $f_\ell$, $h_\ell$, respectively, the number of qubits, full-adders and half-adders of weight $w_\ell$ to be processed.
Moreover, note that, as each adder produces one higher-weight qubit, $b_{\ell+1}=f_\ell + h_\ell$.
As a full adder effectively removes two qubits of the given weight to account for, and is applied as long as more than two weight $w_\ell$ qubits are present: $b_{\ell+1}=\lfloor b_\ell / 2 \rfloor=\lfloor b / 2^{\ell+1} \rfloor$, and a half-adder is only required if $b_\ell$ is even.
Since $\left\lfloor \cdot/2^\ell \right\rfloor$ is a binary shift operator---such that the binary representation of $b_\ell$ is that of $b$ shifted by $\ell$ places---it is convenient to shift to a binary description where $b=\sum_{\ell=0}^{\lfloor \log_2 b\rfloor} \beta_\ell 2^\ell $.
It immediately follows that $h_\ell=1\oplus \beta_\ell$, hence
\begin{align}
    h_{\rm tot}(b,C)&=\sum_{\ell=0}^{m_{Cb}-1} h_\ell = m_{Cb}-\sigma_2(b,C),
    \nonumber \\
    f_{\rm tot}(b,C)&=\sum_{\ell=0}^{m_{Cb}-1} b_{\ell+1}-h_\ell = b- m_{Cb}-\lfloor b 2^{-m_{Cb}} \rfloor,
\end{align}
where evaluation of the latter used the 2-adic valuation of Legendre's formula~\cite{beyerstedt2011_LegendreFormula}.
Then, the total number of adders is
\begin{align}
    N_{\rm add}^{\rm HWP}(b,C) = b-\sigma_2(b,C)-\left\lfloor b2^{-m_{Cb}}\right\rfloor,
\end{align}
exactly matching previous estimates in the absence of the cutoff $C$~\cite{campbell2022_FH}.

To determine the gate count, we note that each full- or half-adder consumes four $T$ states, while the measurement-based inverse adders used during uncomputation are $T$-free.
The $m_{Cb}$ rotations applied to the Hamming-weight register must also be synthesized, while the $\left\lfloor b 2^{-m_{Cb}}\right\rfloor$ residual
carry qubits at the cutoff weight are handled by individual $T$ gates.
The total $T$ count for one batch is therefore
\begin{equation}
    N_T^{\rm HWP}(b,C,\epsilon)
    =
    4N_{\rm add}^{\rm HWP}(b,C)
    +
    \left\lfloor \frac{b}{2^{m_{Cb}}} \right\rfloor
    +
    m_{Cb}N_T^{\rm Tdec}(\epsilon).
    \label{eq:hwp-total-T}
\end{equation}
The Clifford count also follows from the total number of adders used in the Hamming weight computation, and their inverse in the subsequent uncomputation.
We do not write this count explicitly, since it is directly obtained from $f_{\rm tot}$, $h_{\rm tot}$, and the adder gadgets.
The synthesized rotations on the Hamming-weight register add the same average Clifford overhead discussed in~\cref{app:subsec:resources-tsynthesis}.

To determine the spacetime cost, we first consider the maximally parallel implementation.
In the Hamming-weight computation, each adder requires one ancilla, which cannot be re-used once the adder has been implemented, and the simultaneous production and injection of $T$ states, in turn requiring an injection reservoir.
Thus, the overall ancillae requirement is the sum of separate terms
\begin{equation}
    Q_{\rm syn}^{\rm HWP,prl}(b,C,A_{\rm FT}) =
    Q_{\rm add}^{\rm HWP}(b,C) +
    Q_{\rm inj}^{\rm HWP,prl}(b,C,A_{\rm FT}).
    \label{eq:hwp-total-ancillae}
\end{equation}
with
\begin{align}
    Q_{\rm add}^{\rm HWP}(b,C) &= N_{\rm add}^{\rm HWP}(b,C),
    \nonumber \\
    Q_{\rm inj}^{\rm HWP,prl}(b,C, A_{\rm FT}) &= \left\lceil b\,q_{\rm inj}^{\rm HWP}(b,C,A_{\rm FT}) \right\rceil,
    \label{eq:hwp-max-ancillae}
\end{align}
where $q_{\rm inj}^{\rm HWP}(b,C,A_{\rm FT})$ is the number of injection side patches per input qubit needed to keep the adder tree supplied with $T$ states.
This depends on the number of attempts per kept T states $A_{\rm FT}$ introduced in the previous subsection; 
moreover, oppositely to the equivalent quantities for previous methods, it has a (mild) dependence on batch size.
It will now be determined. 
As each full-adder takes three unprocessed input qubits and returns two, we assume that the number of adders per parallel step is
\begin{equation}
    N_{\rm add,s}^{\rm HWP}(b) \approx 
    \frac{b}{3} \left( \frac{2}{3}\right)^s.
\end{equation}
It should be noted that each parallel step $s$ does not coincide with a weight layer $\ell$: a single step can involve adders processing different weights, and, conversely, a weight might require more than a single step to be processed.
Moreover, the adder's ancillae can be used to produce T gates before their use in the adders. 
Therefore, the ancillae available at step $s$ for T states are---omitting their dependences for compactness,
\begin{equation}
    Q_{\rm inj,s}^{\rm HWP,prl} \approx 
    Q_{\rm inj}^{\rm HWP,prl} + Q_{\rm add}^{\rm HWP} - \sum_{t=0}^{s-1} N_{\rm add,t}^{\rm HWP}
\end{equation}
At all steps, the injection-ancillae reservoir must be large enough to supply T states for the injection as well as encode states for the next step---with the possibility of re-using the $4 (A_{\rm FT}-1) N_{\rm add,s}^{\rm HWP}$ logical patches that failed to encode T states.
This requires
\begin{equation}
    Q_{\rm inj,s}^{\rm HWP,prl} \geq 4  N_{\rm add,s}^{\rm HWP}
    \left[ 1 + \max\left(A_{\rm FT}-1, \frac{2}{3}A_{\rm FT}\right)
    \right]
\end{equation}
at all steps. 
The minimum overhead required thus is
\begin{align}
    q_{\rm inj}^{\rm HWP}(b,C,A_{\rm FT}) \approx&\, 
    \frac{1}{3} 
    \begin{cases}
        4 A_{\rm FT} -3
            &\mathrm{if}\; A_{\rm FT} \geq 3 \\[0.4em]
        \frac{8}{3} A_{\rm FT} + 1,
            & \text{otherwise}
    \end{cases}
    \nonumber \\
    &+ 2^{-m_{Cb}} + \sigma_2(b,C)/b.
\end{align}

The circuit depth is well approximated by analytic closed forms obtained by simple heuristic arguments, yet empirically justified---see~\cref{fig:FTstuff-HWP-depth}.
In the absence of cutoff, the protocol terminates once a number of qubits of order unity is left, and, since every full-adder effectively reduces by $2/3$ the number of unprocessed qubits, the depth is roughly $\log_{3/2}b\approx 1.7 \log_{2}b $.
In the presence of a cutoff much smaller than $\log_2 b$, only the first relatively-few layers matter, and, since they are reduced by $1/3$ at each step, the depth scales as $\log_3 b$.
In an intermediate stage, the depth is well approximated by $\log_2 b$ scaling.
Therefore, the cumulative adders' depth is empirically approximated by the following piecewise ansatz
\begin{equation}
    D^{\rm add,prl}(b,C)\sim
    \begin{cases}
        \log_{3/2} b
        &\mathrm{if}\; \lfloor \log_2 b \rfloor \lesssim  C-2
            \\[0.4em]
        \log_{2} b + \text{const.}
        &\mathrm{else\,if}\; \lfloor \log_2 b \rfloor \lesssim 2C
        \\[0.4em]
        \log_3 b + \text{const.} 
        & \mathrm{otherwise}
        \label{eq:hwp:depth-heuristic-ansatz}
    \end{cases}
\end{equation}
where the constants' values depend on the cutoff and are fit to match the boundaries.
The uncomputation of the Hamming weight uses measurements and feedforward, but no T gates, hence has the maximally parallel adder depth, independently of the availability of non-Clifford resources. 
The phasing stage consists of rotation synthesis by means of Clifford+$T$ decomposition, with depth determined as per~\cref{eq:standard-tsyn-depth}.
Therefore, the maximally parallel HWP depth is
\begin{equation}
    D^{\rm HWP, prl}(b,C,\epsilon)
    =
    2D^{\rm add,prl}(b,C) + D^{\rm Tdec,prl}(\epsilon).
    \label{eq:hwp-max-depth}
\end{equation}

The Clifford depth in the computation and uncomputation processes follows the same adder-tree structure, with the forward and inverse adders contributing a constant number of Clifford layers per step.
This is roughly 8 times the feed-forward depth $D^{\rm add}$, as shown in the inset of~\cref{fig:FTstuff-HWP-depth}, and as expected from the structure of the adders and their inverse, depicted in~\cref{fig:FTstuff-building-blocks}.
The synthesized Hamming-weight rotations contribute the same Clifford depth discussed for T-gate synthesis.

Finally, we consider spacetime tradeoffs.
In the near term, limited hardware size is likely to limit the size of the ancillae reservoir to a significantly reduced size $Q_{\rm red} < b$.
Since the number of ancillae needed for the adders approximately matches the batch size, only a "reduced" version of HWP is possible, where the batch $b$ is accordingly split into balanced chunks $b_j$,
\begin{equation}
    b = \sum_{j=1}^{\mathfrak{r}} b_j,
    \qquad
    b_j \in
    \left\{
        \left\lfloor \frac{b}{\mathfrak{r}} \right\rfloor,\,
        \left\lceil \frac{b}{\mathfrak{r}} \right\rceil
    \right\},
    \label{eq:hwp-balanced-chunks}
\end{equation}
with $\mathfrak{r}$ chosen as the smallest integer such that
\begin{equation}
     \max_j Q_{\rm syn}^{\rm HWP, prl}(b_j,C,A_{\rm FT}) \leq
    Q_{\rm red}.
    \label{eq:hwp-reduced-batch-condition}
\end{equation}
These are executed sequentially, and the peak ancilla count is then the maximum chunk footprint.
The novel gate count and depth are obtained by summing the requirements of the chunks,
\begin{align}
    N_T^{\rm HWP, red}(b,C,\epsilon,\mathfrak{r})
    &=
    \sum_{j=1}^{\mathfrak{r}}
    N_T^{\rm HWP}(b_j,C,\epsilon).
    \label{eq:hwp-reduced-total-T}
    \\
    D^{\rm HWP, red}(b,C,\epsilon,\mathfrak{r}) &=
    \sum_{j=1}^{\mathfrak{r}}
    D^{\rm HWP, \rm prl}(b_j,C,\epsilon).
    \label{eq:hwp-reduced-batch-depth}
\end{align}
To get an intuitive understanding of the spacetime tradeoff, we note that in practical implementations in the near-term, the depth can be generally approximated as
\begin{equation}
    D^{\rm HWP, red}(b,\epsilon,\mathfrak{r}) \sim
    \mathfrak{r}\left[2\log_{3/2}(b/\mathfrak{r}) + D^{\rm Tdec,prl}(\epsilon)\right].
    \label{eq:hwp-reduced-batch-depth-approx}
\end{equation}
In fact, $b_j \sim b/\mathfrak{r} < 2^{C-1}$ is expected, due to the combination of relatively small system sizes and high $\mathfrak{r}$.
The resulting depth for different batch sizes and $\mathfrak{r}$ values is portrayed in~\cref{fig:FTstuff-HWP-reduced-depth}, showing excellent agreement between exact numerical simulations and the analytic ansatz above.
The same chunking logic applying to the Clifford depth.
Thus, the Clifford depth in the adders and their inverse remains roughly 8 times the feed-forward depth $D^{\rm add}$, as shown in the inset of~\cref{fig:FTstuff-HWP-reduced-depth}, and the total Clifford depth is approximately
\begin{equation}
    D^{\rm HWP, red}_{\rm Clf}(b,\epsilon,\mathfrak{r}) \sim
    \mathfrak{r}\left[16\log_{3/2}(b/\mathfrak{r}) + D^{\rm Tdec}(\epsilon)\right].
    \label{eq:hwp-reduced-batch-clifford-depth-approx}
\end{equation}

A "grouped" version of HWP also exists~\cite{kivlichan2020_FH}, where the Hamming weight of the whole batch is computed one sub-batch at a time.
Unlike the reduced HWP, the rotation synthesis is still applied only once, to the Hamming weight of the full batch $b$.
However, this approach significantly increases circuit depth and Clifford usage, as it requires repeating weight computation and uncomputation twice for each sub-batch---more details can be found in Ref.~\cite{kivlichan2020_FH}.
Therefore, it is unlikely to be a viable solution in the near-term, when shorter depth is preferable, and the combination of higher error rate and smaller code distance make noise from Clifford gates relevant.

\bibliography{main}

\end{document}